\documentclass[a4paper,11pt]{article}
\pdfoutput=1 
\usepackage{amssymb}
\usepackage{amsmath}
\usepackage{braket}
\usepackage{mathtools}
\usepackage[usenames,dvipsnames,svgnames,table]{xcolor}
\usepackage [utf8] {inputenc}
\usepackage{color}
\usepackage{cite}
\usepackage{lmodern}
\usepackage{footnote}
\usepackage{slashed}
\usepackage[pdftex] {graphicx}
\usepackage{multirow}
\usepackage{jheppub} 
\usepackage{here}
\usepackage{hyperref}
\usepackage[justification=centering, singlelinecheck=false]{caption}
\usepackage[list=true, labelfont=bf, labelformat=brace, position=top]{subcaption}

\newcommand{\eq}{\begin{eqnarray}}
\newcommand{\en}{\end{eqnarray}}

\newcommand{\cM}{\mathcal{M}}
\newcommand{\cT}{\mathcal{T}}
\newcommand{\cK}{\mathcal{K}}
\newcommand{\cV}{\mathcal{V}}

\newcommand{\LRvec}[1]{\overset{\text{$\leftrightarrow$}}{#1}}

\title{Relativistic $N$-particle energy shift in finite volume}
\author[1]{Fernando Romero-L\'opez}
\affiliation[1]{IFIC, CSIC-Universitat de Val\`encia, 46980 Paterna, Spain}
\emailAdd{fernando.romero@uv.es}

\author[2,3]{, Akaki Rusetsky}
\emailAdd{rusetsky@hiskp.uni-bonn.de}
\affiliation[2]{HISKP and BCTP, Rheinische Friedrich-Wilhelms Universit\"at Bonn, 53115 Bonn, Germany}
\affiliation[3]{Tbilisi State University, 0186 Tbilisi, Georgia}

\author[2]{, Nikolas Schlage}
\emailAdd{schlage@hiskp.uni-bonn.de}
\author[2]{ and Carsten Urbach}
\emailAdd{urbach@hiskp.uni-bonn.de}

\abstract{ 
  We present a general method for deriving the energy shift of an interacting system of
  $N$ spinless particles in a finite volume. To this end, we use the
  nonrelativistic effective field theory (NREFT), and match the pertinent low-energy
  constants to the scattering amplitudes. Relativistic corrections are explicitly
  included up to a given order in the $1/L$ expansion. We apply this method to obtain the ground state of $N$ particles, and
  the first excited state of two and three particles to order $L^{-6}$ in terms of the threshold parameters  of the
  two- and three-particle relativistic scattering amplitudes. 
  We use these expressions to analyze the $N$-particle ground state
energy shift in the complex $\varphi^4$ theory.
}

\allowdisplaybreaks
\begin{document}
\maketitle
\flushbottom

\section{Introduction}
\label{sec:intro}

Recent years have witnessed a substantially increased interest in the study of
many-body dynamics (three particles and more) on the lattice. On the one hand, available computing resources already enable one to
carry out calculations of the three-meson spectrum in QCD
at a reasonable accuracy, so that results of lattice
measurements can be used for the extraction of parameters of
many-body interactions in infinite volume~\cite{Beane:2007es,Horz:2019rrn,Culver:2019vvu,Fischer:2020jzp,Blanton:2019vdk,Hansen:2020otl,Alexandru:2020xqf}. On the other hand, there
has been substantial progress in the development of the framework that enables
one to analyze the lattice data.

 These developments have followed different patterns. First,
 Rayleigh-Schr\"odinger perturbation theory has been used to calculate the
ground-state energy shift of $N$ identical bosons in a finite box of size $L$
in a systematic expansion in powers of $1/L$ (modulo logarithms). This
problem has been attracting attention for decades
already (see, e.g.~\cite{Lee:1957zzb,Huang:1957im,Wu:1959zz}) but
in the context of lattice calculations, L\"uscher~\cite{Luscher:1986n2}
was the first to propose using finite-volume two-body energy levels
for the extraction of scattering lengths. In the three-particle sector,
 perturbative calculations have culminated in Refs.~\cite{Beane:2007qr,Detmold:2008gh}, where the result is given
up-to-and-including order  $L^{-6}$ and $L^{-7}$, respectively (see also
Ref.~\cite{Tan:2007bg}). On the other
hand, starting
from 2012 (see Ref.~\cite{Polejaeva:2012ut}), the exact three-body
quantization condition---not based on the perturbative expansion---has been formulated in three different settings, which can be referred to as relativistic
field theory (RFT)~\cite{Hansen:2014eka,Hansen:2015zga}
non-relativistic effective field theory (NREFT)~\cite{Hammer:2017uqm,Hammer:2017kms}
and finite-volume unitarity (FVU)~\cite{Mai:2017bge} approaches, see
also Ref. \cite{Hansen:2019nir} for a review.\footnote{Numerical studies, closely related to
the NREFT approach, were done earlier (see Refs.~\cite{Kreuzer:2010ti,Kreuzer:2009jp,Kreuzer:2008bi,Kreuzer:2012sr}).} In the following years, extensive exploratory
studies and theoretical extensions have been undertaken, addressing various aspects of all three approaches~\cite{Hansen:2015zta,Hansen:2016fzj,Briceno:2017tce,Sharpe:2017jej,Briceno:2018mlh,Briceno:2018aml,Blanton:2019igq,Briceno:2019muc,Romero-Lopez:2019qrt,Hansen:2020zhy,Blanton:2020jnm,Blanton:2020gha,Doring:2018xxx,Pang:2019dfe,Pang:2020pkl,Jackura:2019bmu,Mikhasenko:2019vhk,Jackura:2018xnx,Mai:2018djl,Mai:2019fba,Jackura:2020bsk,Dawid:2020uhn}. The equivalence of
different approaches has been discussed in Refs.~\cite{Hammer:2017kms,Hansen:2019nir}, and demonstrated for RFT and FVU in Ref.~\cite{Blanton:2020jnm}. It should be especially stressed here that the perturbative expansion of
the quantization condition in powers of $1/L$ reproduces the results of Rayleigh-Schr\"odinger perturbation theory for the ground state~\cite{Hansen:2015zta}.
The expansion of the first excited level can be also obtained in this manner~\cite{Pang:2019dfe}. Finally, alternative approaches to study many-body dynamics
are provided by the HAL QCD method~\cite{Kawai:2017goq,Doi:2011gq},
the optical potential method~\cite{Agadjanov:2016mao}, a method of the spectral density
functions~\cite{Bulava:2019kbi} and the variational
method~\cite{Guo:2018ibd,Guo:2020spn,Guo:2020wbl}. The volume dependence of bound states has also been studied in Ref.~\cite{Konig:2017krd}. Moreover, multi-particle interactions in
lower-dimensional systems have been addressed in
Refs.~\cite{Guo:2017crd,Guo:2018xbv}.

It should be stressed that for the extraction of the many-body interaction parameters
(effective couplings) from the measured energy levels on the lattice, the perturbative
expansion in powers of $1/L$ provides a very convenient tool,
whereas for the study of resonances, an exact quantization condition should be used.

In our opinion, despite the recent progress in the field, there still remain issues
requiring further scrutiny. Namely:
\begin{itemize}
\item[i)] Using the exact quantization condition for the derivation of the perturbative
expression for the energy shift is a rather complicated enterprise. On the contrary, the
Rayleigh-Schr\"odinger perturbative expansion provides a straightforward
path to a final result. However, the inclusion of relativistic corrections, up to now,
has been only performed in the ground state.
Since at order $L^{-6}$ several effects (relativistic corrections, two-body effective-range corrections, three-body force) contribute to the ground-state energy, it would be important to extend this to excited states in order to separate these effects from each other.

\item[ii)] The NREFT three-particle formalism provides a very simple derivation of the quantization condition. Thus, it would be interesting to
systematically study the size of the relativistic effects in finite-volume energy
levels.

\item[iii)]
All the above derivations implicitly assume that the corrections,
exponentially suppressed at large $L$, are very small and can be safely neglected.
In reality, the situation is more nuanced. The three-body coupling, which should be
extracted from data, enters first at order $L^{-6}$, i.e., it is suppressed
by three powers of $L$ with respect to the leading-order contribution. Thus, making $L$
very large will render the extraction more difficult, whereas the exponential corrections
are not sufficiently suppressed for smaller values of $L$. Hence, one has to balance between two opposing trends.
It would be interesting to estimate the order of magnitude of exponential corrections,
in order to establish, whether a window in $L$ exists for which the
extraction can be carried out with a small systematic error.

\item[iv)]
The energy shift of the $N$-particle state (both the ground state and the excited ones)
at $O(L^{-6})$ depends only on three independent dynamical
parameters: the two-body scattering length and the effective range, as well as the
three-body non-derivative coupling constant. Consequently, a global fit to the data
with different $N$ enables one to put more stringent constraints on these parameters
that well result in the substantially improved precision.

\item[v)]
It would be very interesting to test the extraction method in models where the answer
is known from the beginning (e.g., where the perturbation theory is applicable).

\end{itemize}

In the present paper, we address these problems in complex scalar
$\varphi^4$ theory
on the lattice, which has been already used for the similar
purpose in~\cite{Romero-Lopez:2018rcb}. A big advantage of this model
as compared to lattice QCD is that it can be simulated with much less computational effort. It allows one to carry out lattice calculations at many different values of $L$ in the
sectors with different number of particles. Moreover, the model is perturbative
for small values of the coupling constant, so that the values of the parameters, extracted on the lattice, can be confronted with results of direct perturbative calculations.
In the present, follow-up work,
we, in addition to Ref.~\cite{Romero-Lopez:2018rcb}, calculate the ground-state energies
of the 4- and 5-particle states. We also
discuss relativistic effects, exponentially-suppressed polarization
effects, and compare to perturbative results in finite volume. Finally, as the present work was
already in progress, Ref.~\cite{Beane:2020ycc} appeared. There, relativistic corrections were perturbatively included at $O(L^{-6})$ in the case of $N$ identical particles, albeit only for the ground state.

The paper is naturally divided in two parts, containing the derivation of the theoretical
framework (Sections \ref{sec:derivation} and \ref{sec:shifts}), and the analysis of the results of numerical calculations (Section \ref{sec:numerics}). More specifically, in Section~\ref{sec:derivation}, we carry out the construction of the effective non-relativistic
Lagrangian and the matching of its couplings in great detail.
Further, in Section~\ref{sec:shifts}, the evaluation of the consecutive terms in Rayleigh-Schr\"odinger perturbation theory order by order is considered. The energy shift
(both the ground state and the first excited state) is calculated up to $O(L^{-6})$,
and relativistic corrections are taken into account at the same order. The calculations
on the lattice (at present, only ground-state levels in the sectors with different $N$),
as well as the extraction of parameters through the fit to the measured
energy shifts are presented in Section~\ref{sec:numerics}. Finally, Section~\ref{sec:conclu}
contains our conclusions.

\section{Matching in the NREFT Lagrangian}
\label{sec:derivation}

Below, we shall present the derivation of the energy shift to the
$N$-body state of identical spinless particles with mass $m$. The derivation
proceeds in two steps. In the first step, we perform the matching of the two- and
three-body effective Lagrangians, taking into account relativistic corrections.
In the second step, we use these effective Lagrangians to perform the calculation of the
energy shift in the framework of the Rayleigh-Schr\"odinger perturbation theory. It will be
seen that this procedure allows one to get the final result with a surprising ease and
elegance.

In this section, we will focus on the first step of this process. The starting point is the
non-relativistic effective Lagrangian containing all two- and three-particle interactions
that are relevant in the calculations up to order $L^{-6}$. In a general
moving frame, it takes the form:
\eq\label{eq:LNR}
\mathcal{L} &=& \psi^\dagger\left(i \partial^0 - m + \frac{\nabla^2}{2m}
  + \frac{\nabla^4}{8m^3}   \right) \psi - \frac{g_1}{4} (\psi^\dagger \psi)^2
- g_2 (\psi^\dagger \psi^\dagger) \nabla^2(\psi \psi)
\nonumber\\[2mm]
&-& g_3 \bigl( (\psi^\dagger \psi^\dagger)\, (\psi {\LRvec{\nabla}^2}\!\!\psi)
+  \text{ h.c.}\bigr)\, - \frac{\eta_3}{6} (\psi^\dagger \psi)^3 ,
\en
with $a \LRvec{\nabla} b=\frac{1}{2}\,(a\nabla b-b\nabla a)$. Hence,
\eq
 \psi \LRvec{\nabla}^2\psi  = \frac{1}{4} \left( \psi (\nabla^2 \psi) +  (\nabla^2 \psi)  \psi  -2 (\nabla \psi) (\nabla \psi) \right) = \frac{1}{2} \left( \psi (\nabla^2 \psi) - (\nabla \psi) (\nabla \psi) \right).
 \en
As will be seen in Section \ref{subsec:groundN}, terms with higher derivatives,
as well as four- and more particle interactions do not contribute to the
energy shifts at order $L^{-6}$.
 
 In general, the matching of relativistic and non-relativistic theories proceeds as follows.
 Let $\cM_N$ denote the relativistic $N$-particle scattering amplitude. The corresponding non-relativistic amplitude will
 be denoted by $\cT_N$. The matching condition states that
 \eq\label{eq:matching}
 \cT_N =  \prod_{i=1}^N \frac{1}{(2 w'_i)^{1/2}}\,  \cM_N
 \prod_{i=1}^N \frac{1}{(2 w_i )^{1/2}}\, , 
\en
with $w_i=\sqrt{m^2 + {\bf p}_i^2}$ for incoming particles and, similarly,
$w'_i=\sqrt{m^2 + {{\bf p}_i'}^2}$ for outgoing particles. More precisely, it is
understood that both sides of Eq.~(\ref{eq:matching}) are expanded in a Taylor series
in the three-momenta ${\bf p}_i,{\bf p}'_i$ around threshold, and the coefficients of this
expansion are set equal to each other up to a given order in ${\bf p}$. Then, the
matching condition expresses the non-relativistic effective couplings $g_1,g_2,g_3,\ldots$
in terms of the relativistic amplitudes. Note that both relativistic and non-relativistic
amplitudes contain also terms that are not analytic in external momenta
${\bf p}_i,{\bf p}'_i$. However, the structure of such terms is exactly the same in both
theories, and they automatically drop from the matching condition without imposing
additional constraints on the effective couplings. An explicit example of such a
cancellation will be given below. Note also that due to particle number conservation
in the non-relativistic theory, the sectors containing different number of particles do
not talk to each other. For this reason, the matching in the two- and three-particle sectors
can be done without considering other sectors. For a detailed review on the
matching of non-relativistic and relativistic theories, see, e.g.,
Ref.~\cite{Gasser:2007zt}.

In the remainder of this section,  we will express the effective couplings,
$g_1,\, g_2,\, g_3$ and $\eta_3$, in terms of observable quantities, such as the
scattering length, $a_0$, the effective range $r_0$ and the (subtracted) three-particle
scattering amplitude at threshold.

\subsection{Matching in the two-body sector}

Let us start with the matching in the two-body sector. It should be pointed out that
it is not enough to perform matching in the center-of-mass (CM) frame of two-particles,
because in many-body systems two-particle subsystems have, in general, a
non-zero CM momentum. 

Let ${\bf p}$ be a generic three-momentum of a particle, and $|{\bf p}|\ll m$.
As seen from Eq.~(\ref{eq:LNR}), the kinetic term contains the expansion of the
relativistic one-particle energy up to and including $O({\bf p}^4)$. Further, the terms
up to $O({\bf p}^2)$ are retained in the expansion of the two-body interaction
Lagrangian. Note that the term with $g_2$ is absent if the matching is carried out in the
CM frame. When working in an arbitrary moving frame this term must be included. As
it will become clear later, the coupling $g_2$ is not an independent physical quantity---it is
uniquely fixed by the requirement of relativistic invariance.

Let us now recall that the expression of the relativistic $S$-wave amplitude\footnote{
  $P$-waves are absent due to Bose-symmetry. The presence of $D$-waves requires operators with four derivatives, which can contribute at most to order $L^{-7}$. Hence, at order $L^{-6}$, it is justified to assume that only the $S$-wave phase
  shift is different from zero.} in all frames is given by the Lorentz-invariant expression:
\eq\label{eq:M2kcot}
\cM_2(s)  = \frac{16 \pi \sqrt{s}}{k \cot \delta(k) -i k}, \quad\quad  k \cot \delta(k)= -\frac{1}{a_0} + \frac{1}{2} r_0 k^2 + O(k^4)\, .
\en
Here, $s$ is the usual Mandelstam variable related to the relative three-momentum in the
CM frame as $k=(s/4-m^2)^{1/2}$. Further, $\delta(k)$ denotes the $S$-wave phase shift
and $a_0,r_0$ are the pertinent scattering length and the effective range, respectively. In addition, the scattering $K$-matrix can be expressed in terms of the amplitude given in Eq.~(\ref{eq:M2kcot}):
\eq\label{eq:K2cot}
\cK_2(s)  = \frac{16 \pi \sqrt{s}}{k}\,\tan\delta(k)=-32\pi a_0m\biggl\{
1+\frac{1}{2}\,k^2\biggl(\frac{1}{m^2}+a_0r_0\biggr)+O(k^4)\biggr\}.
\en
As noted above, the matching condition for the scattering amplitudes contains both analytic
and non-analytic terms at small momenta. Only the former are relevant for carrying out
the matching, and the latter should match automatically. For this reason, it would be
technically convenient to ``purify'' the matching condition from these non-analytic terms
(we shall check {\em a posteriori} that the non-analytic terms are indeed the same
in the relativistic and in the non-relativistic theories and cancel in the matching condition).
The goal can be easily achieved in the two-body sector, where the sole source of the
non-analytic contribution is rescattering in the $s$-channel (the so-called
bubble diagrams). One could remove this contribution by considering the
quantity 
$\cK_2$ and its non-relativistic counterpart $\cV_2$ which, in contrast to the pertinent
amplitudes, are analytic at threshold. In terms of the $K$-matrices, the matching
condition is written down in a form identical to Eq.~(\ref{eq:matching}):
\eq\label{eq:matching-K}
 \cV_2 =  \prod_{i=1}^2 \frac{1}{(2 w'_i)^{1/2}}\,  \cK_2
 \prod_{i=1}^2 \frac{1}{(2 w_i )^{1/2}}\, . 
 \en
  Namely,
in dimensional regularization, which will be used throughout this paper,
$\cV_2$ is given only by the tree-level diagrams, produced by the non-relativistic effective Lagrangian~(\ref{eq:LNR}).

In order to do the matching, one has to perform the non-relativistic expansion of the
relativistic variable $k^2$:
\eq\label{eq:k2}
k^2&=&\frac{s}{4}-m^2=\frac{1}{4}\,(P^0-\sqrt{{\bf P}^2+4m^2})(P^0+\sqrt{{\bf P}^2+4m^2})
\nonumber\\[2mm]
&=&
\tilde q_0^2\biggl(1+\frac{{\bf P}^2}{8m^2}+\frac{\tilde q_0^2}{4m^2}+
\frac{{\bf P}^4}{64m^2\tilde q_0^2}+O({\bf p}^4)\biggr)\,,
\en
where ${\bf P}={\bf p}_1+{\bf p}_2={\bf p}_1'+{\bf p}_2'$,
\eq
P^0=p_1^0+p_2^0={p_1'}^0+{p_2'}^0=2m+\frac{1}{4m}\,({\bf p}_1^2+{\bf p}_2^2+{{\bf p}_1'}^2+{{\bf p}_2'}^2)+\cdots\, ,
\en
and
\eq
\tilde q_0^2=m\biggl(P^0-2m-\frac{{\bf P}^2}{4m}\biggr)
=\frac{1}{8}\,\bigl(
({\bf p}'_1-{\bf p}'_2)^2+({\bf p}_1-{\bf p}_2)^2\bigr)+\cdots\, .
\en
Expanding the right-hand side of the matching condition in Eq.~(\ref{eq:matching-K}) up to and including terms of order ${\bf p}^2$, we get:
 \eq\label{eq:matching-K-rel}
&& \prod_{i=1}^2 \frac{1}{(2 w'_i)^{1/2}}\,  \cK_2
 \prod_{i=1}^2 \frac{1}{(2 w_i )^{1/2}}
\nonumber\\[2mm]
 &=&-\frac{8\pi a_0}{m}\biggl\{1+\frac{1}{16}\,
\biggl(-\frac{1}{m^2}+a_0r_0\biggr)\bigl(
({\bf p}'_1-{\bf p}'_2)^2+({\bf p}_1-{\bf p}_2)^2\bigr)
-\frac{1}{4m^2}\,{\bf P}^2\biggr\}+\cdots \, . 
\en
On the other hand,in the non-relativistic effective theory we obtain:
\eq\label{eq:V2}
\cV_2=-g_1+4g_2{\bf P}^2+g_3\bigl(
({\bf p}'_1-{\bf p}'_2)^2+({\bf p}_1-{\bf p}_2)^2\bigr)+\cdots\, .
\en
From the above two equations, one can readily fix the values for the constants
$g_1,g_2,g_3$:
\eq\label{eq:g123}
g_1&=&\frac{8\pi a_0}{m}\, ,
 \ \ \
g_2=\frac{\pi a_0}{2m^3}\, ,
\nonumber\\[2mm]
g_3&=&-\frac{\pi a_0^2}{2m}\biggl(-\frac{1}{a_0m^2}+r_0\biggr)\,.
\en
Note also that three effective couplings $g_1,g_2,g_3$ are expressed in terms of two
dynamical quantities $a_0,r_0$ only. This confirms the statement that was made above, namely
that one of the couplings is not independent and is fixed through relativistic invariance.

\begin{figure}[t]
  \begin{center}
    \includegraphics[width=14.cm]{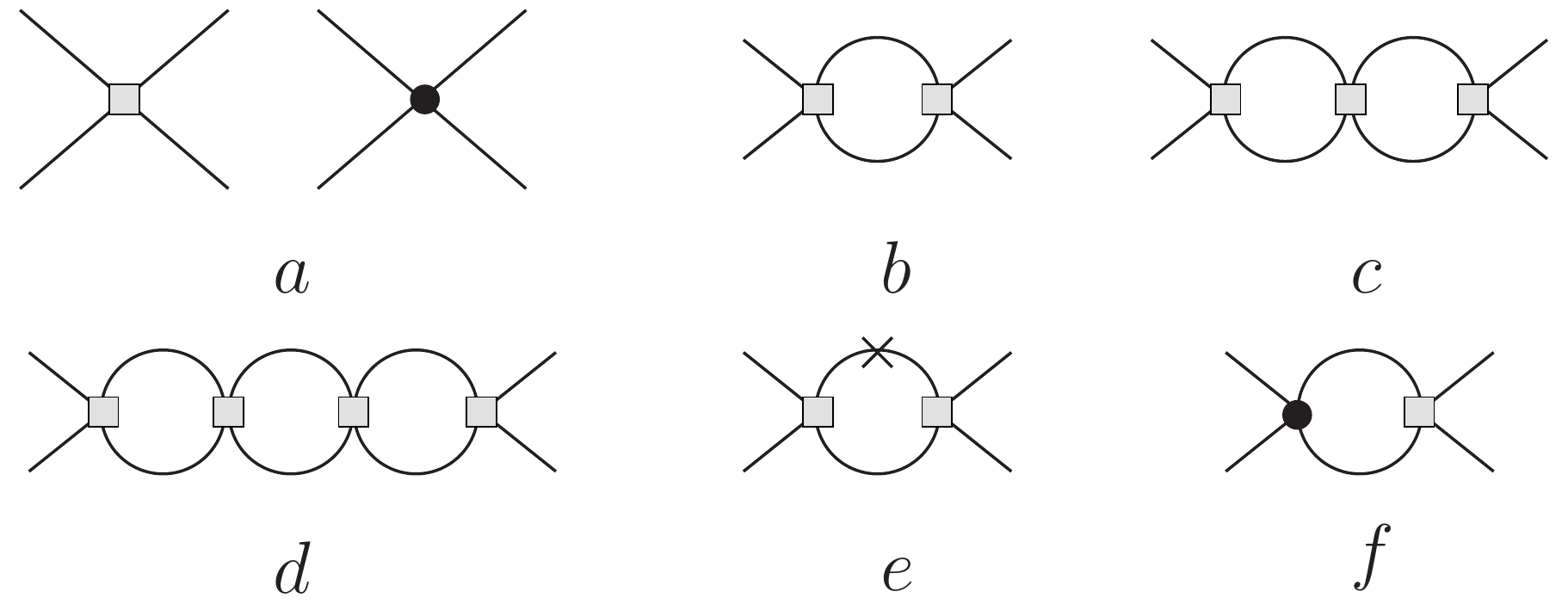}
    \caption{Diagrams contributing to the non-relativistic scattering amplitude
      $\cT(p_1',p_2';p_1,p_2)$
      up to and including $O(p^3)$. The shaded squares stand for vertices with the
     non-derivative coupling $g_1$, the filled circles for the vertices with derivative
couplings $g_2,g_3$, and the cross corresponds to the relativistic insertion ${\bf p}^4/(8m)$.    }
    \label{fig:twobody}
  \end{center}
  \end{figure}

\subsection{Loops in the two-body sector}

Now, we shall explicitly demonstrate that the matching of the couplings, given in Eq.~(\ref{eq:g123}), ensures the matching of the scattering amplitudes up to and including
$O({\bf p}^3)$ (i.e., the loops
that contribute to the amplitude, are automatically the same in the relativistic and non-relativistic theories). This is because additional
couplings first emerge at $O({\bf p}^4)$. All diagrams, contributing to the non-relativistic amplitude at $O({\bf p}^3)$, are depicted in Fig.~\ref{fig:twobody}.
Apart from the tree-level contributions, shown in Fig.~\ref{fig:twobody}a, at this order one has the one-, two- and three-loop diagrams with the non-derivative vertex $g_1$
shown in Fig.~\ref{fig:twobody}b,c,d, the one-loop vertex with the insertion of the
relativistic correction (Fig.~\ref{fig:twobody}e) and
the one-loop vertex with the insertion of the
derivative vertices (Fig.~\ref{fig:twobody}f). We shall calculate all of them separately.
The tree-level contribution coincides with the quantity $\cV_2$ given in Eq.~(\ref{eq:V2}):
\eq
\cT_2^{(a)}=-g_1+4g_2{\bf P}^2+g_3\bigl(
({\bf p}'_1-{\bf p}'_2)^2+({\bf p}_1-{\bf p}_2)^2\bigr)\, .
\en
Further, using dimensional regularization, we can calculate the one-loop diagram in 
Fig.~\ref{fig:twobody}b:
\eq
\cT_2^{(b)}&=&\frac{1}{2}\,g_1^2\int \frac{d^Dl}{(2\pi)^Di}\,
\frac{1}{\biggl(m-l^0+\dfrac{{\bf l}^2}{2m}-i\varepsilon\biggr)
  \biggl(m-P^0+l^0+\dfrac{({\bf P}-{\bf l})^2}{2m}-i\varepsilon\biggr)}
\nonumber \\[2mm]
&=&\frac{img_1^2}{8\pi}\,\tilde q_0\, ,
\en
where $P_\mu=(p_1+p_2)_\mu$ is the total four-momentum of the pair and $D$ is
the number of the space-time dimensions ($D\to 4$ at the end of calculations).
Similarly, we have:
\eq
\cT_2^{(c)}&=&-g_1^3\biggl(\frac{im}{8\pi}\,\tilde q_0\biggr)^2\, ,
\nonumber\\[2mm]
\cT_2^{(d)}&=&g_1^4\biggl(\frac{im}{8\pi}\,\tilde q_0\biggr)^3\, ,
\en
The insertion of the relativistic correction in the internal line gives:
\eq
\cT_2^{(e)}&=&g_1^2\int \frac{d^Dl}{(2\pi)^Di}\,\frac{{\bf l}^4}{8m^3}\,
\frac{1}{\biggl(m-l^0+\dfrac{{\bf l}^2}{2m}-i\varepsilon\biggr)^2
  \biggl(m-P^0+l^0+\dfrac{({\bf P}-{\bf l})^2}{2m}-i\varepsilon\biggr)}
\nonumber\\[2mm]
&=&\frac{ig_1^2\tilde q_0}{64\pi m}\,\biggl(5\tilde q_0^2+\frac{5}{2}\,{\bf P}^2
+\frac{{\bf P}^4}{16\tilde q_0^2}\biggr).
\en
Finally,
\eq
\cT_2^{(f)}&=&-\frac{1}{2}\,g_1\int \frac{d^Dl}{(2\pi)^Di}\,
\frac{8g_2{\bf P}^2+2g_3(2{\bf l}-{\bf P})^2
  +g_3\bigl(({\bf p}'_1-{\bf p}'_2)^2+({\bf p}_1-{\bf p}_2)^2\bigr)}
{\biggl(m-l^0+\dfrac{{\bf l}^2}{2m}-i\varepsilon\biggr)\biggl(m-P^0+l^0+\dfrac{({\bf P}-{\bf l})^2}{2m}-i\varepsilon\biggr)}
\nonumber\\[2mm]
&=&-\frac{img_1\tilde q_0}{8\pi}\,\biggl(8g_2{\bf P}^2
+8g_3\tilde q_0^2+g_3\bigl(({\bf p}'_1-{\bf p}'_2)^2+({\bf p}_1-{\bf p}_2)^2\bigr)\biggr)\, .
\en
Adding everything together up to and including $O({\bf p}^3)$, we get:
\eq\label{eq:T2exp}
\cT_2&=&-g_1+4g_2{\bf P}^2+8g_3\tilde q_0^2
+\frac{img_1^2\tilde q_0}{8\pi}\,\biggl(1+\frac{5\tilde q_0^2}{8m^2}+\frac{5{\bf P}^2}{16m^2}+\frac{{\bf P}^4}{128m^2\tilde q_0^2}\biggr)
\nonumber\\[2mm]
&-&\frac{img_1\tilde q_0}{\pi}\,\biggl(g_2{\bf P}^2
+2g_3\tilde q_0^2\biggr)
-g_1^3\biggl(\frac{im}{8\pi}\,\tilde q_0\biggr)^2
+g_1^4\biggl(\frac{im}{8\pi}\,\tilde q_0\biggr)^3\, .
\en
Now, let us expand the quantity relativistic amplitude up to and including $O({\bf p}^3)$:
\eq\label{eq:M2exp}
\frac{\cM_2}{4(w_1'w_2'w_1w_2)^{1/2}}
=\frac{-16\pi a_0\sqrt{s}}{4(w_1'w_2'w_1w_2)^{1/2}}
\biggl(1-ika_0+(ika_0)^2-(ika_0)^3+\frac{a_0r_0}{2}\,k^2-ia_0^2r_0k^3\biggr)\, .
\nonumber\\
\en
Taking into account the relation between the quantities $k^2$ and $\tilde q_0^2$,
which is given by Eq.~(\ref{eq:k2}), one can finally verify that the expansion of the
expression  in Eq.~(\ref{eq:M2exp}) exactly coincides with Eq.~(\ref{eq:T2exp}) term
by term. This is a nice check of the matching procedure, carried out in a moving frame
and including leading-order relativistic corrections.

\subsection{Matching in the three-particle sector}

At the order we are working, it suffices to
include a single operator that describes the short-range non-derivative 
three-particle interactions. Note also that the coupling
$\eta_3$ should be ultraviolet divergent in order  to cancel the divergences
emerging from the loops. When the dimensional 
regularization is used, then $\eta_3$ contains the term proportional to
$(D-4)^{-1}$ and a finite piece.
The matching condition that allows one to determine this finite piece,
is given by:
\eq\label{eq:matching3}
 \cT_3 =  \prod_{i=1}^3 \frac{1}{(2 w'_i)^{1/2}}\,  \cM_3
\prod_{i=1}^3 \frac{1}{(2 w_i )^{1/2}}\, , 
\en
Again, both left- and right-hand sides of the above equation
contain non-polynomial contributions, which should be
separated before the matching for the regular part is carried out.
In the two-particle case, one has achieved this goal by rewriting the matching condition
in terms of the $K$-matrix, which does not contain non-polynomial contributions
at all. Here, the situation is more complicated.
It is important to realize that there is a certain ambiguity in the definition of the
regular part of the three-body amplitude at threshold, in contrast to the two-body
sector, where the parameterization of the amplitude through the effective-range
expansion is commonly accepted. Below, we give a simple prescription that operates on the on-shell amplitudes only. The advantage of this prescription is
that such amplitudes are physical observables that can be defined both in relativistic and non-relativistic theories.\footnote{Note that this prescription differs slightly from the
  one used in Ref.~\cite{Pang:2019dfe}, which operated with an off-energy-shell
  non-relativistic scattering amplitude at an intermediate step. While the prescription is
  well-defined in the non-relativistic case, one encounters difficulties in properly defining
  it for the relativistic amplitudes. Of course, the choice made in Ref.~\cite{Pang:2019dfe}
  does not lead to any physical consequences, since in that paper one has stayed exclusively
  within the non-relativistic framework.}

One should also note here that, once the relation between $\eta_3$ and the threshold
amplitude is established, one can use either of these in the fit of the three-particle
energy levels. From this point of view, the matching procedure described below
is superfluous. Still, we shall present it here in order to shed light on the physical meaning
of the three-body coupling $\eta_3$ (a three-body force, in the terminology of the
non-relativistic scattering theory).

\begin{figure}[t]
  \begin{center}
    \includegraphics[width=6.cm]{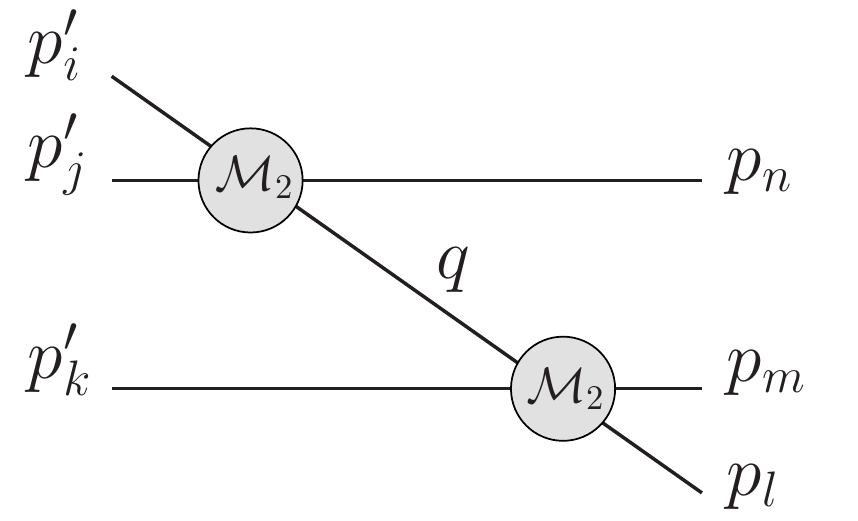}
    \caption{The most singular contribution to the relativistic three-body scattering
      amplitude. The four-momentum $q=p_i'+p_j'-p_n$.}
    \label{fig:most_singular}
  \end{center}
\end{figure}

Let us start by identifying the most singular piece in the relativistic amplitude. It is
given by the (infinite) sum of all one-particle reducible diagrams, shown in
Fig.~\ref{fig:most_singular}. In total, there are 9 diagrams of this type, differing by permutations of the initial $p_l,p_m,p_n$ and final $p_i',p_j',p_k'$ momenta. The contribution of
all such diagrams can be written down in the following form:
\eq\label{eq:1PI}
{\cal M}_3^{1PI}(p_1',p_2',p_3';p_1,p_2,p_3)=\sum_{(ijk)}\sum_{(lmn)}
\frac{{\cal M}_2(p_i',p_j';p_n,q){\cal M}_2(p_k',q;p_l,p_m)}{m^2-q^2-i\varepsilon}\, .
\en
The two-body scattering amplitude, ${\cal M}_2$, entering the above expression
depend on three four momenta as $q=p_i'+p_j'-p_n=p_l+p_m-p_k'$. These amplitudes are
in general off-mass-shell, since $(p_i'+p_j'-p_n)^2\neq m^2$ and
$(p_l+p_m-p_k')^2\neq m^2$. However, the singular contribution at $q^2=m^2$
contains only on-shell amplitudes. In order to define the on-mass-shell amplitudes, say,
for the first amplitude in the numerator of Eq.~(\ref{eq:1PI}) we perform a Lorentz boost
to the center-of-mass (CM) frame of $p_i',p_j'$ and denote the boosted momenta
${p_i'}^*,{p_j'}^*,{p_n}^*$ and ${q}^*={p_i'}^*+{p_j'}^*-{p_n}^*$. In this
frame, the amplitude depends of four variables. Two of them are related to the total
energy in the initial and final state:
\eq
s&=&4\bigl(({{\bf p}_n}^*)^2+m^2\bigr)\, ,
\nonumber\\[2mm]
s'&=&4\bigl(({{\bf p}_i'}^*)^2+m^2\bigr)=4\bigl(({{\bf p}_j'}^*)^2+m^2\bigr)
=(p_i'+p_j')^2\, ,
\en
and two are unit vectors:
\eq
{\bf n}^*=\frac{{{\bf p}_n}^*}{|{{\bf p}_n}^*|}\, ,\quad\quad
{{\bf n}'}^*=\frac{{{\bf p}_i'}^*}{|{{\bf p}_i'}^*|}=-\frac{{{\bf p}_j'}^*}{|{{\bf p}_j'}^*|},
\en
The partial-wave expansion of the two-body amplitude takes the form:
\eq\label{eq:pw}
{\cal M}_2({p_i'}^*,{p_j'}^*;{p_n}^*,{q}^*)
=4\pi\sum_{\ell m}Y_{\ell m}({{\bf n}'}^*){\cal M}_2^\ell(s',s)Y^*_{\ell m}({{\bf n}}^*)\, .
\en
The on-shell two-body amplitude is then {\em defined} at $s=s'$:
\eq
\overline{{\cal M}}_2({p_i'}^*,{p_j'}^*;{p_n}^*,{q}^*)
=4\pi\sum_{\ell m}Y_{\ell m}({{\bf n}'}^*){\cal M}_2^\ell(s',s')Y^*_{\ell m}({{\bf n}}^*)\, .
\en
The second amplitude in the numerator of Eq.~(\ref{eq:1PI}) can be treated similarly.
It is evaluated at $s=(p_l+p_m)^2$.

Finally, taking into account the identity
\eq
&&\hspace*{-.4cm}\frac{1}{m^2-q^2}=\frac{1}{2w({\bf q})}\,\biggl\{
\frac{1}{w({\bf q})+w({\bf p}_n)-w({\bf p}_i')-w({\bf p}_j')}+
\frac{1}{w({\bf q})-w({\bf p}_n)+w({\bf p}_i')+w({\bf p}_j')}\biggr\}\, ,
\nonumber\\
&&\hspace*{-.4cm}
\en
where only the first term is singular, one could finally {\em define} the most singular (pole)
piece of the three-body amplitude as:
\eq\label{eq:pole}
{\cal M}_3^{(pole)}(p_1',p_2',p_3';p_1,p_2,p_3)=\sum_{(ijk)}\sum_{(lmn)}
\frac{\overline{{\cal M}}_2({p_i'}^*,{p_j'}^*;{p_n}^*,{q}^*)
  \overline{{\cal M}}_2({p_k'}^*,{q}^*;{p_l}^*,{p_m}^*)}
{2w({\bf q})(w({\bf q})+w({\bf p}_n)-w({\bf p}_i')-w({\bf p}_j'))}\, .
\nonumber\\
\en
The advantage of this definition is that the singular part is defined solely in terms
of the observable on-shell two-body amplitudes.
Note also that at the order we are working, it suffices to retain only the $S$-wave
contribution
in the partial-wave expansion, Eq.~(\ref{eq:pw}). Consequently, the numerator
in Eq.~(\ref{eq:pole}) simplifies to  $ {\cal M}_2(s') {\cal M}_2(s)$,
with the on-shell $S$-wave amplitudes given in Eq.~(\ref{eq:M2kcot}).

The difference ${\cal M}_3-{\cal M}_3^{(pole)}$ is still singular. Moreover,
the singularity depends on the configuration of the initial and final three-momenta. In order
to define the threshold amplitude in general, one should first perform the partial-wave
expansion and then consider the limit where the magnitudes of external momenta
tend to zero. However, to invoke such a cumbersome framework for matching a single
coupling $\eta_3$ seems to be an overkill, since this goal can be achieved much easier.
For example, we could agree to perform the limit ${\bf p}\to 0$ for a particular
configuration of the initial and final momenta. Namely, let ${\bf e}_x,{\bf e}_y,{\bf e}_z$
denote unit vectors in the direction of axes in the momentum space. Then, choose
the configuration, for example, as:
\eq
\begin{array}{l}
  {\bf p}_1=\lambda {\bf e}_y\, ,\\[2mm]
  {\bf p}_2=\lambda\biggl(\dfrac{\sqrt{3}}{2}\,{\bf e}_x-\dfrac{1}{2}\,{\bf e}_y\biggr)\, , \\[2mm]
   {\bf p}_3=-\lambda\biggl(\dfrac{\sqrt{3}}{2}\,{\bf e}_x+\dfrac{1}{2}\,{\bf e}_y\biggr)\, ,
\end{array}
\en
and $ {\bf p}_i = -  {\bf p}'_i$. This means that
\eq\label{eq:scalarproducts}
{\bf p}_k{\bf p}_n={\bf p}_k'{\bf p}_n'
=-{\bf p}'_k {\bf p}_n = -\frac{1}{2} \lambda^2  + \frac{3}{2}  \lambda^2 \delta_{kn},
\en
For this configuration of momenta, both ${\cal M}_3$ and ${\cal M}_3^{(pole)}$ become
functions of a single variable $\lambda$. In the vicinity of the three-particle threshold,
their difference can be expanded in a Lorraine series:
\eq\label{eq:allsing}
\mbox{Re}\,\biggl({\cal M}_3(\lambda)-{\cal M}_3^{(pole)}(\lambda)\biggr)
=\frac{1}{\lambda}\,{\cal M}_3^{(-1)}
+\ln\frac{\lambda}{m}\,{\cal M}_3^{(l)}+{\cal M}_3^{(0)}+\cdots\, ,
\en
where ellipses stand for the terms that vanish as $\lambda\to 0$, and hence do not
contribute to the matching condition at threshold at this order. Here, we have arbitrarily
chosen $m$ to be a scale appearing in the logarithm.
Any other choice would change the definition of the regular part of the threshold
amplitude ${\cal M}_3^{(0)}$ by an additive contribution. Note also that
${\cal M}_3^{(pole)}(\lambda)$ behaves like $\lambda^{-2}$ near threshold. Thus,
the subtraction of the pole part indeed eliminates the leading singularity in the
relativistic three-particle amplitude\footnote{Our definition of the regular part of the
  threshold amplitude is very similar in spirit to the procedure considered in Refs.~\cite{Hansen:2014eka,Hansen:2015zga}, but differs in details. For example, in difference to those papers, our threshold amplitude does not depend on the ultraviolet cutoff. On the other hand, the infrared cutoff, introduced in those papers, enables one to set all external momenta
to zero in a trivial fashion. However, it is very inconvenient to use such a cutoff in the context of the dimensional regularization.}.

\begin{figure}[t]
  \begin{center}
    \includegraphics[width=14cm]{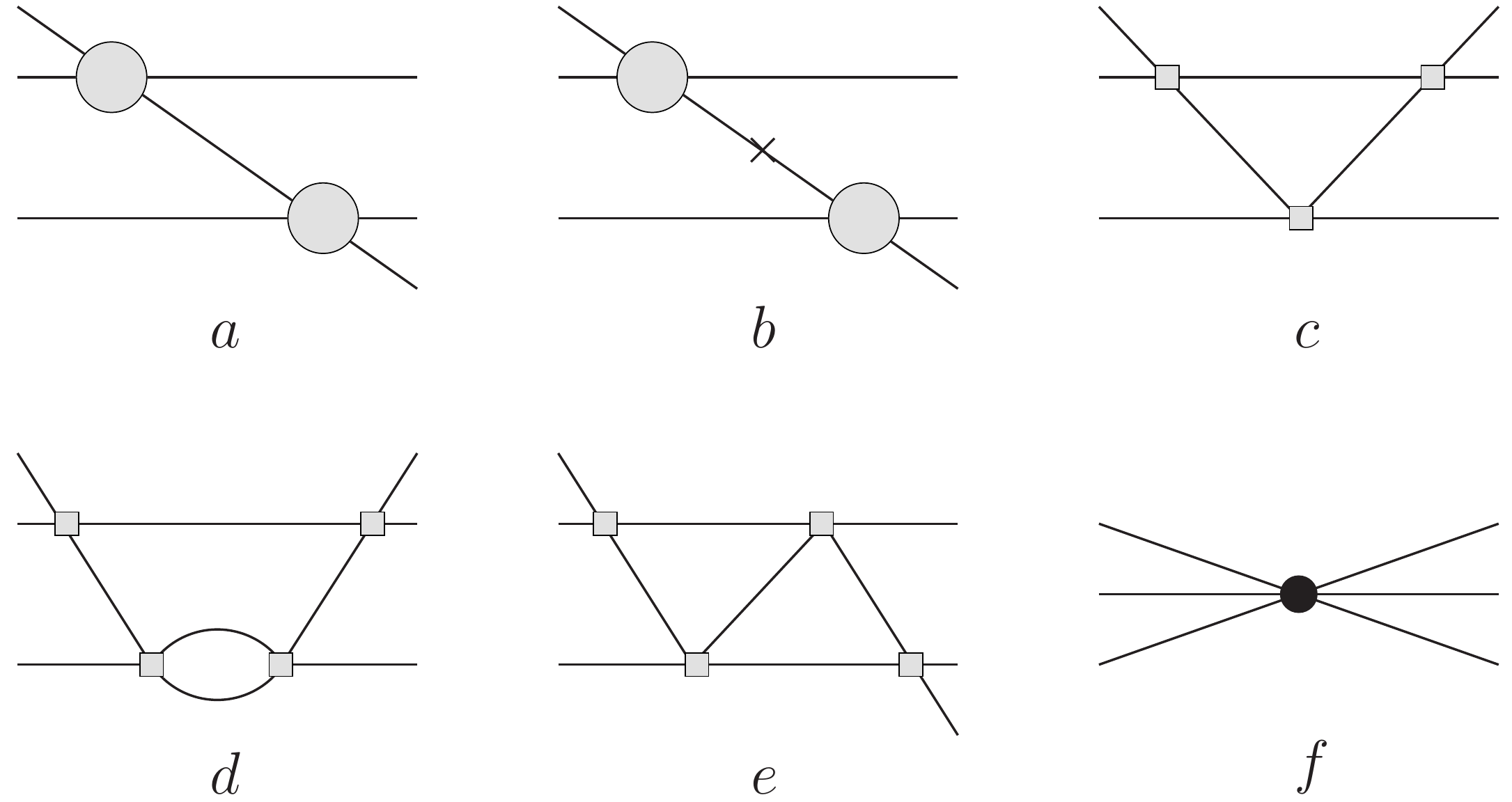}
    \caption{Diagrams that contribute to the non-vanishing three-particle scattering amplitude at threshold. The large shaded blobs in the diagram {\em a} depict the two-body scattering amplitude at $O({\bf p}^2)$, the shaded squares correspond to the vertices with the non-derivative coupling $g_1$, the cross in the diagram {\em b} depicts the relativistic insertion ${\bf p}^4/(8m)$, and the large filled circle in the diagram {\em f} corresponds to the three-particle coupling $\eta_3$.}
    \label{fig:nonrelativistic_threshold}
  \end{center}
  \end{figure}

Next, we turn to the non-relativistic amplitude and carry out an extraction of the singular
pieces in a similar fashion.
It is important to note that if the dimensional regularization is used, only a finite
number of diagrams, shown in Fig.~\ref{fig:nonrelativistic_threshold},
will contribute to the regular part of the threshold amplitude. In the following, we shall examine all these diagrams one by one.

\subsubsection{Singular diagrams in the non-relativistic theory}

  We start from the diagrams Fig.~\ref{fig:nonrelativistic_threshold}a and
  Fig.~\ref{fig:nonrelativistic_threshold}b. In fact, one could resum all relativistic insertions
  in the internal line---this obviously results in replacing the non-relativistic energies
  in the energy denominator by their relativistic counterparts. Thus, we get:\footnote{The permutations refer to different choices of the spectator particle in the initial and final state.
  There are nine choices in total.}
  \eq
\cT_3^{(a)}+\cT_3^{(b)}=\sum_{\text{perm}(ijk)}\sum_{\text{perm}(lmn)}
\frac{\cT_2({\bf p}'_i,{\bf p}'_j;{\bf p}_n,{\bf q})\cT_2({\bf p}'_k,{\bf q};{\bf p}_l,{\bf p}_m)}{w({\bf q})-w({\bf p}'_i)-w({\bf p}'_j)+w({\bf p}_n)-i\varepsilon}\,
\en
where ${\bf q}={\bf p}'_i+{\bf p}'_j-{\bf p}_n={\bf p}_l+{\bf p}_m-{\bf p}'_k$.
The calculations are performed in the CM frame of three particles,
i.e., ${\bf p}'_1+{\bf p}'_2+{\bf p}'_3={\bf p}_1+{\bf p}_2+{\bf p}_3=0$. The
quantity $\cT_2$ denotes the two-point amplitude in this particular kinematics.
For our purposes, it suffices to evaluate this amplitude at order ${\bf p}^2$,
since the higher-order terms will lead to the contributions that vanish
at threshold (we remind the reader that the denominator in the above equation
is of order ${\bf p}^2$). Using Eq.~(\ref{eq:T2exp}) at this order, we get:
\eq\label{eq:T2-off}
\cT_2({\bf p}'_i,{\bf p}'_j;{\bf p}_n,{\bf q})&=&-g_1+4g_2{\bf P}^2
+g_3\bigl(({\bf p}'_i-{\bf p}'_j)^2+({\bf p}_n-{\bf q})^2\bigr)
\nonumber\\[2mm]
&+&i\frac{mg_1^2}{8\pi}\,q'_{ij}+\frac{m^2g_1^3}{64\pi^2}\,(q'_{ij})^2+O({\bf p}^3)\, ,
\en
 where
\eq
(q'_{ij})^2=\frac{s'_{ij}}{4}-m^2\, ,\quad\quad
s'_{ij}=(w({\bf p}'_i)+w({\bf p}'_j))^2-({\bf p}'_i+{\bf p}'_j)^2\, .
\en
The amplitude $\cT_2({\bf p}'_k,{\bf q};{\bf p}_l,{\bf p}_m)$ is defined in a similar way.

The one-loop diagram, shown in Fig.~\ref{fig:nonrelativistic_threshold}c, is given by
\eq
\cT_3^{(c)}&=&-g_1^3\sum_{\text{perm}(ijk)}\sum_{\text{perm}(lmn)}
\int \frac{d^Dq}{(2\pi)^Di}\,
\frac{1}{m-q^0+\dfrac{{\bf q}^2}{2m}-i\varepsilon}
\nonumber\\[2mm]
&\times&
\frac{1}{m-{p'_i}^0-{p'_j}^0+q^0+\dfrac{({\bf p}'_i+{\bf p}'_j-{\bf q})^2}{2m}-i\varepsilon}
\nonumber\\[2mm]
&\times&
\frac{1}{m-p_l^0-p_m^0+q^0+\dfrac{({\bf p}_l+{\bf p}_m-{\bf q})^2}{2m}-i\varepsilon}\, .
\en
whereas the two-loop diagrams in Fig.~\ref{fig:nonrelativistic_threshold}d,e take the following form:
\eq
\cT_3^{(d)}&=&\frac{1}{2}\,g_1^4\sum_{\text{perm}(ijk)}\sum_{\text{perm}(lmn)}
\int \frac{d^Dq}{(2\pi)^Di}\,\int \frac{d^Dk}{(2\pi)^Di}\,
\frac{1}{m-q^0+\dfrac{{\bf q}^2}{2m}-i\varepsilon}
\nonumber\\[2mm]
&\times&
\frac{1}{m-{p'_i}^0-{p'_j}^0+q^0+\dfrac{({\bf p}'_i+{\bf p}'_j-{\bf q})^2}{2m}-i\varepsilon}
\nonumber\\[2mm]
&\times&
\frac{1}{m-p_l^0-p_m^0+q^0+\dfrac{({\bf p}_l+{\bf p}_m-{\bf q})^2}{2m}-i\varepsilon}\,
\frac{1}{m-k^0+\dfrac{{\bf k}^2}{2m}-i\varepsilon}
\nonumber\\[2mm]
&\times&\frac{1}{m-p_l^0-p_m^0-p_n^0+q^0+k^0+\dfrac{({\bf p}_l+{\bf p}_m+{\bf p}_n-{\bf q}-{\bf k})^2}{2m}-i\varepsilon}\, ,
\en
and
\eq
\cT_3^{(e)}&=&g_1^4\sum_{\text{perm}(ijk)}\sum_{\text{perm}(lmn)}
\int \frac{d^Dq}{(2\pi)^Di}\,\int \frac{d^Dk}{(2\pi)^Di}\,
\frac{1}{m-q^0+\dfrac{{\bf q}^2}{2m}-i\varepsilon}
\nonumber\\[2mm]
&\times&
\frac{1}{m-{p'_i}^0-{p'_j}^0+q^0+\dfrac{({\bf p}'_i+{\bf p}'_j-{\bf q})^2}{2m}-i\varepsilon}
\nonumber\\[2mm]
&\times&
\frac{1}{m-p_l^0-p_m^0+k^0+\dfrac{({\bf p}_l+{\bf p}_m-{\bf k})^2}{2m}-i\varepsilon}\,
\frac{1}{m-k^0+\dfrac{{\bf k}^2}{2m}-i\varepsilon}
\nonumber\\[2mm]
&\times&\frac{1}{m-p_l^0-p_m^0-p_n^0+q^0+k^0+\dfrac{({\bf p}_l+{\bf p}_m+{\bf p}_n-{\bf q}-{\bf k})^2}{2m}-i\varepsilon}\, .
\en
Finally, the diagram in Fig.~\ref{fig:nonrelativistic_threshold}f gives:
\eq
\cT_3^{(f)}=-6\eta_3\, .
\en

\subsubsection{Evaluation of the singular diagrams}

We start with the diagrams in Fig.~\ref{fig:nonrelativistic_threshold}a,b.
As already mentioned, there is a subtlety here, which consists in the fact
that $\cT_2$, given by Eq.~(\ref{eq:T2-off}) is not on shell,
i.e., $w({\bf p}'_i)+w({\bf p}'_j)\neq w({\bf p}_n)+w({\bf q})$. For the on-shell amplitude,
the Mandelstam variable $s=(p'_i+p'_j)^2$. Since both particles $i,j$ are on the mass
shell, we get also $s=4m^2+({\bf p}'_i-{\bf p}'_j)^2$, but
$s\neq 4m^2+({\bf p}_n-{\bf q})^2$. Thus, the on-shell amplitude is given by:
\eq\label{eq:barT2}
\bar{\cT}_2({\bf p}'_i,{\bf p}'_j;{\bf p}_n,{\bf q})&=&-g_1+4g_2{\bf P}^2
+2g_3({\bf p}'_i-{\bf p}'_j)^2
\nonumber\\[2mm]
&+&i\frac{mg_1^2}{8\pi}\,q'_{ij}+\frac{m^2g_1^3}{64\pi^2}\,(q'_{ij})^2+O({\bf p}^3)=
\cT_2+\delta \cT_2\, ,
\en
where
\eq\label{eq:deltaT2}
\delta \cT_2=g_3\bigl(({\bf p}'_i-{\bf p}'_j)^2-({\bf p}_n-{\bf q})^2\bigr)\, .
\en
For the second amplitude, we have the similar expression.

Further, in the matching condition, we have the factor $\bigl(w({\bf p}'_i)w({\bf p}'_j) \bigr)^{1/2}$ for the outgoing particles and the similar factor for the ingoing particles. In the off-shell kinematics, the second factor will not be equal to  $\bigl(w({\bf p}_n)w({\bf q}) \bigr)^{1/2}$ anymore. Our aim is to establish, which factor should be present here. To this end, let us start from the outgoing state. In the CM frame of the pair $(i,j)$, the above factor is replaced
by $\bigl(w({\bf p}')w(-{\bf p}')\bigr)^{1/2}$, where ${\bf p}'$ is the relative momentum
of the pair. Performing now the boost to the frame where the total momentum of the pair is ${\bf P}'_{ij}={\bf p}'_i+{\bf p}'_j$, we get
\eq
w({\bf p}'_i)&=&\sqrt{1-v^2}\bigl(w({\bf p}')+v({\bf p}'{\bf u})\bigr)\, ,
\nonumber\\[2mm]
w({\bf p}'_j)&=&\sqrt{1-v^2}\bigl(w(-{\bf p}')-v({\bf p}'{\bf u})\bigr)\, ,
\nonumber\\[2mm]
v&=&\frac{|{\bf P}'_{ij}|}{\sqrt{s+{{\bf P}'_{ij}}^2}}\, ,
\en
where ${\bf u}$ denotes the unit vector in the direction of ${\bf P}'_{ij}$.

Let now the initial pair have the relative momentum ${\bf p}$ on shell that means that
$|{\bf p}'|=|{\bf p}|$, but the direction can be arbitrary. Consequently, under the
Lorentz transformation,
\eq
w({\bf p})\to \bar w_1&=&\sqrt{1-v^2}\bigl(w({\bf p})+v({\bf p}\,{\bf u})\bigr)\, ,
\nonumber\\[2mm]
w(-{\bf p})\to \bar w_2&=&\sqrt{1-v^2}\bigl(w(-{\bf p})-v({\bf p}\,{\bf u})\bigr)\, .
\en
Taking into account that we are dealing with the identical particles, i.e., $w({\bf p}) w(-{\bf p})$, we finally get:
\eq
w({\bf p}'_i)w({\bf p}'_j)-\bar w_1\bar w_2=v^2({\bf p}'{\bf u})^2-v^2({\bf p}{\bf u})^2=O({\bf p}^4)\, ,
\en
since $v^2=O({\bf p}^2)$.

The brief summary of this a bit lengthy discussion is simple: at $O({\bf p}^2)$, the matching condition for the on-shell amplitudes looks as follows
\eq
\bar{\cT}_2({\bf p}'_i,{\bf p}'_j;{\bf p}_n,{\bf q})
=\frac{1}{2\bigl(w({\bf p}'_i)w({\bf p}'_j)\bigr)^{1/2}}\,\cM_2(s'_{ij})
\frac{1}{2(\bar w_1\bar w_2)^{1/2}}=
\frac{\cM_2(s'_{ij})}{4w({\bf p}'_i)w({\bf p}'_j)}\,.
\en
Now, the evaluation of the finite part in $\cT_3^{(a)}+\cT_3^{(b)}$ is straightforward:
\eq\label{eq:matching_a_b}
\cT_3^{(a)}+\cT_3^{(b)}=\bar{\cT}_3^{(a+b)}
-\sum_{\text{perm}(ijk)}\sum_{\text{perm}(lmn)}
\frac{\bar{\cT}'_2\delta \cT_2+\delta \cT'_2\bar \cT_2+\bar{\cT}'_2\bar{\cT}_2(F-1)}
{w({\bf q})-w({\bf p}'_i)-w({\bf p}'_j)+w({\bf p}_n)-i\varepsilon}\, ,
\en
where
\eq
\bar{\cT}_3^{(a+b)}
&=&\prod_{i=1}^3 \frac{1}{(2 w'_i)^{1/2}}\,
\prod_{i=1}^3 \frac{1}{(2 w_i )^{1/2}}
\nonumber\\[2mm]
&\times&\biggl(\sum_{\text{perm}(ijk)}\sum_{\text{perm}(lmn)}
\frac{\cM_2(s'_{ij})\cM_2(s_{lm})}{2w({\bf q})(w({\bf q})-w({\bf p}'_i)-w({\bf p}'_j)+w({\bf p}_n)-i\varepsilon)}\biggr)\, ,
\en
and
\eq
F=
\frac{(w({\bf p}'_i) w({\bf p}'_j))^{1/2}}{(w({\bf p}_n)w({\bf q}))^{1/2}}\,
\frac{(w({\bf p}_l)w({\bf p}_m))^{1/2}}{(w({\bf p}'_k)w({\bf q}))^{1/2}}\, .
\en
The second term in Eq.~(\ref{eq:matching_a_b}) can be simplified by expanding both the numerator and denominator in momenta. The denominator cancels, and we get:
  \eq\label{eq:matching1_a_b}
\cT_3^{a}+\cT_3^{(b)}-\bar{\cT}_3^{(a+b)}
&=&-72mg_1g_3+\frac{9g_1^2}{m}+\cdots
\nonumber\\[2mm]
&=&\frac{288\pi^2a_0^3}{m}\,
\biggl(-\frac{1}{a_0m^2}+r_0\biggr)+\frac{576\pi^2a_0^2}{m^3}+\cdots\, ,
\en
As one sees, the difference is a regular function at threshold, albeit
both ${\cal T}_3^{a}+{\cal T}_3^{(b)}$ and $\bar {\cal T}_3^{(a+b)}$ scale as $1/\lambda^2$ at small momenta.

The rest of calculations is straightforward, because one can consider non-relativistic
two-body amplitudes at $O({\bf p}^0)$ (i.e., replace these with the scattering length).
The diagram in Fig.~\ref{fig:nonrelativistic_threshold}c
is ultraviolet-finite in the dimensional regularization. Its explicit expression
is not needed. It suffices to note that, in a given configuration of external momenta,
\eq
 \mbox{Re } {\cal T}_3^{(c)}=\frac{\mbox{const}}{\lambda}\, ,
\en
so that the constant piece is absent. Further,
the diagrams in Fig.~\ref{fig:nonrelativistic_threshold}d,e are ultraviolet-divergent:
\begin{align}
\begin{split} \label{eq:T3de}
\mbox{Re }\cT_3^{(d)}&=9\sqrt{3}g_1^4m^3\biggl(L_2+\frac{1}{(4\pi)^3}\,\ln\frac{\lambda^2}{m^2}+\frac{\delta^{(d)}}{(4\pi)^3}\biggr)\,  = 9\sqrt{3}g_1^4m^3 L_2 + \mbox{Re }\bar{\cT}_3^{(d)},  \\
\mbox{Re }\cT_3^{(e)}&=-9 \frac{4\pi}{3}\,g_1^4m^3\biggl(L_2+\frac{1}{(4\pi)^3}\,\ln\frac{\lambda^2}{m^2}+\frac{\delta^{(e)}}{(4\pi)^3}\biggr)\, = -9\, \frac{4\pi}{3}\,g_1^4m^3 L_2 + \mbox{Re }\bar{\cT}_3^{(e)} ,
\end{split}
\end{align}
where
\eq
L_2=\frac{(\mu^2)^{D-4}}{(4\pi)^3}\,\biggl(\frac{1}{D-4}-\Gamma'(1)-\ln 4\pi+\ln\frac{m^2}{\mu^2}\biggr)\, ,
\en
$\mu$ denotes the scale of the dimensional regularization, and factor 9 comes from summing all permutations. The finite parts of the above diagrams $\delta^{(d)},\delta^{(e)}$
are pure numbers, given by the integrals over Feynman parameters, see
Appendix~\ref{app:delta}.

Using now the matching condition for the coupling $g_1$, it can be seen that the three-body
coupling should contain the following divergent part that cancels the divergences from the loops:
\eq
\eta_3=\frac{32\pi a_0^4}{m}\,(3\sqrt{3}-4\pi)(\mu^2)^{D-4}\biggl(\frac{1}{D-4}
-\Gamma'(1)-\ln 4\pi\biggr)+\eta_3^r(\mu)\, ,
\en
with $\eta_3^r(\mu)$ being the renormalized coupling. Choosing, for simplicity, the scale $\mu=m$ in this low energy constant, we may rewrite the matching condition as:
\eq\label{eq:eta3r}
-6\eta_3^r(m)=-\frac{288\pi^2a_0^3}{m}\,
\biggl(-\frac{1}{a_0m^2}+r_0\biggr)-\frac{576\pi^2a_0^2}{m^3}+\bar \cT\, ,
\en
whereas at an arbitrary scale
\eq
\eta_3^r(\mu)=\eta_3^r(m)-\frac{32\pi a_0^4}{m}\,(4\pi-3\sqrt{3})\,\ln\frac{m^2}{\mu^2}\, .
\en
Here, the threshold amplitude $\bar {\cal T}$ is defined as:
\eq\label{eq:barT}
\bar \cT=\lim_{\lambda\to 0} \mbox{Re}\,\biggl(\prod_{i=1}^3 \frac{1}{(2 w'_i)^{1/2}}\,  \cM_3
\prod_{i=1}^3 \frac{1}{(2 w_i )^{1/2}}-\bar \cT_3^{(a+b)}-\cT_3^{(c)}-\bar \cT_3^{(d)}
-\bar \cT_3^{(e)}\biggr)\, .
\en
Finally, using Eq.~(\ref{eq:allsing}), one may express $\bar\cT$ through the relativistic
threshold amplitude:
\eq\label{eq:barTT}
\bar \cT=\frac{1}{(2m)^3}\,{\cal M}_3^{(0)}-\frac{192\pi a_0^4}{m}\,(3\sqrt{3}\delta^{(d)}-4\pi\delta^{(e)})\, ,
\en
with $\delta^{(d)},\delta^{(e)}$ as in Appendix \ref{app:delta}. In principle, Eqs.~(\ref{eq:eta3r}) and (\ref{eq:barTT}) solve the problem completely:
they express the coupling $\eta_3^r(m)$ in terms of physical observables in the two- and
three-particle sectors. A non-trivial statement, which is implicit in these equations is that threshold singularities of the three-particle amplitude are produced by two-body rescattering processes only---that is, by subtracting diagrams that describe exactly these processes, one has to arrive at a non-singular expression. Mathematically, this is equivalent to the statement that the limit $\lambda\to 0$ in Eq.~(\ref{eq:barT}) exists.

\section{Energy shifts in the ground and excited states}
\label{sec:shifts}

In this section, we aim to derive energy shifts of the ground and
excited states of multiparticle systems. Following Ref.~\cite{Beane:2007qr},
this calculation can be easily done
within the framework of Schr\"odinger perturbation theory. 

Let us first
set up the notations.
The energy levels of $N$ identical particles in a free, non-relativistic
theory in finite volume are given by:
\eq\label{eq:E-unperturbed}
  E_n ( \mathbf{p}_1, \hdots  ,\mathbf{p}_N) = \frac{1}{2m}\,\sum_{i=1}^N{\bf p}_i^2\,
  \quad\quad
  \mathbf{p}_i = \frac{2\pi}{L}  \mathbf{n}_i, \quad  \mathbf{n}_i \in \mathbb{Z}^3,
\en
where $\mathbf{p}_i$ is the momentum of the $i$-th particle. The set
$n=\{ \mathbf{n}_i\}$ labels the unperturbed levels. For example, the ground
state corresponds to $\mathbf{n}_i=0,~i=1,\ldots N$. 

We shall further see that in a regime, in which $|a_0/L| \ll 1$, and $mL\gg 1$, the effect of the interactions
on the finite-volume spectrum in the relativistic theory can be treated as
a perturbation to the free
non-relativistic energy levels.  
Using the canonical procedure,
one can construct the Hamiltonian of the system from
the Lagrangian density, given in Eq.~(\ref{eq:LNR}):
\eq
\mathbf{H}&=&\mathbf{H}_0+\mathbf{H}_r+\mathbf{H}_1+\mathbf{H}_2
+\mathbf{H}_3+\mathbf{H}_\eta\, ,
\nonumber\\[2mm]
\mathbf{H}_X&=&\int_L d^3{\bf x}\,{\cal H}_X({\bf x},0)\, ,\quad X \equiv 0,r,1,2,3,\eta\, , \label{eq:allH}
\en
where the space integral is taken over a finite cube with side $L$, and 
\eq
{\cal H}_0&=&-\frac{1}{2m}\,\psi^\dagger\nabla^2\psi\, ,\quad\quad
{\cal H}_r=-\frac{1}{8m^3}\,\psi^\dagger\nabla^4\psi\, ,\quad\quad
{\cal H}_1=\frac{g_1}{4}\,(\psi^\dagger\psi)^2\, ,
\nonumber\\[2mm]
{\cal H}_2&=&g_2(\psi^\dagger\psi^\dagger)\nabla^2(\psi\psi)\, ,\quad\quad
{\cal H}_3=g_3\bigl((\psi^\dagger\psi^\dagger)(\psi{\stackrel{\leftrightarrow}{\nabla}}^2\psi)+\mbox{h.c}\bigr)\, ,\quad\quad
{\cal H}_\eta=\frac{\eta_3}{6}\,(\psi^\dagger\psi)^3\, .\quad\quad
\en
In the above expressions, $\mathbf{H}_0$ denotes the free Hamiltonian, 
whereas
$\mathbf{H}_\text{I}=\mathbf{H}_r+\mathbf{H}_1+\mathbf{H}_2+\mathbf{H}_3
+\mathbf{H}_\eta$ is treated
as a perturbation.

In a finite volume, free fields can be expanded in a Fourier series over the creation and annihilation operators $a^\dagger({\bf p}),a({\bf p})$:
\eq
  \psi({\bf x},t) =\frac{1}{L^3}\,\sum_{\bf p} e^{-ip^0t+i{\bf p}{\bf x}}
  a({\bf p})\, ,\quad\quad
  \psi^\dagger({\bf x},t) =\frac{1}{L^3}\,\sum_{\bf p} e^{ip^0t-i{\bf p}{\bf x}}
  a^\dagger({\bf p})\, . \label{eq:fieldsL}
  \en
  Here, the momentum ${\bf p}$ runs over discrete values
  ${\bf p}=2\pi{\bf n}/L\, ,~{\bf n}\in\mathbb{Z}^3$, and the creation/annihilation operators obey the following commutation relations:
  \eq
  [a({\bf p}),a^\dagger({\bf q})]=L^3\delta_{{\bf p}{\bf q}}\, .
  \en
  The properly normalized unperturbed eigenstates are given by:   
  \eq
\ket{\mathbf p_1,\hdots,\mathbf p_N} = \frac{1}{\sqrt{N!}\, L^{3N/2}}\, a^\dagger(\mathbf p_1) \hdots a^\dagger(\mathbf p_N) \ket{0}\, .  \label{eq:statesL}
\en
These eigenstates obey the eigenvalue equation:
\eq
\mathbf{H}_0\ket{\mathbf p_1,\hdots,\mathbf p_N}=
E_n ( \mathbf{p}_1, \hdots  ,\mathbf{p}_N)\ket{\mathbf p_1,\hdots,\mathbf p_N}\, .
\en
One may also define the matrix elements of the interaction Hamiltonian
between the unperturbed states:
\eq
V_{qp}=\bra{\mathbf q_1,\hdots,\mathbf q_N}\mathbf{H}_\text{I}\ket{\mathbf p_1,\hdots,\mathbf p_N}\, .
\en
The Rayleigh-Schr\"odinger perturbation theory enables one to calculate
the energy shift of a state $n$ using the interaction Hamiltonian
$\mathbf{H}_\text{I}$. A standard formula, which can be found in the quantum
mechanics textbooks (see, e.g.,~\cite{Landau-Lifshits}), reads:
\eq\label{eq:RS}
\Delta E_n=V_{nn}+\sum_{p\neq n}\frac{|V_{np}|^2}{E_n-E_p}
+\sum_{p,q\neq n}\frac{V_{np}V_{pq}V_{qn}}{(E_n-E_p)(E_n-E_q)}
-V_{nn}\sum_{p\neq n}\frac{|V_{np}|^2}{(E_n-E_p)^2}+O(V^4)\, .
\nonumber\\
\en
Note that we decided not to display here a (rather cumbersome)
fourth-order formula, which is also needed in the calculations at order $L^{-6}$.

\subsection{Two-particle levels}

As a warm-up example, we consider the well-known case of two particles in the center-of-mass (CM)
frame. The $1/L$ expansion of the energy shift for the two-particle ground
state has been known for a long time up to and including
$O(L^{-5})$~\cite{Huang:1957im,Luscher:1986n2}. At this order, the effects
of relativity are irrelevant and one can work in a non-relativistic setup.
The next order was worked out in the NREFT
approach~\cite{Beane:2007qr}, and in a fully relativistic
setup~\cite{Hansen:2015zta,Hansen:2016fzj}. As pointed out in
Ref.~\cite{Hansen:2015zta}, these two calculations differ by a term that
vanishes in the limit $m \to \infty$, and can be referred to as ``relativistic corrections''. Recently, in Ref.~\cite{Beane:2020ycc},
relativistic corrections to the ground state were derived in the NREFT with
a method, which is identical to the one used here.
Below, for illustrative purpose, we shall rederive $1/L$ expansion of the
energy shift up to and including $O(L^{-6})$ by using
Rayleigh-Schr\"odinger perturbation theory.

The matrix element to the interaction Hamiltonian between the
two-particle unperturbed states is given by:
\eq
V_{qp} = \braket{\mathbf q_1,\mathbf q_2 | \mathbf{ H}_\text{I} | \mathbf p_1,\mathbf p_2} \, .
\en
Calculating the matrix element, one gets:
\eq\label{eq:Vqp}
V_{qp}&=& \frac{1}{2L^3} \left[ g_1 -g_3 (\mathbf q_1 -\mathbf q_2)^2-g_3 (\mathbf p_1 -\mathbf p_2)^2  \right]  \delta_{ \mathbf q_1 +\mathbf q_2,  \mathbf p_1 +\mathbf p_2}
\nonumber\\[2mm]
 &-& \frac{1}{16 m^3} \left[ \mathbf q_1^4 +\mathbf q_2^4  +\mathbf p_1^4+\mathbf p_2^4 \right]\,   \frac{1}{2!}\, \sum_{(ij) \in\,\mathcal{S}_2} \delta_{  \mathbf q_1  \mathbf p_i } \delta_{  \mathbf q_2  \mathbf p_j }\, , \label{eq:pseudopotential2}
\en
where the sum in the last term runs over the two-element permutation
group $\mathcal{S}_2$ for the indices $i,j$.
The relation between $g_{1,3}$ and $a_0, r_0$ is given in Eq.~(\ref{eq:g123}).
The term with $g_2$ does not contribute in the CM frame and is therefore omitted here.

Now, using Eq.~(\ref{eq:RS}), we can calculate the energy shift of the ground
state order by order in perturbation theory. It is easy to verify that,
at lowest order, only terms with the non-derivative coupling
$g_1$ from the interaction Hamiltonian, $\mathbf{H}_\text{I}$, contribute:
\eq
\Delta E_0^{(1)}
= V_{00} = \frac{g_1}{2L^3}  = \frac{4 \pi a_0}{m L^3}\, .
\en
In order to evaluate the next-order result, we use Eq.~(\ref{eq:Vqp}) and get:
\eq
V_{0p}=-\frac{1}{2L^3}\,(g_1-4g_3{\bf p}^2)\delta_{{\bf 0},{\bf p}_1+{\bf p}_2}\, ,
\quad\quad {\bf p}_1=-{\bf p}_2={\bf p}\, .
\en
Substituting this expression into the second-order energy shift,
Eq.~(\ref{eq:RS}), and using $E_0=0$, $E_p={\bf p}^2/m$, we get:
\eq\label{eq:second}
\Delta E_0^{(2)} &=& \frac{1}{4L^6}\,
\sum_{p \neq 0}\frac{(g_1-4g_3{\bf p}^2)^2 }{ E_0 - E_p  } = - \frac{g_1^2 m}{16 \pi^2 L^4} \sum_{\mathbf n \neq \mathbf 0} \frac{1}{\mathbf n^2}
+\frac{2g_1g_3m}{L^6}\,\sum_{\mathbf n \neq \mathbf 0}1
+O(L^{-8})
\nonumber\\[2mm]
&=& - \frac{4 \pi a_0 }{m L^3} \frac{a_0}{\pi L} \mathcal{I}
+\frac{4\pi a_0}{mL^3}\,\frac{2\pi a_0^2}{L^3}\,\biggl(r_0-\frac{1}{a_0m^2}\biggr)
+O(L^{-8})\, .
\en
Here, one has used ${\bf p}=2\pi {\bf n}/L$ in order to display the powers
of $L$ explicitly. It is already clear that one has a double perturbative
expansion in powers of $1/L$. First, each order in Rayleigh-Schr\"odinger
perturbation theory brings a factor $L^{-3}$ associated to the momentum sum
and a factor $L^2$, emerging from the propagator, $L^{-1}$ in total. The pertinent expansion parameter here is $a_0/L$. Further,
each derivative vertex is suppressed by powers of $L$, since the momentum
${\bf p}$ scales as $1/L$. In this case, the pertinent expansion parameter
 is $1/(mL)$. Consequently, in order to obtain an expression
of the energy shift to a given order in $1/L$, it suffices to retain finite
number of terms in the effective Lagrangian and truncate the perturbative
series at a fixed order.

Furthermore, the momentum sums emerging in Eq.~(\ref{eq:second}), are formally
ultraviolet divergent. In order to be consistent with the matching condition,
this divergence should be subtracted in the same way as in the infinite volume,
i.e., by using the $\overline{\text{MS}}$ scheme in dimensional regularization.
In this particular case, it is very simple. Namely, one might rewrite
the first sum as:
\eq
{\cal I}=\sum_{\mathbf n \neq \mathbf 0} \frac{1}{\mathbf n^2}=
\sum_{\mathbf n \neq \mathbf 0} \frac{1}{\mathbf n^2(\mathbf n^2+1)}-1
+\sum_{\mathbf n}\frac{1}{\mathbf n^2+1}\, .
\en
Here, we have added and subtracted the term with $\mathbf n=\mathbf 0$ in the last sum.
Now, the first term is perfectly convergent, and the last term can be calculated by means of the Poisson summation formula, applying the $\overline{\text{MS}}$ scheme to the $|\mathbf n| =0$ term:
\begin{equation}
 \sum_{\mathbf n}\frac{1}{\mathbf n^2+1} = \int d^dx \frac{1}{x^2+1} + \sum_{n \neq 0} \frac{\pi}{| \mathbf n |} e^{- 2 \pi |\mathbf n |}.
\end{equation}
This way, one gets:
\eq
   {\cal I}=\sum_{\mathbf n \neq \mathbf 0} \frac{1}{\mathbf n^2(\mathbf n^2+1)}-1-2\pi^2+\sum_{\mathbf n \neq \mathbf 0}\frac{\pi}{|{\bf n}|}\,e^{-2\pi|{\bf n}|}\simeq -8.9136329.
   \en
   Acting in the same way, one can easily see that
   \eq
   \sum_{\mathbf n \neq \mathbf 0}1=-1\, ,
   \en
   and finally, using the matching condition for $g_{1,3}$, one arrives at the last
   line of Eq.~(\ref{eq:second}).

   Continuing in the same manner, one can calculate corrections up to
   and including order $L^{-6}$. The result is given by the following
   expression:
\eq
\Delta E_0 &=& \frac{4\pi a_0}{m L^3} \Bigg\{ 1 - \frac{a_0}{\pi L} \mathcal I + \left( \frac{a_0}{\pi L}\right)^2 \left[ \mathcal I ^2 - \mathcal J\right] - \frac{\pi a_0}{m^2 L^3} 
+\frac{2\pi a_0^2 r_0}{L^3}
\nonumber\\[2mm]
&-& \left( \frac{a_0}{\pi L}\right)^3\left[ \mathcal I^3-3\mathcal I\mathcal J +\mathcal K \right]         \Bigg\}\, ,  \label{eq:deltaE2}
\en
where
\eq
\mathcal{J} =  \sum_{\mathbf n \neq \mathbf 0} \frac{1}{\mathbf n^4} \simeq 16.5323\,  ,
\quad\quad
\mathcal{K} =  \sum_{\mathbf n \neq \mathbf 0} \frac{1}{\mathbf n^6} \simeq 8.40192.
\en
This, of course, agrees with the result of
Refs.~\cite{Hansen:2015zta,Hansen:2016fzj} as well as Ref.~\cite{Beane:2020ycc}.
Moreover, relativistic corrections are easy to identify. They
emerge from the contributions from $\mathbf{H}_3$ and $\mathbf{H}_r$.
If they are dropped,
the result in Ref.~\cite{Beane:2007qr} is recovered.

Next, we turn to the relativistic energy shift for the first excited state
in the two-particle sector. We start by defining the unperturbed
states. In the non-interacting theory and for identical particles,
this state is threefold degenerate in the relativistic, as well in the
non-relativistic case and has unit back-to-back momentum in the units
of $2\pi/L$, directed along the $x,y$ or $z$-axis.
These three states can be classified in irreducible representations
of the cubic group. There is a one-dimensional irrep $A_1^+$, and a
two-dimensional irrep $E^+$. The two-particle $A^+_1$ state is given by:
\eq\label{eq:A1}
\ket{({\bf p},-{\bf p})_{A_1^+}} = \frac{1}{\sqrt{6}} \sum_{a=x,y,z} \biggl( \ket{\mathbf p_a,-\mathbf  p_a} +  \ket{-\mathbf  p_a,\mathbf  p_a} \biggr),
\en
where $\mathbf  p_x = \dfrac{2\pi}{L}\,(1,0,0)$, and so on.
The basis vectors of the irrep $E^+$ are given by:
\eq\label{eq:E}
  \ket{({\bf p},-{\bf p})_{E^+},1} &=& \frac{1}{2} \biggl(
    \ket{\mathbf  p_x,-\mathbf p_x} +  \ket{-\mathbf p_x,\mathbf p_x}
    -  \ket{\mathbf p_y,-\mathbf p_y} -  \ket{-\mathbf p_y,\mathbf p_y}
     \biggr)\, , \nonumber\\[2mm]
     \ket{({\bf p},-{\bf p})_{E^+},2} &=& \frac{1}{\sqrt{12}} \biggl(
     \ket{\mathbf p_x,-\mathbf p_x} +  \ket{-\mathbf p_x,\mathbf p_x}
     +\ket{\mathbf p_y,-\mathbf p_y} +  \ket{-\mathbf p_y,\mathbf p_y}
\nonumber\\[2mm]
     &-& 2 \ket{\mathbf p_z,-\mathbf p_z} -2  \ket{-\mathbf p_z,\mathbf p_z}
     \biggr)\, .
\en
Because of the cubic symmetry, any of these two states can be used
for the calculation of the energy shift.

The calculation of the energy shifts proceeds analogously to the case
of ground state. We do not display the intermediate steps here. Note that
the energy shifts are calculated with respect to the relativistic unperturbed
energy of the first unperturbed state
$\displaystyle E_1=2\sqrt{m^2+(2\pi/L)^2}-2m$.
It is easy to see that the leading contribution to the energy
shift in the irrep $E^+$ arises from the $D$-wave interaction and is beyond
the accuracy we are calculating:
\begin{equation}
\Delta E_1^{E^+} = O(L^{-7}).
\end{equation}
The shift in the irrep $A_1^+$ at order $L^{-5}$ was calculated in
Ref.~\cite{Luscher:1986n2}. In the present paper this result is extended
up to and including order $L^{-6}$, where the method, used in Ref.~\cite{Luscher:1986n2}, does not directly apply for the calculation of the relativistic corrections. The final answer reads:
\begin{align}
\begin{split}
\Delta E^{A_1^+}_1 = \frac{24 \pi a_0}{m L^3} \Bigg\{&1 - \mathcal{I}_1\frac{a_0}{\pi L} + (\mathcal{I}_1^2 - 6\mathcal{J}_1) \left( \frac{a_0}{\pi L} \right)^2 + ( -\mathcal{I}_1^3 + 18 \mathcal{I}_1 \mathcal{J}_1 -36 \mathcal{K}_1) \left( \frac{a_0}{\pi L} \right)^3  \\ &+ \frac{2 \pi^2 a_0 r_0}{L^2}\left[ 1   - \mathcal{I}_1 \frac{a_0}{\pi L}  \right]   -  \frac{2 \pi^2 }{m^2 L^2} \left[ 1   + \frac{ \mathcal{J}_1 -1}{2} \frac{a_0}{\pi L}  \right]    \Bigg\}\, . \label{eq:shiftexc2}
\end{split}
\end{align}
Here,
\eq
\mathcal{I}_1&=&\sum_{|\mathbf n|\neq 1}\frac{1}{{\bf n}^2-1}
\doteq\sum_{|\mathbf n| \neq 1} \frac{2}{(\mathbf n^2-1)(\mathbf n^2+1)}-3-2\pi^2+\sum_{\mathbf n \neq \mathbf 0}\frac{\pi}{|{\bf n}|}\,e^{-2\pi|{\bf n}|}\simeq -1.2113357\, ,
\nonumber\\[2mm]
\mathcal{J}_1&=&\sum_{|\mathbf n|\neq 1}\frac{1}{({\bf n}^2-1)^2}\simeq 23.2432219\, ,
\nonumber\\[2mm]
\mathcal{K}_1&=&\sum_{|\mathbf n|\neq 1}\frac{1}{({\bf n}^2-1)^3}\simeq 13.0593768\,
\en

\subsection{Ground state of $N$ particles\label{subsec:groundN}}

If we are dealing with more than two particles, two new effects arise.
First, the two-particle subsystems can be in the moving frames, even if
the total momentum of the $N$-particle system vanishes. Therefore,
$\mathbf{H}_2$ in Eq.~(\ref{eq:allH}) starts to contribute (still, as we will see, this contribution is only non-vanishing for excited states ). Second, one has to take
into account the contribution from the non-derivative three-particle
interaction, which is parameterized by the coupling constant $\eta_3$.
Four- and more particle interactions, as well as the operators with higher
derivatives in the two- and three-particle sectors do not contribute at the
order we are working. In particular, from Eq.~(\ref{eq:statesL}) it is seen that the
normalization of the $N$-particle wave function depends only on $L$. Then,
on dimensional grounds, a generic operator in the Lagrangian,
$\delta{\cal L}=c_k \ \mathcal O_k$, where $c_k$ has the mass dimension
$[c_k]=M^{k}$, starts to contribute to the energy shift at
$\Delta E \sim c_k  L^{k-1}$. For example, one can see that the operator
of the type $(\psi\psi^\dagger)^N$ contributes to order $L^{-3(N-1)}$.

The expression in Eq.~(\ref{eq:pseudopotential2}) can be easily generalized
to more than two particles. For instance, for three particles, we have:
\eq
V_{qp} &=& \sum_{k = 1,2,3} \frac{1}{2L^3} \left[ g_1 -4g_2 (\mathbf q_i +\mathbf q_j)^2 -g_3 (\mathbf q_i -\mathbf q_j)^2 -g_3 (\mathbf p_i -\mathbf p_j)^2  \right]  \delta_{ \mathbf q_i +\mathbf q_j,  \mathbf p_i +\mathbf p_j} \delta_{\mathbf q_k, \mathbf p_k}
\nonumber\\[2mm]
&+& \frac{\eta_3}{L^6} \,\delta_{ \mathbf q_1 +\mathbf q_2 + \mathbf q_3,  \mathbf p_1 +\mathbf p_2 + \mathbf p_3}
\nonumber\\[2mm]
&-& \frac{1}{16 m^3}\,
\left[ \mathbf q_1^4 +\mathbf q_2^4+\mathbf q_3^4
      +\mathbf p_1^4+\mathbf p_2^4+\mathbf p_4^4 \right]\,
\frac{1}{3!}\, \sum_{(ijk)\, \in\,\mathcal{S}_3}
\delta_{  \mathbf q_1  \mathbf p_i }
\delta_{  \mathbf q_2  \mathbf p_j }
\delta_{  \mathbf q_3  \mathbf p_k }\, ,
\label{eq:pseudopotential3}
\en
where $i\neq j\neq k$. This expression is easy to understand. In the first line, there are three
pairwise interactions, in which $k$ labels the spectator particle;
the second line stems from the three-particle contact interaction;
and the third line corresponds to the kinetic term at $O({\bf p}^4)$.
Note that, in this case, the indices in the last sum run over the
three-element permutation group $\mathcal{S}_3$. For more than three
particles, the generalizations are straightforward at this order and
include only additional combinatorial factors.
In particular, there exist $ \binom{N}{2}$
 pairwise interactions, $\binom{N}{3}$ 
three-particle interactions, and $N!$ permutations in the kinetic term.

Further, we shall take an advantage of the fact that the expression
without relativistic corrections has been already derived previously
exactly with the same method~\cite{Beane:2007qr}. The remainder can be
calculated relatively easy. The final result for the $N$-particle
ground state, expressed in terms of the
threshold amplitude $\bar\cT$, has the form:
\begin{align}
\begin{split}
\Delta E_0& =\binom{N}{2} \frac{4\pi a_0}{mL^3} \Bigg\{ 1 - \frac{a_0}{\pi L} \mathcal I + \left( \frac{a_0}{\pi L}\right)^2 \left[ \mathcal I ^2 + (2N-5)\mathcal J\right]  \\&   - \left( \frac{a_0}{\pi L}\right)^3\left[\mathcal I^3 +(2N-7)\mathcal I\mathcal J +(5 N^2 - 41N + 63)\mathcal K  + 8 (N-2) (2\mathcal{Q} + \mathcal{R}) \right]          \\&  
+ (4N-9) \frac{\pi a_0}{m^2 L^3} +(4N-6)\frac{\pi a_0^2 r_0}{L^3}  \Bigg\}
\\ &+\binom{N}{3} \Bigg\{ \frac{32 \pi a_0^4}{m L^6} \left( 3 \sqrt{3} - 4 \pi \right) (2\ln (mL)-\Gamma'(1)-\ln 4\pi) - \frac{\bar{\cT}}{6 L^6 }    \Bigg\}\, . \label{eq:deltaEn}
\end{split}
\end{align}
Here, the quantities ${\cal I},{\cal J},{\cal K}$ were defined above, whereas
the quantities ${\cal Q}\simeq -102.1556055$ and ${\cal R}\simeq 19.186903$
are the quantities defined in Ref.~\cite{Beane:2007qr}\footnote{Note that there
  was a small numerical error in the value of ${\cal Q}$ in Ref.~\cite{Beane:2007qr}, which was corrected in Ref.~\cite{Detmold:2008gh}. The corrected
value agrees exactly with the number given in Ref.~\cite{Pang:2019dfe}.}.
The factor $(\Gamma'(1)+\ln 4\pi)$ emerges, because the quantities
${\cal Q},{\cal R}$ in Ref.~\cite{Beane:2007qr} are defined in the MS
renormalization scheme rather than $\overline{\text{MS}}$, which is used
in the present paper. Further, we confirm that
the above equation agrees with Eq.~(42)
of Ref.~\cite{Beane:2020ycc}, if expressed in terms of the renormalized
coupling $\eta_3^r$. In order to compare with the result of
Refs.~\cite{Hansen:2015zta,Hansen:2016fzj}, one has to carry out the matching
of the threshold amplitude, defined in those papers, to the quantity
$\bar\cT$. We do not do this (rather straightforward but tedious)
exercise here.

\subsection{Energy shift for the three-particle excited states. }

In this section we discuss the calculation of relativistic
corrections to the excited state levels in the three-particle
state. To the best of our knowledge, this issue has not been
addressed so far in the literature and appears here for the first time.
Note also that the combinatorics in the excited states becomes
rather cumbersome. For this reason, we first focus on three particles. 

Let us start with the definition of the state vectors. In the three-particle
system in the CM frame, the first excited unperturbed state contains one
particle at rest, whereas two other particles have back-to-back momenta
of unit magnitude, in units of $2\pi/L$. Again, the first excited states
contains only the $A_1^+$ and $E^+$ irreps of the cubic group. The pertinent
basis functions are:
\eq
\ket{A_1^+}&=&\frac{1}{\sqrt{3}}\,\biggl(
\ket{({\bf p}_1,{\bf p}_2)_{A_1^+},{\bf p}_3=0}
+\ket{({\bf p}_3,{\bf p}_1)_{A_1^+},{\bf p}_2=0}
+\ket{({\bf p}_2,{\bf p}_3)_{A_1^+},{\bf p}_1=0}\biggr)\, ,
\nonumber\\[2mm]
\ket{E^+,\alpha}&=&\frac{1}{\sqrt{3}}\,\biggl(
\ket{(({\bf p}_1,{\bf p}_2)_{E^+},\alpha),{\bf p}_3=0}
+\ket{(({\bf p}_3,{\bf p}_1)_{E^+},\alpha),{\bf p}_2=0}
\nonumber\\[2mm]
&&\hspace*{5.cm}+\ket{(({\bf p}_2,{\bf p}_3)_{E^+},\alpha),{\bf p}_1=0}
\biggr)\, ,
\en
where $\alpha=1,2$ and
the basis vectors $\ket{(p_i,p_j)_{A_1^+}}$ and
$\ket{(p_i,p_j)_{E^+},\alpha}$ with back-to-back three-momenta ${\bf p}_i$
and ${\bf p}_j$ are given by Eqs.~(\ref{eq:A1}) and (\ref{eq:E}), respectively.

As a simple check, let us calculate the energy level shift at leading order:
\eq
\Delta E_1^{A_1^+}&=&\braket{A_1^+|\mathbf{H}_1|A_1^+}=\frac{40\pi a_0}{mL^3}
+O(L^{-4})\, ,
\nonumber\\[2mm]
\Delta E_1^{E^+}&=&\braket{E^+,\alpha|\mathbf{H}_1|E^+,\alpha}=\frac{16\pi a_0}{mL^3}
+O(L^{-4})\, .\label{eq:leading3}
\en
It is seen that the leading-order result of Ref.~\cite{Pang:2019dfe} is readily
reproduced. One might continue along the same path and evaluate all
contributions up to and including $O(L^{-6})$. However, since the result in
the non-relativistic limit is already known~\cite{Pang:2019dfe},
one can achieve the same goal with a significantly less effort, evaluating
only the relativistic corrections to this result. Below we shall demonstrate,
how this can be done.

Let us start from the energy shift in the irrep $E^+$, which does not contain
the three-body coupling $\eta_3^r$. The non-relativistic result, given
in Ref.~\cite{Pang:2019dfe}, reads:
\eq
\Delta E_1^{E^+}(\text{non-rel})=\frac{16\pi a_0}{mL^3}\,\biggl(1+\frac{h_1'}{L}+\frac{h_2'}{L^2}+\frac{h_4'}{L^3}+\cdots\biggr)\, . \label{eq:Enonrel}
\en
Here,
\eq
h_1'&=&y_1'a_0\, ,
\nonumber\\[2mm]
h_2'&=&y_2'a_0^2+\frac{\pi^2}{2}\,r_0a_0\, ,
\nonumber\\[2mm]
h_4'&=&y_4'a_0^3+\pi(11-{\cal I}_2)r_0a_0^2\, ,
\en
where the numerical constants $y_1'=2.984094$, $y_2'=3.001706$, $y_4'=-28.89478538$
are complicated combinations of the infinite sums. The quantity
${\cal I}_2$ is defined as:
\begin{align}
\begin{split}
   {\cal I}_2&=\sum_{{\bf n}\neq {\bf 0},-{\bf n}_0}\frac{1}{{\bf n}^2+{\bf n}{\bf n}_0}\doteq
   \sum_{{\bf n}\neq {\bf 0},-{\bf n}_0}\biggl(\frac{1}{{\bf n}^2+{\bf n}{\bf n}_0}
   -\frac{1}{{\bf n}^2+1}\biggr)-\frac{3}{2}
\\
   &-2\pi^2+\sum_{\mathbf n \neq \mathbf 0}\frac{\pi}{|{\bf n}|}\,e^{-2\pi|{\bf n}|}\simeq -6.37480912\, ,
   \end{split}
\end{align}
   where ${\bf n}_0=(0,0,1)$ is a unit vector~\cite{Pang:2019dfe}.

   Relativistic corrections to this result emerge from three different sources:
   \begin{enumerate}
   \item
     shifting the effective range $r_0\to r_0-1/(a_0m^2)$, according to Eq.~(\ref{eq:g123}):
     \eq
     \Delta E_1^{E^+}(\text{eff.range})=
     \frac{16\pi a_0}{mL^3}\,\biggl(-\frac{\pi^2}{2m^2L^2}-\frac{\pi a_0}{m^2L^3}\,(11-{\cal I}_2)\biggr)\, ;
     \en
   \item
     contributions, containing the coupling $g_2$, which vanishes in the
     non-relativistic limit. This contribution describes the CM motion of
     the two-particle subsystem in the three-particle system at rest:
     \eq
     \Delta E_1^{E^+}(\text{CM})&=&-\frac{32\pi^2}{L^5}\,g_2+\frac{8mg_1g_2}{L^6}\,
          (-3+{\cal I}_2)
     \nonumber\\[2mm]
&=&\frac{16\pi a_0}{mL^3}\,
     \biggl(-\frac{\pi^2}{m^2L^2}+\frac{2\pi a_0}{m^2L^3}\,({\cal I}_2-3)\biggr)\, ;
     \en
   \item
     contributions, emerging from the $O({\bf p}^4)$ kinetic term
     $\mathbf{H}_r$:
      \eq
     \Delta E_1^{E^+}(\text{kinetic})
     =\frac{16\pi a_0}{mL^3}\,\frac{\pi a_0}{2m^2L^3}\,(15-4{\cal S})\, ,
     \en
     where
     \eq
        {\cal S}&=&\sum_{\mathbf n\neq\mathbf 0,\mathbf n_0}
        \frac{\mathbf n^4+(\mathbf n-\mathbf n_0)^4+1}
             {(\mathbf n^2+(\mathbf n-\mathbf n_0)^2-1)^2}
\nonumber\\[2mm]
             &=&-\frac{13}{8}
             +\frac{1}{2}\,\sum_{\mathbf n\neq\mathbf 0,\mathbf n_0}
             (\mathbf n^2+(\mathbf n\mathbf n_0)^2-2\mathbf n\mathbf n_0+1)
             \biggl(
             \frac{1}{(\mathbf n^2-\mathbf n\mathbf n_0)^2}
             -\frac{1}{(\mathbf n^2+1)^2}\biggr)
\nonumber\\[2mm]
&-&\frac{1}{6}\,\sum_{\mathbf n}\frac{1}{(\mathbf n^2+1)^2}
-\frac{4\pi^2}{3}
               +\sum_{\mathbf n \neq 0}\frac{2\pi}{3|\mathbf n|}\,e^{-2\pi|\mathbf n|}
               \simeq 1.849866\, .
          \en

\end{enumerate}
Bringing everything together, we get:
\eq
\Delta E_1^{E^+}=\Delta E_1^{E^+}(\text{non-rel})+\Delta E_1^{E^+}(\text{rel})\, ,
\en
where   
\eq
\Delta E_1^{E^+}(\text{rel})=\frac{16\pi a_0}{mL^3}\,
\biggl(-\frac{3\pi^2}{2(mL)^2}+\frac{\pi a_0}{2m^2L^3}\,
(-19+6{\cal I}_2-4{\cal S})\biggr)\, ,
\en
and $\Delta E_1^{E^+}(\text{non-rel})$ as in Eq.~(\ref{eq:Enonrel}).

Next, we turn to the energy shift in the irrep $A_1^+$, which, in contrast
to the $E^+$, contains contribution from the three-body coupling. We shall
again be using a shortcut, calculating only the relativistic corrections
to the result given in Ref.~\cite{Pang:2019dfe}. To this end,
comparing first the expressions for the ground-state energy shift in
the non-relativistic limit, we read
off the relation between the three-body coupling $\eta_3^r(m)$
and the threshold amplitude $\hat{\cal M}$ from Ref.~\cite{Pang:2019dfe}:
\eq\label{eq:eta3-M}
\eta_3^r(m)=-\frac{96\pi^2a_0^2}{m}\,\hat{\cal M}+\frac{48\pi^2a_0^3}{m}\,r_0-\frac{12a_0^4}{\pi^2m}\,A\, ,
\en
where $A$ collects all numerical factors that account for the different regularization and conventions, used in Ref.~\cite{Pang:2019dfe} and the present article:
\begin{align}
\begin{split}
A&=\frac{1}{3}\,(2\pi)^3(3\sqrt{3}-4\pi)(2\ln 2\pi-\Gamma'(1)-\ln 4\pi)
+\pi\delta+16(\tilde{\cal Q}-{\cal Q})+8(\tilde{\cal R}-{\cal R})\\&\simeq 1278.49567\,. 
\end{split}
\end{align}
Here, $\delta\simeq 887.65392$ and $\tilde{\cal Q}\simeq -105.0597779$,
$\tilde{\cal R}\simeq -32.60560475$ are the counterparts of ${\cal Q},{\cal R}$
in the cutoff regularization. All these quantities have been defined
in Ref.~\cite{Pang:2019dfe}.

Next, let us consider the expression for the energy shift of the first excited
state in the $A_1^+$ irrep in the non-relativistic theory, given in Ref.~\cite{Pang:2019dfe}:
\eq
\Delta E_1^{A_1^+}(\text{non-rel})=\frac{40\pi a_0}{mL^3}\,
\biggl(1+\frac{h_1}{L}+\frac{h_2}{L^2}+\frac{h_3}{L^3}\,\ln\frac{mL}{2\pi}
+\frac{h_4}{L^3}\biggr)\, ,
\en
where
\eq
h_1&=&y_1a_0\, ,
\nonumber\\[2mm]
h_2&=&y_2a_0^2+\frac{7\pi^2}{5}\,a_0r_0\, ,
\nonumber\\[2mm]
h_3&=&\frac{144}{5}\,(3\sqrt{3}-4\pi)a_0^3\, ,
\nonumber\\[2mm]
h_4&=&y_4a_0^3+\frac{\pi}{10}\, a_0^2r_0(212-24{\cal I}_1-4{\cal I}_2)-\frac{27}{5}\,8\pi a_0\hat{\cal M}\, ,
\en
where $y_1\simeq 0.279070$, $y_2\simeq 8.494802$ and $y_4\simeq -172.001650$
are numerical constants, defined in Ref.~\cite{Pang:2019dfe}. Using
now the relation~(\ref{eq:eta3-M}), one may rewrite last of the above equations
as:
\eq
h_4=\biggl(y_4+\frac{27}{5\pi^3}\,A\biggr)a_0^3-\frac{2\pi}{5}\,a_0^2r_0(1+6{\cal I}_1+{\cal I}_2)+\frac{9m}{20\pi a_0}\,\eta_3^r(m)\, .
\en
Now, we can readily evaluate the relativistic corrections to this result from
three different sources, listed above:
\begin{enumerate}
\item
  shifting the effective range $r_0\to r_0-1/(a_0m^2)$:
  \eq
  \Delta E_1^{A_1^+}(\text{eff.range})=
  \frac{40\pi a_0}{mL^3}\,\biggl(-\frac{7\pi^2}{5m^2L^2}+\frac{2\pi a_0}{5m^2L^3}\,
  (1+6{\cal I}_1+{\cal I}_2)\biggr)\, ;
  \en
\item
  the CM motion of the two-particle subsystems:
  \eq
  \Delta E_1^{A_1^+}(\text{CM})&=&-\frac{32\pi^2}{L^5}\,g_2
  +\frac{8mg_1g_2}{L^6}\,(9+{\cal I}_2)
\nonumber\\[2mm]
  &=&\frac{40\pi a_0}{mL^3}\,\biggl(
  -\frac{2\pi^2}{5m^2L^2}+\frac{4\pi a_0}{5m^2L^3}\,(9+{\cal I}_2)\biggr);
  \en
  
\item
  the kinetic term:
  \eq
  \Delta E_1^{A_1^+}(\text{kinetic})=\frac{40\pi a_0}{mL^3}\,\frac{-\pi a_0}{5m^2L^3}\,
  (54+6{\cal I}_1+3{\cal J}_1+4{\cal S})\, ,
  \en
  where
  \eq
     {\cal J}_1=\sum_{|\mathbf n|\neq 1}\frac{1}{(\mathbf n^2-1)^2}\simeq 23.2432219\, ;
     \en
     Finally, expressing the three-body coupling in terms of the threshold amplitude
     $\bar \cT$, we arrive at the following relation:
     \eq\label{eq:A1-final}
 \Delta E_1^{A_1^+}&=& \frac{40\pi a_0}{mL^3}\,
\biggl(1+\frac{\tilde h_1}{L}+\frac{\tilde h_2}{L^2}+\frac{\tilde h_3}{L^3}\,\ln\frac{mL}{2\pi}
+\frac{\tilde h_4}{L^3}
\nonumber\\[2mm]
&-&\frac{9\pi^2}{5m^2L^2}+\frac{\pi a_0}{5m^2L^3}\,(92+6{\cal I}_1-3{\cal J}_1
+6{\cal I}_2-4{\cal S})\biggr)
\, ,
\en
where
\eq
\tilde h_1&=&h_1,\quad\quad
\tilde h_2=h_2,\quad\quad
\tilde h_3=h_3,
\nonumber\\[2mm]
\tilde h_4&=&\biggl(y_4+\frac{27}{5\pi^3}\biggr)a_0^3
-\frac{2\pi}{5}\,a_0^2r_0(-53+6{\cal I}_1+{\cal I}_2)-\frac{3m}{40\pi a_0}\,\bar\cT\, .
\en
The equation~(\ref{eq:A1-final}) represents our final result for the energy shift in the first excited state, irrep $A_1^+$. The last two terms in the brackets are the relativistic corrections.

  \end{enumerate}

\subsection{Leading contribution to $N$-particle excited states}

To conclude this section, we focus on the leading contribution to the first $N$-particle excited states. For this, we need the $N$-particle generalization of the $A^+_1$ and $E^+$ states, which must contain all possible pairs with
back-to-back momentum in one specific irrep:
\begin{equation}
\ket{\Gamma=A_1^+,E^+} = \sqrt{\frac{2}{N(N-1)}} \sum_{n_1, n_2 \in \text{comb($N$,2)}} \ket{({\bf p}_{n_1},{\bf p}_{n_2})_{\Gamma},{\bf p}_{n_3}=0, \hdots,{\bf p}_{n_N}=0},
\end{equation}
where ``comb($N$,2)" stands for all elements in the combination of $N$ particles in pairs, with the pair states given in Eqs.~(\ref{eq:A1}) and (\ref{eq:E}). For instance, in the case of $N=4$, the possible combinations are
\begin{equation}
\text{comb($N$,2)} =  \bigg \{ (1,2) ,\ (1,3),\  (1,4),\ (2,3),\ (2,4),\ (3,4) \bigg \}.
\end{equation}

Now, using the $N$-particle generalization of Eq.~(\ref{eq:pseudopotential3}), we calculate the leading effect to the $N$-particle energies with $\mathbf n^2=1$:
\begin{align}
\Delta E_1^{N, A_1^+} &= \frac{2 \pi a_0}{m L^3} \left( N^2 + 3N +2 \right) + O(L^{-4}), \\
\Delta E_1^{N, E^+} &= \frac{2 \pi a_0}{m L^3} \left( N^2 + 3N -10 \right) + O(L^{-4}).
\end{align}
The extension to higher orders in the previous equation is straightforward but rather tedious.

\section{Application: $N$-particle energy levels in $\varphi^4$ theory}

\label{sec:numerics}

In this section, we investigate the extraction of the scattering parameters on the lattice in the complex $\varphi^4$ theory. For the extraction, Eq.~(\ref{eq:deltaEn})
will be used. The main goal is to explore strategies and challenges in the determination of the three-particle
coupling from lattice simulations, using the fit to the ground-state
energy spectrum in the sectors for
two, three, four and five particles. Note that, to this end, in Ref. \cite{Romero-Lopez:2018rcb} only the two- and three-particle energies were used. In addition, we study the impact of
including relativistic and exponentially-suppressed corrections. Moreover, a comparison between the obtained results and perturbative predictions is made. In this manner, we confront the lattice description of the $\varphi^4$ theory to the continuum perturbation theory. For additional fits, we refer the reader to the following repository \cite{phi4:2020}.

\subsection{Perturbative results for the $\varphi^4$ theory in the continuum}
\label{sec:PT}
\begin{figure}[t]
\centering 
\includegraphics[width=.8\textwidth]{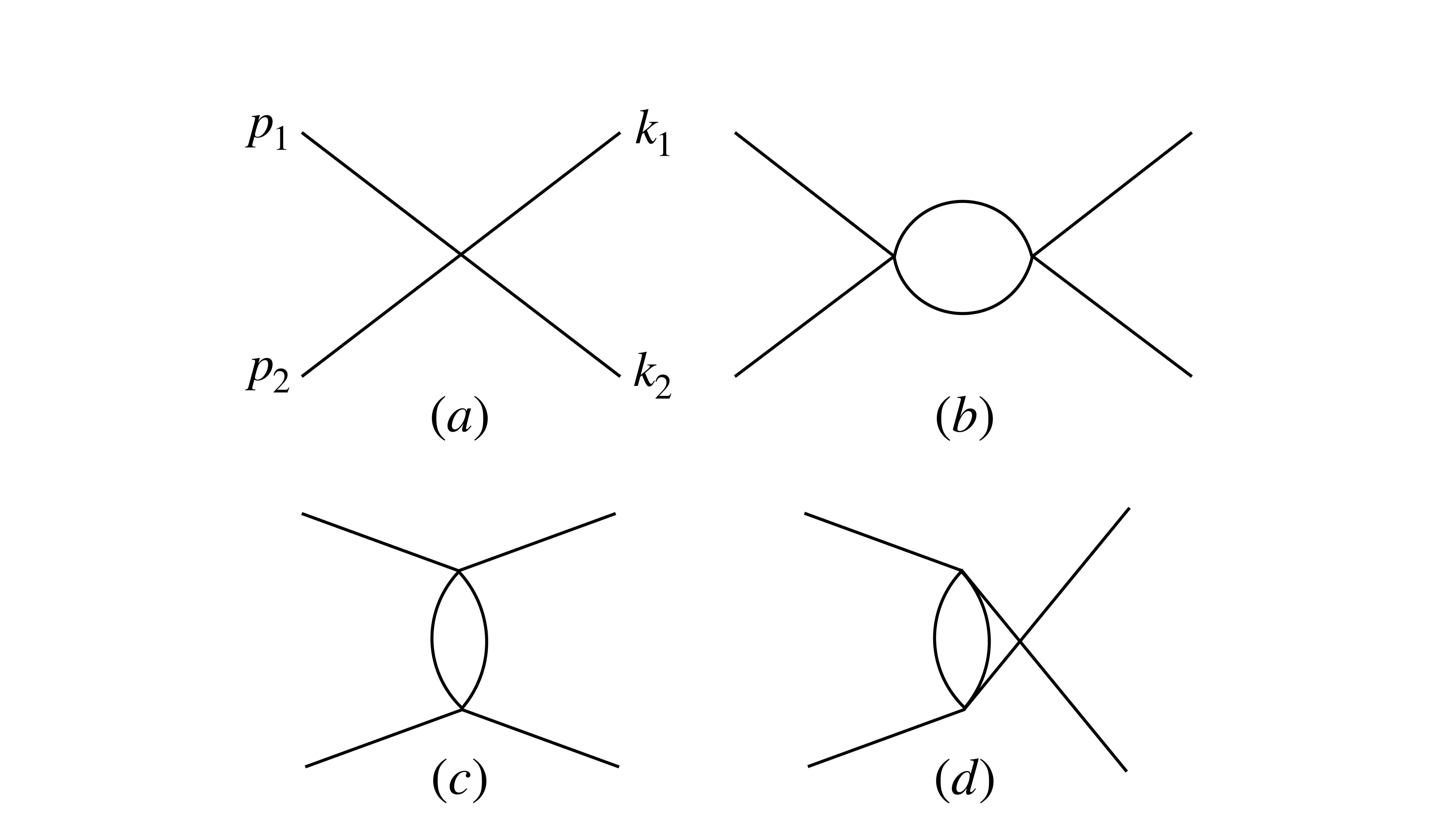}
\caption{\label{fig:diag2phi4}  Diagrams contributing to the one-loop two-particle scattering amplitude, $\mathcal{M}_2$.}
\end{figure}

Before turning to the lattice simulations, we discuss here perturbative results for the $\varphi^4$ theory in the continuum. This will provide  useful check for the scattering parameters, obtained from lattice simulations in this work. The starting point is the continuum Lagrangian of the complex $\varphi^4$ theory:
\begin{equation}
\mathcal{L} = \partial_\mu \varphi^*   \partial^\mu \varphi - m_0^2 \varphi^*  \varphi - \lambda_0 (\varphi^*  \varphi)^2. \label{eq:continuumlag}
\end{equation}
Here $m_0$ denotes the bare particle mass and $\lambda_0$ the dimensionless coupling constant.
The first quantity of interest is the two-particle scattering amplitude. Up to one loop, the contributing diagrams are shown in Fig. \ref{fig:diag2phi4}. The amplitude is then given by
\begin{equation}
\mathcal{M}_2 = -4 \lambda_0 + (4 \lambda_0)^2 \left[ \frac{1}{2} J(p_1+p_2) + J(p_1-k_1) + J(p_1-k_2) \right],
\end{equation}
with the choice of momenta as in Fig. \ref{fig:diag2phi4}, and the loop integral being
\begin{equation}
J(q) =  \int \frac{d^D k}{(2\pi)^Di} \frac{1}{m^2-k^2}\,
\frac{1}{m^2-(k+q)^2}. \label{eq:loopint}
\end{equation}
We shall use the dimensional regularization and $\overline{\text{MS}}$ scheme for renormalization:
\eq
J(q)=-2\,\frac{\mu^{D-4}}{16\pi^2}\,\biggl(\frac{1}{D-4}-\frac{1}{2}\,(\Gamma'(1)+\ln 4\pi)\biggr)+J_r(q)\, .
\en
In the two-particle sector, the two quantities of interest are the scattering length and the effective range. In order to compute them, we just need to compute the expansion up to $O(\mathbf p^2)$
in Eq.~(\ref{eq:loopint}):
\begin{align}
J_r(p_1+p_2) \equiv J_r(s)&= \frac{1}{16 \pi^2} \left[ - \ln \frac{m^2}{\mu^2} + 2  + i \pi \frac{|\mathbf p\, |}{m} - 2 \frac{ \mathbf p^2  }{m^2} +  O(p^3)\right], \\
J_r(p_1 - k_1)  \equiv J_r(t)&= -\frac{1}{16 \pi^2}\left[  \ln \frac{m^2}{\mu^2}  - \frac{ t}{6m^2} \right]  +  O(p^4), \ \ t = -2\mathbf p^2 (1- \cos \theta),  \\
J_r(p_1 - k_2)  \equiv J_r(u)&= -\frac{1}{16 \pi^2}\left[  \ln \frac{m^2}{\mu^2} -\frac{ u}{6m^2}\right]  +  O(p^4), \ \ u = -2\mathbf p^2 (1+ \cos \theta)\, . 
\end{align}
Carrying out the renormalization, $\lambda_0$ is replaced by the renormalized coupling
$\lambda_r$. The two-body scattering amplitude is then given by:
\begin{equation}
\mathcal{M}_2 = -4 \lambda_r +  \frac{(4 \lambda_r)^2}{16 \pi^2} \left[ - \frac{5}{2} \ln \frac{m^2}{\mu^2} + 1 + i \pi \frac{|\mathbf p\, |}{2m} - \frac{5}{3}\frac{\mathbf p^2}{m^2} + O(\mathbf p^3)\right] + O(\lambda^3) ,
\end{equation}
Now, by comparing to the expansion of Eq.~(\ref{eq:M2kcot}):
\begin{equation}
\mathcal{M}_2 = -32 \pi m a_0 \left[ 1  - i a  |\mathbf p|  + \frac{\mathbf p^2}{2m^2}\left( 1+ m^2 a_0 r_0 \right) \right] + O(\mathbf p^3) + O(a_0^3),
\end{equation}
we can obtain the one-loop values of the scattering length and the effective range:
\begin{equation}
 \lambda_r(m)=8\pi m a_0\biggl(1+\frac{2m a_0}{\pi}\biggr), \quad\quad
mr_0 =  - \frac{1}{m a_0} + \frac{20}{3 \pi}.
\label{eq:lambda_r}
\end{equation}
Note here that in  the running coupling $\lambda_r(m)$ the scale $\mu$ is set equal to $m$.

We now turn to the computation of the subtracted  three-particle scattering amplitude at threshold, $\bar{\cT}$. At leading order, the only diagram that contributes is like diagram (a) in Fig. \ref{fig:most_singular} (including all nine combinations of incoming and outgoing particles). The tree-level result is:
\begin{equation}
\bar{\cT}  = \frac{9 \lambda_r^2}{2m^5}  = \frac{288 \pi^2 a_0^2}{m^3}.
\label{eq:Tau}
\end{equation}

\subsubsection{Exponentially-suppressed corrections}

The Eq.~(\ref{eq:deltaEn}), which is used for the extraction of the scattering parameters
from lattice simulations, contains only terms that vanish like powers of $L$.
There are, however, exponentially-suppressed effects,
which might be relevant. In order to take them into account, one can apply perturbation theory in the continuum and evaluate the corrections
in the parameters that enter  Eq.~(\ref{eq:deltaEn}).
In this manner, it is possible to explicitly include the leading exponential corrections into
the fit.

We start from the single-particle mass, where the leading-order result of Ref. \cite{Gasser:1986vb}
will be used:
\begin{equation}
M_\varphi(L) = m + c_M \cdot \frac{K_1 (m L)}{m L} \to  m + c_M'\frac{e^{- mL}}{(mL)^{3/2}} .
\label{eq:M_phi}
\end{equation}
In this formula, $M_\varphi(L)$ denotes the finite-volume mass and $m$ the infinite-volume one. Further, $c_M, c'_M$ are constants whose exact value is not of interest at this stage,
and $K_\nu (z)$ are the modified Bessel functions of second kind.

The leading exponentially-suppressed finite-volume correction to the scattering length can be calculated, following the strategy explained in Ref. \cite{Romero-Lopez:2018rcb}. The result is given by:
\begin{equation}
a_0(L) = a_0\left( 1 - \frac{24 m a_0}{\pi} K_0(mL)  \right) \rightarrow a_0\left( 1 -  \frac{24 m a_0}{\sqrt{2 \pi m L}} e^{- mL}  \right). \label{eq:expa}
\end{equation}

\subsection{Lattice simulation}

For the numerical approach to complex scalar $\varphi^4$ theory, the discretization of
space-time
is required. This can be realized by introducing a finite hypercubic lattice of a
volume $T \cdot L^3$. Here $T$ denotes the Euclidean time length of the lattice
and $L$ is the
spatial extension. Such a four-dimensional lattice can be described by means of
discrete points $x^\mu$, separated by the lattice spacing $a$. In addition,
periodic boundary conditions
are assumed in all four directions. These are implemented by requiring
\eq
  \varphi \left( x^\mu +  Le^\mu  \right) = \varphi \left( x^\mu  \right)\,,\qquad&&\mbox{for}~\mu=1,2,3,
\nonumber\\[2mm]
  \varphi \left( x^\mu + Te^\mu  \right) = \varphi \left( x^\mu  \right)\,,\qquad&&\mbox{for}~\mu=4,
  \en
  where $\varphi \left(  x^\mu \right) \in \mathbb{C}$ is the complex scalar field and
  $e^\mu$ denotes a unit vector in the direction of axis $\mu$.

For the numerical simulation of field configurations, a discretization of the action is mandatory. This is realized by replacing the four-dimensional integral by a sum, as well as the partial derivatives by finite differences. Furthermore, introducing a hopping parameter $\kappa$, the fields are rescaled, according to
\begin{equation}
a\, \varphi \left( x^\mu \right) = \sqrt{\kappa}\, \varphi_x.
\end{equation}
Similarly, the dimensionful mass $m$ is removed from the action by converting the terms,
depending on absolute values of $\varphi_x$ to a binary quadratic form. For that purpose, we use the following relations:
\begin{align}
\label{eq:lambda}
  \lambda_0 = \frac{\lambda'}{\kappa^2}\, ,\quad\quad
  (a m)^2 = \frac{1 - 2\lambda' - 8\kappa}{\kappa}\, .
\end{align}
Taking this into account, the discretized expression for the action becomes:
\begin{equation}
  S = \sum_{x^\mu} \left[
    \Bigg( -\kappa \sum_\mu \varphi_x^* \, \varphi_{x+\mu}+ c.c. \Bigg)
    + \lambda' \, \left(|\varphi_x|^2 - 1\right)^2 + |\varphi_x|^2 \right].
\label{eq:phi4_action_2}
\end{equation}
In the simulations, particle mass and coupling constant are set to $m^2 = -4.9$ and $\lambda_0 = 10.0$, respectively. This leads to $\lambda^\prime = 0.253308$, and $\kappa = 0.159156$ in the discretized action, Eq.~(\ref{eq:phi4_action_2}). This choice---identical to Ref.~\cite{Romero-Lopez:2018rcb}---has the advantage that it leads to a small
renormalized coupling constant, which strongly suppresses excited states.

By means of Metropolis-Hastings simulations, ensembles are generated with constant $T = 48$ and varying $L$ in the interval $L \in [6, 24]$ (from now on, $L$ and $T$ will be given
in lattice units). More extended lattices are not studied,
because of the increased numerical cost. Likewise, smaller lattices with $L < 6$ are not taken into account, as they are dominated by the boundaries, and marginally incorporate physical information. The different volumes and the corresponding number of produced configurations, as well as the volume-dependent number of Metropolis updates per thread are summarized in Tab.~\ref{tab:configs}.
\begin{table}[H]
	\centering
	\begin{tabular}{rrr|rrr}
		\hline\hline
		$L$ & $n_\text{conf}$ & ud$/$thr & $L$ & $n_\text{conf}$ & ud$/$thr\\
		\hline \hline
		6 & 100,000 & 800 & 16 & 100,000 & 1880\\
		7 & 100,000 & 800 & 17 & 100,000 & 2240\\
		8 & 100,000 & 800 & 18 & 100,000 & 2800\\
		9 & 100,000 & 800 & 19 & 100,000 & 3128\\
		10 & 100,000 & 480 & 20 & 80,000 & 3800\\
		11 & 100,000 & 800 & 21 & 70,000 & 4224\\
		12 & 100,000 & 800 & 22 & 50,000 & 5000\\
		13 & 100,000 & 1000 & 23 & 50,000 & 5544\\
		14 & 100,000 & 1280 & 24 & 30,000 & 6400\\
		15 & 100,000 & 1600 &&&\\
	\hline\hline
	\end{tabular}
	\caption{Summary of simulations used in this work. $T=48$ is kept throughout. $n_\text{conf}$ labels the number of field configurations. The number of updates per thread, ud/thr, is provided as well. In order to decrease correlation, only every $1000^\text{th}$ configuration is saved.}
	\label{tab:configs}
\end{table}

\subsubsection{Effective potential}
\label{sec:effective_pot}

The $\varphi^4$ theory has two distinct phases: the symmetric and the broken one.
Before performing simulations, one wants to know, which phase is realized by a given
choice of parameters. Our framework, used for the analysis of the multi-particle
energy levels, can be used in the symmetric phase only, where the particle,
described by the field $\varphi$, has a non-zero mass. In order to answer the above question, note that measuring the effective potential allows one to distinguish between the two
phases: a monotonically growing potential is associated with the symmetric phase, whereas a ``Mexican-hat" potential characterizes the broken phase.

\begin{figure}[t]
	\centering
	\includegraphics[width=0.6\textwidth]{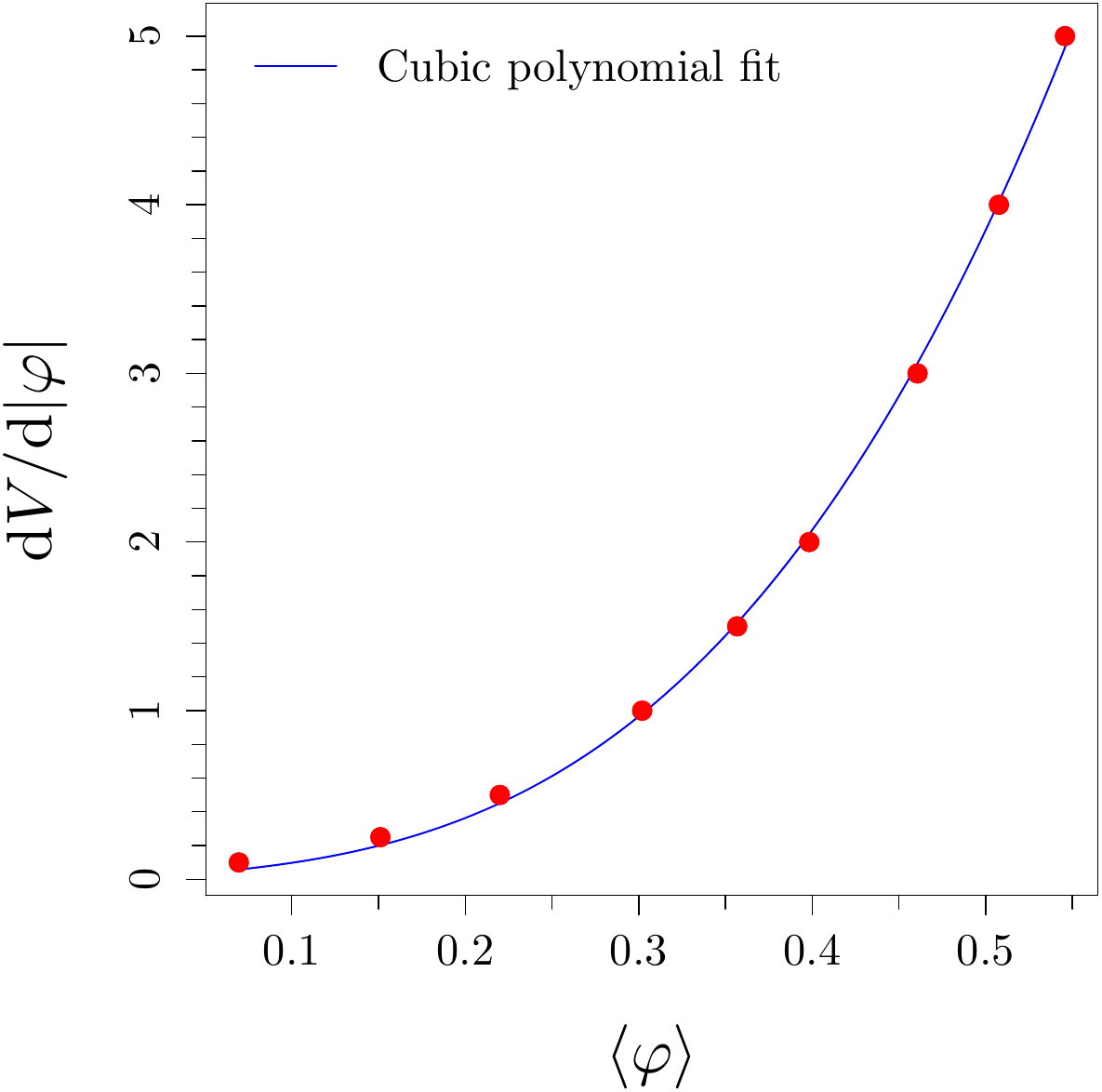}
	\caption{Derivative of the effective potential with respect to $|\varphi|$ as a function of $\langle\varphi\rangle$. The existing statistical errors on $\langle\varphi\rangle$ are too small to be visible. The blue curve corresponds to a cubic polynomial fit with Eq.~(\ref{eq:J}).}
	\label{fig:eff_pot}
\end{figure}

The effective potential of the $\varphi^4$ can be computed by including a term in the action---with the coefficient $J$---that explicitly breaks the $U(1)$ global symmetry.
The continuum Lagrangian in the presence of the symmetry-breaking term has the form:
\begin{equation}
\mathcal{L} = \partial_\mu \varphi^*   \partial^\mu \varphi - m_0^2 \varphi^*  \varphi - \lambda_0 (\varphi^*  \varphi)^2 - J | \varphi |. \label{eq:Jlag}
\end{equation}
This way, the effective (renormalized) potential
\begin{equation}
U(|\varphi|) = V(|\varphi|)  + J | \varphi | =  m^2 \varphi^*  \varphi + \lambda_r (\varphi^*  \varphi)^2 + J | \varphi |,
\end{equation}
will have a minimum at 
\begin{equation}
\frac{\text{d}U}{\text{d}|\varphi|} \Bigg \rvert_{\varphi = \langle \varphi \rangle} = 0 \longrightarrow J( \langle \varphi \rangle) = \frac{\text{d}V}{\text{d}|\varphi|}  \Bigg \rvert_{\varphi = \langle \varphi \rangle} .
\end{equation}
Now, by measuring $ \langle \varphi \rangle $ as a function of $J$, we could infer the shape of the potential:
\begin{equation}
{
\frac{\text{d}V}{\text{d}|\varphi|}  \Bigg \rvert_{\varphi = \langle \varphi \rangle} = a_1 \,\langle\varphi\rangle + a_2 \,\langle\varphi\rangle^3.
}
\label{eq:J}
\end{equation}
We have performed the simulations on a $16^4$ lattice for different values of $J$, using the same $m_0$ and $\lambda_0$ as in the ``main'' simulations. The results are shown in Fig.~\ref{fig:eff_pot}.
{The parameters obtained from the corresponding cubic polynomial fit read}
\begin{equation}
{
a_1 = 0.70(13), ~~~a_2 = 28.01(58).
}
\end{equation}
As can be seen, the shape of $\frac{\text{d}V}{\text{d}|\varphi|}$ is that of a positive monotonically growing function. This implies a monotonically growing effective potential, $V(\varphi)$. Therefore, we can confirm that the set of parameters, chosen for this work, corresponds to the symmetric phase of the $\varphi^4$ theory.

\subsection{The operators and correlation functions}

With the configurations from Tab.~\ref{tab:configs}, the one- to five-particle correlators are computed. The errors are determined from the corresponding bootstrap samples (standard deviation).  Then, particle energies are extracted by fitting the analytic form of the respective correlation function to the data. The $N$-particle correlation function can be written as:
\begin{equation}
C_N (t_\text{in}, t_\text{out}) = \Braket{\hat{\mathcal{O}}_{N\varphi} (t_\text{out}) \, \hat{\mathcal{O}}_{N\varphi}^\dagger (t_\text{in})},
\label{eq:Corr1}
\end{equation}
with 
\begin{equation}
\hat{\mathcal{O}}_{N\varphi} =\left( \sum_x  \varphi(x) \right)^N.
\end{equation}
We restrict ourselves to $N = 1, \dots, 5$ henceforth. In Euclidean time, the temporal evolution of an operator generally reads
\begin{equation}
\hat{\mathcal{O}} (t) = e^{\hat{H}\,t} \, \hat{\mathcal{O}}(0) \, e^{-\hat{H}\,t}.
\end{equation}
Considering this definition, with the source operator at the initial time $t_\text{in} = 0$, and the sink operator at the final time $t_\text{out} = t$, Eq.~(\ref{eq:Corr1}) can be expressed as
\begin{equation}
C_N (t) = \frac{1}{Z} \, \text{Tr}\, \Bigg[ e^{-\hat{H}\,(T-t)} \, \hat{\mathcal{O}}_{N\varphi}(0) \, e^{-\hat{H}\,t} \, \hat{\mathcal{O}}_{N\varphi}^\dagger (0) \Bigg].
\label{eq:Corr2}
\end{equation}
The trace occurring here is due to the periodic boundary conditions. It is the origin of the undesirable thermal pollutions of the correlation functions, that is, effects from particles propagating across the boundary backwards in time. In order to find the analytic form of the correlation function, Eq.~(\ref{eq:Corr1}) has to be expanded in an infinite sum of eigenstates, resulting in:
\begin{align}
C_N (t) &= \frac{1}{Z} \, \sum_{m,n} \left|\braket{n|\hat{\mathcal{O}}_{N\varphi}|m}\right|^2 \, e^{-E_n \,(T-t)} \, e^{-E_m \, t}\nonumber\\
&= \frac{1}{Z} \, \sum_{m,n} \left|\braket{n|\hat{\mathcal{O}}_{N\varphi}|m}\right|^2 \, e^{-(E_m + E_n)\, \frac{T}{2}} \, \cosh\left( (E_m - E_n) \, (t - \tfrac{T}{2}) \right).
\label{eq:Corr3}
\end{align}
Keeping only the leading terms, the explicit expressions of the one- to five-particle correlation function follows are:
\begin{align}
\label{eq:C_1}
C_1 (t) &= \big| A^{(1)}_{1 \leftrightarrow 0} \big|^2 \exp \left( -M_\varphi \tfrac{T}{2} \right) \, \cosh \left( M_\varphi (t - \tfrac{T}{2}) \right)\\[2mm]
C_2 (t) &= \big| A^{(2)}_{2 \leftrightarrow 0} \big|^2 \exp \left( -E_2 \tfrac{T}{2} \right) \, \cosh \left( E_2 (t - \tfrac{T}{2})\right) + \big| A^{(2)}_{1 \leftrightarrow 1} \big|^2 \exp \left( -M_\varphi T \right)\\[2mm]
C_3 (t) &= \big| A^{(3)}_{3 \leftrightarrow 0} \big|^2 \exp \left( -E_3 \tfrac{T}{2} \right) \, \cosh \left( E_3 (t - \tfrac{T}{2}) \right)\nonumber\\[2mm]
&\quad + \big| A^{(3)}_{2 \leftrightarrow 1} \big|^2 \exp \left( -(E_2 + M_\varphi) \tfrac{T}{2} \right) \, \cosh \left( (E_2 - M_\varphi) (t - \tfrac{T}{2}) \right)\\[2mm]
C_4 (t) &= \big| A^{(4)}_{4 \leftrightarrow 0} \big|^2 \exp \left( -E_4 \tfrac{T}{2} \right) \, \cosh \left( E_4 (t - \tfrac{T}{2}) \right)\nonumber\\[2mm]
&\quad + \big| A^{(4)}_{3 \leftrightarrow 1} \big|^2 \exp \left( -(E_3 + M_\varphi) \tfrac{T}{2} \right) \, \cosh \left( (E_3 - M_\varphi) (t - \tfrac{T}{2}) \right)\nonumber\\[2mm]
&\quad + \big| A^{(4)}_{2 \leftrightarrow 2} \big|^2 \exp \left( -E_2 T \right)\\[2mm]
C_5 (t) &= \big| A^{(5)}_{5 \leftrightarrow 0} \big|^2 \exp \left( -E_5 \tfrac{T}{2} \right) \, \cosh \left( E_5 (t - \tfrac{T}{2}) \right)\nonumber\\[2mm]
&\quad + \big| A^{(5)}_{4 \leftrightarrow 1} \big|^2 \exp \left( -(E_4 + M_\varphi) \tfrac{T}{2} \right) \, \cosh \left( (E_4 - M_\varphi) (t - \tfrac{T}{2}) \right)\nonumber\\[2mm]
&\quad + \big| A^{(5)}_{3 \leftrightarrow 2} \big|^2 \exp \left( -(E_3 + E_2) \tfrac{T}{2} \right) \, \cosh \left( (E_3 - E_2) (t - \tfrac{T}{2}) \right)
\label{eq:C_5}
\end{align}
The first term in each of the previous equations includes the energy to be determined. Except for the single-particle correlator $C_1$ all functions exhibit thermal-pollution terms in addition to the ``cosh'' signal. In case of $C_2$, this term is time-independent, whereas the remaining ones contain time-dependent thermal effects. Although these vanish in the large $T$-limit, they will taken into account, as the analysis is restricted to the finite lattice volumes. For instance, time-independent terms can be removed by shifting the affected functions:
\begin{equation}
C_N^\prime (t+1/2) = C_N (t) - C_N (t+1).
\label{eq:corr_shift}
\end{equation}
This also allows to reduce the correlation between different time slices. The analytical form of the shifted correlators will be as in Eqs.~(\ref{eq:C_1})-(\ref{eq:C_5}), with $\cosh$ replaced by $\sinh$.

\subsection{Fit strategies and parameter extraction}
\label{subsec:fits}

For the parameter extraction, the correlated fits are performed. When the correlation functions $C_N^\prime (t)$ with $N > 2$ are examined, energy levels corresponding to lower numbers of particles, contribute to the time-dependent thermal pollution terms, cf. Eq.~(\ref{eq:C_1}) to (\ref{eq:C_5}). Those terms will be taken into account in the fit. In order to increase the fit sensitivity to the $N$-particle energy $E_N$, lower lying energy levels, already determined in previous fits, may be used as priors for constraining the fit. This procedure is shown in the following scheme (\ref{eq:priors_scheme}):
\begin{equation}
C_2^\prime (t) \xrightarrow{E_2 - M_\varphi} C_3^\prime (t) \xrightarrow{E_3 - M_\varphi} C_4^\prime (t) \xrightarrow{E_4 - M_\varphi, E_3 - E_2} C_5^\prime (t)
\label{eq:priors_scheme}
\end{equation}
Hence, the best fit parameters obtained previously---shown on top of the arrows---enter the current fit as priors. The $\chi^2$-function has to be modified accordingly:
\begin{equation}
\chi^2 \longrightarrow \chi^2 + \sum_{i=1}^{k-1} \left( \frac{P_{p_i} - p_i}{\Delta p_i} \right)^2.
\label{eq_priors1}
\end{equation}
Here, $P_{p_i}$ denotes the prior applied to constrain the $i^\text{th}$ parameter $p_i$ and $\Delta p_i$ the corresponding standard error. Finally, the energy shifts with respect to the free energies are
computed as:
\begin{equation}
\Delta E_N (L) = E_N (L) - N \cdot M_\varphi (L) ~~~\text{with}~~~ M_\varphi (L) \equiv E_1 (L),
\label{eq:DeltaE_k}
\end{equation}
including the corresponding bootstrap samples. Note that the volume-dependent mass is used for the computation of the shift, as it has been shown to cancel
the leading exponentially-suppressed effects \cite{Romero-Lopez:2018rcb}.

The first quantity to be studied is the single-particle mass $M_\varphi(L)$ and its volume-dependence. This is relevant because the mass enters in the threshold expansion in Eq.~(\ref{eq:deltaEn}). We will use the infinite-volume mass, $m$, obtained by fitting Eq.~(\ref{eq:M_phi}) to the $M_\varphi(L)$  data. In practice, using $m$ or $M_\varphi(L)$ in Eq.~(\ref{eq:deltaEn}) induces only higher-order exponentially-suppressed effects.

Once the infinite-volume mass is determined, different strategies are carried out to fit the energy shifts:
\begin{enumerate}
\item Fixed-$N$ fits. Here the energy shifts of $N$ particles are fitted separately. For $N>2$, we also employ $a_0$ and $r_0$ as priors. For this, the $a_0$ and $r_0$, extracted from the two-body sector, are used as input for the analysis of the three- to five-body sector.  Priors in $r_0$ are crucial to disentangle $r_0$ and $\bar{\cT}$, as the two parameters enter at the same order. This also increases the fit sensitivity for the threshold amplitude $\bar{\mathcal{T}}$. According to the prescription in Eq.(\ref{eq_priors1}), the modification of the $\chi^2$-function is given by
\begin{equation}
  \chi^2 \rightarrow \chi^2 + \left( \frac{P_{a_0} - a_0}{\Delta a_0} \right)^2 +
  \left( \frac{P_{r_0} - r_0}{\Delta r_0} \right)^2.
\end{equation}
\item Unconstrained global fits. This fit includes all data with $N=2-5$. It is based on the determination of the values of $\{a_0, r_0, \bar{\mathcal{T}}\}$ that minimize the sum of all contributing $\chi^2$-functions, i.e.,
\begin{equation}
\chi^2_\text{global} (\tilde{p}) = \sum_{N=2}^5\, \sum_{i \in D} \frac{\left[ \Delta E_{N,\text{d}} (L_i) - \Delta E_N (L_i, \tilde{p}) \right]^2}{\left[ \Delta (\Delta E_{N, \text{d}} (L_i)) \right]^2}.
\end{equation}
Here, $\tilde{p}$ denotes a vector containing all fit parameters. Furthermore, $d$ labels the data, $L_i$ denote the sizes of different boxes used in the calculations, and $D$ represents a complete set of ensembles.
\item Prior-constrained global fits. In this case, the data with $N=3-5$ is considered, and we include priors in $a_0, r_0$ from the $\Delta E_2$-fits.
\end{enumerate}

Finally, it is very important to check the convergence of the $1/L$ expansion of the energy
shifts. For this purpose, we study the phase shift, extracted from the L\"uscher formalism~\cite{Luscher:1991n1}, and compare it to the scattering parameters, obtained from the various fits to Eq.~(\ref{eq:deltaEn}). According to \cite{Luscher:1991n1}, the $S$-wave phase shift is determined by
\begin{equation}
\cot (\delta) = \frac{Z_{00}(1,q^2)}{\pi^{3/2} \, q} \qquad \text{with} \qquad q = \frac{L \, k}{2 \pi}
\label{eq:delta}
\end{equation}
and the L\"uscher zeta function $Z_{00}$. The momentum $k$ can be calculated as
\begin{equation}
E_2 = 2\sqrt{k^2 + M_\varphi(L)^2}.
\label{eq:k}
\end{equation}
Furthermore, the effective-range expansion at order $k^6$ reads
\begin{equation}
k\cot(\delta) = -\frac{1}{a_0} + \frac{r_0 k^2}{2} - P r_0^3 k^4 + \mathcal{O}(k^6),
\end{equation}
where $P$ denotes the $S$-wave shape parameter. Starting from this expansion, four different models are set up, as summarized in Tab.~\ref{tab:Swave_models}. They are fitted to the phase-shift data, calculated from Eq.~(\ref{eq:delta}) using bootstrap samples. This allows to extract both $a_0$ and $r_0$ independently from the $N$-particle ground-state fits. In case of model (IIa) and (IIb), the $S$-wave shape parameter $P$ is obtained in addition. The ``a'' and ``b'' models should be equivalent, and they allow us to explore the systematics of different fitting techniques.
\begin{table}[H]
	\centering
	\begin{tabular}{ll}
		\hline\hline\\
		(Ia) & $\delta (k) = \arctan \left[ \left(-\frac{1}{a_0 k} + \frac{r_0 k}{2} \right)^{-1} \right]$\vspace{0.5cm}\\
		(Ib) & $k \cot\left[\delta(k)\right] = -\frac{1}{a_0} + \frac{r_0 k^2}{2}$\vspace{0.5cm}\\
		(IIa) & $\delta (k) = \arctan \left[ \left(-\frac{1}{a_0 k} + \frac{r_0 k}{2} - P r_0^3 k^3 \right)^{-1} \right]$\vspace{0.5cm}\\
		(IIb) & $k\, \cot\left[\delta(k)\right] = -\frac{1}{a_0} + \frac{r_0 k^2}{2} - P r_0^3k^4$\vspace{0.5cm}\\
		\hline\hline
	\end{tabular}
	\caption{$S$-wave phase-shift fit models.}
	\label{tab:Swave_models}
\end{table}

\subsection{Results}

In this section, the numerical results are presented. The material is organized as follows.
First, the volume-dependence of the single-particle mass is analyzed. Then, the energy shifts are investigated at order $L^{-5}$.  After that, the results, obtained from different types of fits at $O(L^{-6})$---described in Section~\ref{subsec:fits}---are presented.
The effects of relativistic and exponentially-suppressed corrections are studied separately.
Finally, the results, obtained from the lattice simulations, are compared to the corresponding perturbative predictions.

\subsubsection{Volume dependence of the mass}

\begin{figure}[H]
	\centering
	\includegraphics[width=0.6\textwidth]{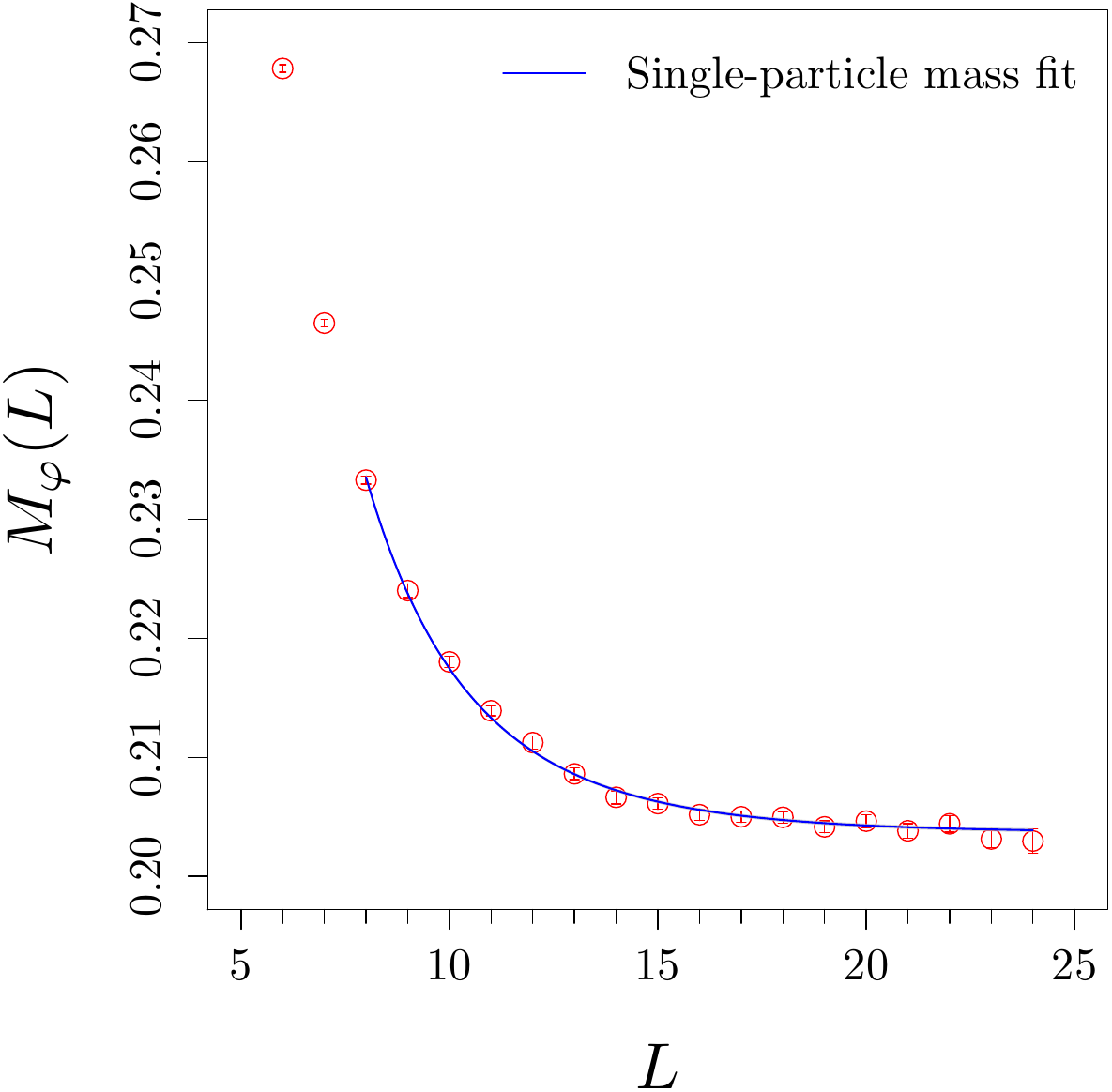}
	\caption{Volume dependence of the single-particle mass. The lattices with temporal extension $T = 48$ are considered. The blue curve corresponds to the fit in the range [8,24]
          for $L$.}
	\label{fig:massfit}
\end{figure}

The results from the fits of the mass to Eq.~(\ref{eq:M_phi}) for different ranges
of $L$ are summarized in Tab.~\ref{tab:massfit}. The best result for the infinite-volume mass is obtained from the range $[8, 24]$:
\begin{equation}
m = 0.20368(17), ~~~\chi^2 /\text{ndof} = 13.13 / 15, ~~~p\text{-value} = 0.59
\label{eq:mass}
\end{equation}
The corresponding curve is shown in Fig.~\ref{fig:massfit}, suggesting that a rough approximation $M_\varphi (L) \approx m$ is valid for $L \geq 20$, or equivalently,
$M_\varphi(L)L\geq 4$. In the forthcoming analysis, the numerical result in
Eq.~(\ref{eq:mass}) will be used, when fitting to Eq.~(\ref{eq:deltaEn}).
\begin{table}[H]
	\centering
	\begin{tabular}{ccccc}
		\hline\hline
		fit interval & $M_\infty$ & $c_M$ & $\chi^2 / \text{ndof}$ & $p$-value\\
		\hline\hline
		$[6, 24]$ & 0.20441(15) & 0.19079(84) & $134.17 / 17 = 7.89$ & 0.00\\
		$[7, 24]$ & 0.20393(16) & 0.2007(13) & $42.87 / 16 = 2.68$ & 2.93e-04\\
		$[8, 24]$ & 0.20368(17) & 0.2107(22) & $13.13 / 15 = 0.88$ & 0.59\\
		$[9, 24]$ & 0.20352(18) & 0.2201(48) & $8.20 / 14 = 0.59$ & 0.88\\
		$[10, 24]$ & 0.20345(19) & 0.2251(61) & $6.52 / 13 = 0.50$ & 0.93\\
		\hline\hline
	\end{tabular}
	\caption{Results, obtained from the single-particle mass fits.}
	\label{tab:massfit}
\end{table}
\subsubsection{$N$-particle ground state to order $L^{-5}$}
\label{subsec:Lmin5_fits}
\begin{figure}[h!]
	\centering
	\includegraphics[width=0.6\textwidth]{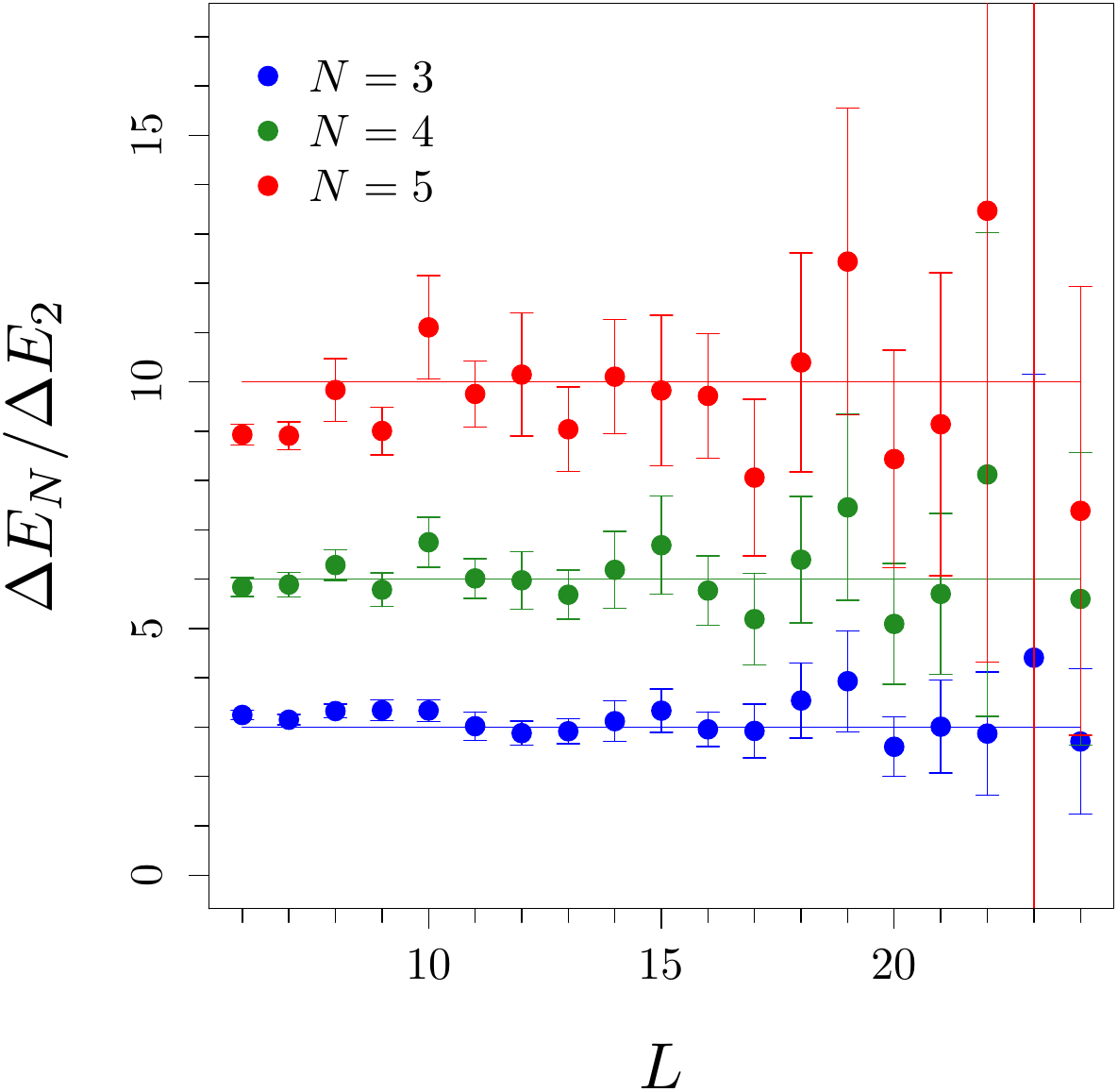}
	\caption{Energy shift ratios in dependency of the lattice parameter $L$. For orientation the colored straight lines mark the ratio values $3, 6, 10$.}
	\label{fig:DEk_DE2}
\end{figure}

We start by studying the energy shifts at order $L^{-5}$. At this order, all energies are solely described by the scattering length. First, the ratios of energy shifts are explored. Because of the binomial coefficients, occurring in the $N$-particle threshold expansion, Eq.~(\ref{eq:deltaEn}), these ratios behave as:
\begin{equation}
\frac{\Delta E_N}{\Delta E_2} = \binom{N}{2} + \mathcal{O}\left( L^{-5} \right).
\end{equation}
The results are shown in Fig.~\ref{fig:DEk_DE2}, where it can be seen that the
values are in a right ballpark. The accuracy of the energy shift ratios, however,
does not suffice for reliable parameter extraction. For this reason,
Fig.~\ref{fig:DEk_DE2} should only be considered as a rough quality test of the simulation.
\begin{table}[h!]
	\centering
	\begin{tabular}{ccccc}
		\hline\hline 
		fit & EC & $a_0$ & $\chi^2 / \text{ndof}$ & $p$-value\\
		\hline\hline
		$\Delta E_2$ & $\times$ & 0.421(18) & $7.42 / 10 = 0.74$ & 0.69\\
		& $\checkmark$ & 0.428(18) & $7.17 / 10 = 0.72$ & 0.71\\
		\hline
		$\Delta E_3$ & $\times$ & 0.441(16) & $4.66 / 10 = 0.47$ & 0.91\\
		& $\checkmark$ & 0.449(16) & $4.29 / 10 = 0.43$ & 0.93\\
		\hline
		$\Delta E_4$ & $\times$ & 0.425(15) & $8.18 / 10 = 0.82$ & 0.61\\
		& $\checkmark$ & 0.432(15) & $7.90 / 10 = 0.79$ & 0.64\\
		\hline
		$\Delta E_5$ & $\times$ & 0.410(14) & $8.77 / 10 = 0.88$ & 0.55\\
		& $\checkmark$ & 0.418(15) & $8.46 / 10 = 0.85$ & 0.58\\
		\hline\hline
	\end{tabular}
	\caption{Comparison of the results obtained from the energy shift fits to order $L^{-5}$ in the interval $[14, 24]$. We provide the results with as well as without EC}
	\label{tab:Eshift_results}
\end{table}

Next, we perform fits to the threshold expansion up to the same order. To this end,
we truncate Eq.~(\ref{eq:deltaEn}) at $O(L^{-5})$, and include only the
data at larger values of $L$,  for which this approximation should be valid. We separate
fits for $N=2,3,4$ and $5$, and obtain $a_0$. The best fit range is determined by
varying the lower bound of the fit interval, while the upper one is kept fixed at $L = 24$.
The best choice turns out to be $[14, 24]$, for which the results are summarized in
Tab.~\ref{tab:Eshift_results}. We also perform a fit, including the exponentially-suppressed corrections (EC) of Eq.~(\ref{eq:expa}). As expected, the results with
and without EC match reliably. This means that, for the values of $L$, considered
here, the EC are weak, and vanish within the range of statistical errors. Moreover,
the fits for different values of $N$ agree nicely within the uncertainty.

\subsubsection{Fixed-$N$ fits to order $L^{-6}$}
\label{subsec:Lmin6_fits}

In order to extract information on the three-body interactions---parameterized by
the quantity $\bar{\cT}$---the fits at order $O(L^{-6})$ should be performed.
We start with the fixed-$N$ fits as described in Section \ref{subsec:fits}. We first fit $\Delta E_2$, and use $a_0, r_0$ as priors for $N>2$. The selected results come from the interval $[9, 24]$. We shall try various fit models to explore the effects of EC and relativistic corrections (RC). The latter are the $O(L^{-6})$ terms in Eq.~(\ref{eq:deltaEn}) that vanish, when $m\to \infty$. All results are displayed Tab.~\ref{tab:Eshift_resultsLmin6}.
\begin{table}[h!]
	\centering
	\begin{tabular}{cccccccc}
		\hline\hline 
		fit & EC & RC & $a_0$ & $r_0$ & $\bar{\mathcal{T}}$ & $\chi^2 / \text{ndof}$ & $p$-value\\
		\hline\hline
		$\Delta E_2$ & $\times$ & $\times$ & 0.430(16) & $-319(17)$ & $-$ & $12.31 / 14$ & 0.58\\
		& $\times$ & $\checkmark$ & 0.430(16) & $-291(17)$ & $-$ & $12.31 / 14$ & 0.58\\
		& $\checkmark$ & $\times$ & 0.432(16) & $-$256(17) & $-$ & $11.69 / 14$ & 0.63\\
		& $\checkmark$ & $\checkmark$ & 0.432(16) & $-$228(18) & $-$ & $11.69 / 14$ & 0.63\\
		\hline
		$\Delta E_3$ & $\times$ & $\times$ & 0.426(14) & $-314(16)$ & $-354647(40323)$ & $20.08 / 15$ & 0.17\\
		& $\times$ & $\checkmark$ & 0.426(14) & $-286(17)$ & $-262938(35766)$ & $20.08 / 15$ & 0.17\\
		& $\checkmark$ & $\times$ & 0.428(14) & $-251(16)$ & $-289575(36963)$ & $19.18 / 15$ & 0.21\\
		& $\checkmark$ & $\checkmark$ & 0.428(14) & $-222(17)$ & $-196974(32676)$ & $19.18 / 15$ & 0.21\\
		\hline
		$\Delta E_4$ & $\times$ & $\times$ & 0.432(13) & $-315(16)$ & $-348072(38974)$ & $18.12 / 15$ & 0.26\\
		& $\times$ & $\checkmark$ & 0.432(13) & $-287(16)$ & $-253866(34474)$ & $18.12 / 15$ & 0.26\\
		& $\checkmark$ & $\times$ & 0.434(13) & $-252(16)$ & $-280922(35690)$ & $17.63 / 15$ & 0.28\\
		& $\checkmark$ & $\checkmark$ & 0.434(13) & $-224(17)$ & $-185831(31390)$ & $17.63 / 15$ & 0.28\\
		\hline
		$\Delta E_5$ & $\times$ & $\times$ & 0.431(13) & $-316(17)$ & $-341000(38374)$ & $16.75 / 15$ & 0.33\\
		& $\times$ & $\checkmark$ & 0.431(13) & $-288(17)$ & $-247376(33891)$ & $16.75 / 15$ & 0.33\\
		& $\checkmark$ & $\times$ & 0.433(13) & $-253(17)$ & $-274134(35092)$ & $16.41 / 15$ & 0.36\\
		& $\checkmark$ & $\checkmark$ & 0.433(13) & $-225(17)$ & $-179652(30784)$ & $16.41 / 15$ & 0.36\\
		\hline\hline
	\end{tabular}
	\caption{The results, obtained from the fixed-$N$ energy shift fits at order $L^{-6}$ in
          the interval $[9, 24]$ with and without EC and RC. The results for $a_0$ and
          $r_0$ obtained from the $\Delta E_2$-fit serve as priors in the remaining fits.}
	\label{tab:Eshift_resultsLmin6}
\end{table}

The fit results for $a_0$ are in agreement with the findings from the fits at order $L^{-5}$, cf. Tab.~\ref{tab:Eshift_results}. As can be seen, the fit quality---represented by $\chi^2 / \text{ndof}$ and $p$-value---is not affected by the RC. Yet, it slightly improves when EC is included. On the other hand, both RC and EC modify significantly the best fit parameters. Since the change is comparable to---or even larger than---the statistical error,
It is important to include these effects in the fit.

\subsubsection{Global $N$-particle ground-state fits}
\label{subsec:global_fits}

\begin{figure}[h!]
	\centering
	\includegraphics[width=0.6\textwidth]{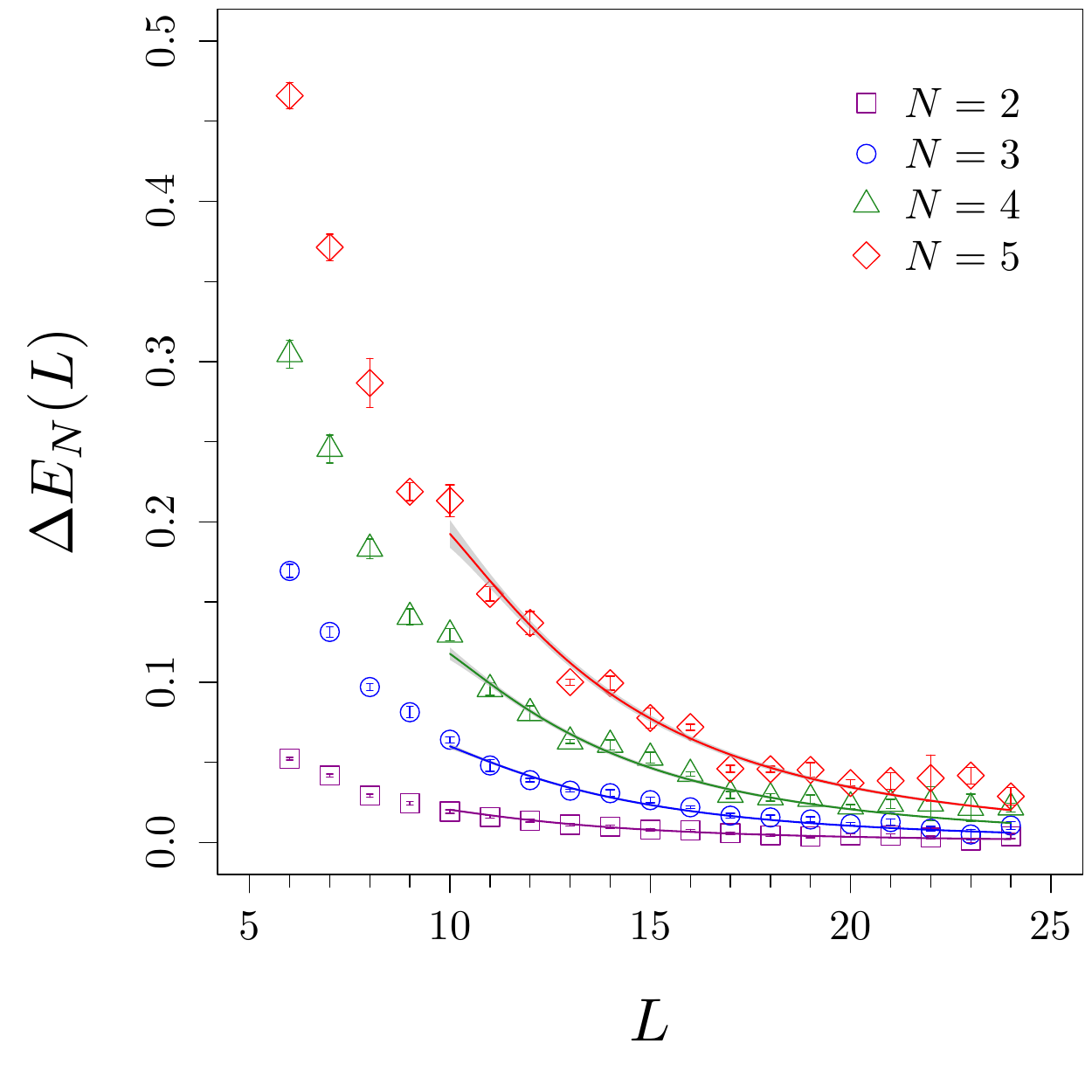}
	\caption{Global fit without priors in the interval $[10, 24]$ for $L$.
          Both EC and RC are included.}
	\label{fig:global_fit}
\end{figure}

In order to increase the fit sensitivity for $\bar{\mathcal{T}}$, it is appropriate to try
another strategy, namely, a combined fit of Eq.~(\ref{eq:deltaEn}) to
the energy shifts $\Delta E_{N=2,3,4,5}$.
Furthermore, global fits to the $\Delta E_{N=3,4,5}$ data with
prior-constrained scattering length (or both $a_0$ and $r_0$), determined from the $\Delta E_2$-fit to order $L^{-6}$, are conducted. Again, the lower fit interval limit is varied, while the upper one stays fixed at $L = 24$. The best results are obtained from the fit interval $[10, 24]$. They are summarized in Tab.~\ref{tab:global_Eshift_resultsLmin6}.
\begin{table}[h!]
	\centering
	\begin{tabular}{cccccccc}
		\hline\hline 
		priors & EC & RC & $a_0$ & $r_0$ & $\bar{\mathcal{T}}$ & $\chi^2 / \text{ndof}$ & $p$-value\\
		\hline\hline
		$-$ & $\times$ & $\times$ & 0.438(15) & $-320(21)$ & $-362327(52137)$ & $52.83 / 57$ & 0.63\\
		$-$ & $\times$ & $\checkmark$ & 0.438(15) & $-292(21)$ & $-265422(46637)$ & $52.83 / 57$ & 0.63\\
		$-$ & $\checkmark$ & $\times$ & 0.439(15) & $-255(22)$ & $-286929(47814)$ & $52.29 / 57$ & 0.65\\
		$-$ & $\checkmark$ & $\checkmark$ & 0.439(15) & $-227(22)$ & $-189799(42507)$ & $52.29 / 57$ & 0.65\\
		\hline
		$a_0$ & $\checkmark$ & $\checkmark$ & 0.443(15) & $-207(25)$ & $-165483(45346)$ & $37.19 / 43$ & 0.72\\
		$a_0$, $r_0$ & $\checkmark$ & $\checkmark$ & 0.450(12) & $-234(18)$ & $-216760(31039)$ & $42.05 / 44$ & 0.56\\
		\hline\hline
	\end{tabular}
	\caption{Comparison of the results, obtained from the global energy shift fits to order $L^{-6}$ in the interval $[10, 24]$ with and without using priors, EC and RC.
          The four results above the horizontal line correspond to the fits without priors.
          The last two results are fits, for which the parameter(s) $a_0$ ($a_0$ and $r_0$) obtained from the corresponding $\Delta E_2$-fit within the interval $[10,24]$ serve(s) as prior(s). }
	\label{tab:global_Eshift_resultsLmin6}
\end{table}
For illustrative purposes, a global fit with EC as well as RC without any prior knowledge is shown in Fig.~\ref{fig:global_fit}.

The results without priors are highly consistent with those from the previous fixed-$N$ fits to order $L^{-6}$, cf. Tab.~\ref{tab:Eshift_resultsLmin6}. Similarly to the fixed-$N$ fits, EC slightly improve the fit quality, and both RC and EC affect the best fit parameters beyond the size of the statistical error. If only $a_0$ is used as prior, the fit is consistent but errors remain comparable. Yet, the inclusion of priors in $a_0$ and $r_0$ reduces notably the statistical errors, while changing the best fit parameters only by order the statistical error. Remarkably, in this last fit we observe that $\bar{\cT}\neq 0$ with $7\sigma$ considering only the statistical error. 

\subsubsection{Phase-shift fits}
\label{subsec:phase_shift_fits}

In order to check the convergence of the $1/L$ expansion, we will compare the previous fit results to the ones from the full L\"uscher formalism---see Eq.~(\ref{eq:delta}). All fit models, used in the phase-shift analysis, are listed in Tab.~\ref{tab:Swave_models}. In general, the best results are obtained from the fits to the complete set of data points. This is particularly important for the fit models that include the shape parameter, $P$. The results are summarized in Tab.~\ref{tab:delta_fit_results}.

\begin{figure}[H]
	\centering
	\includegraphics[width=0.6\textwidth]{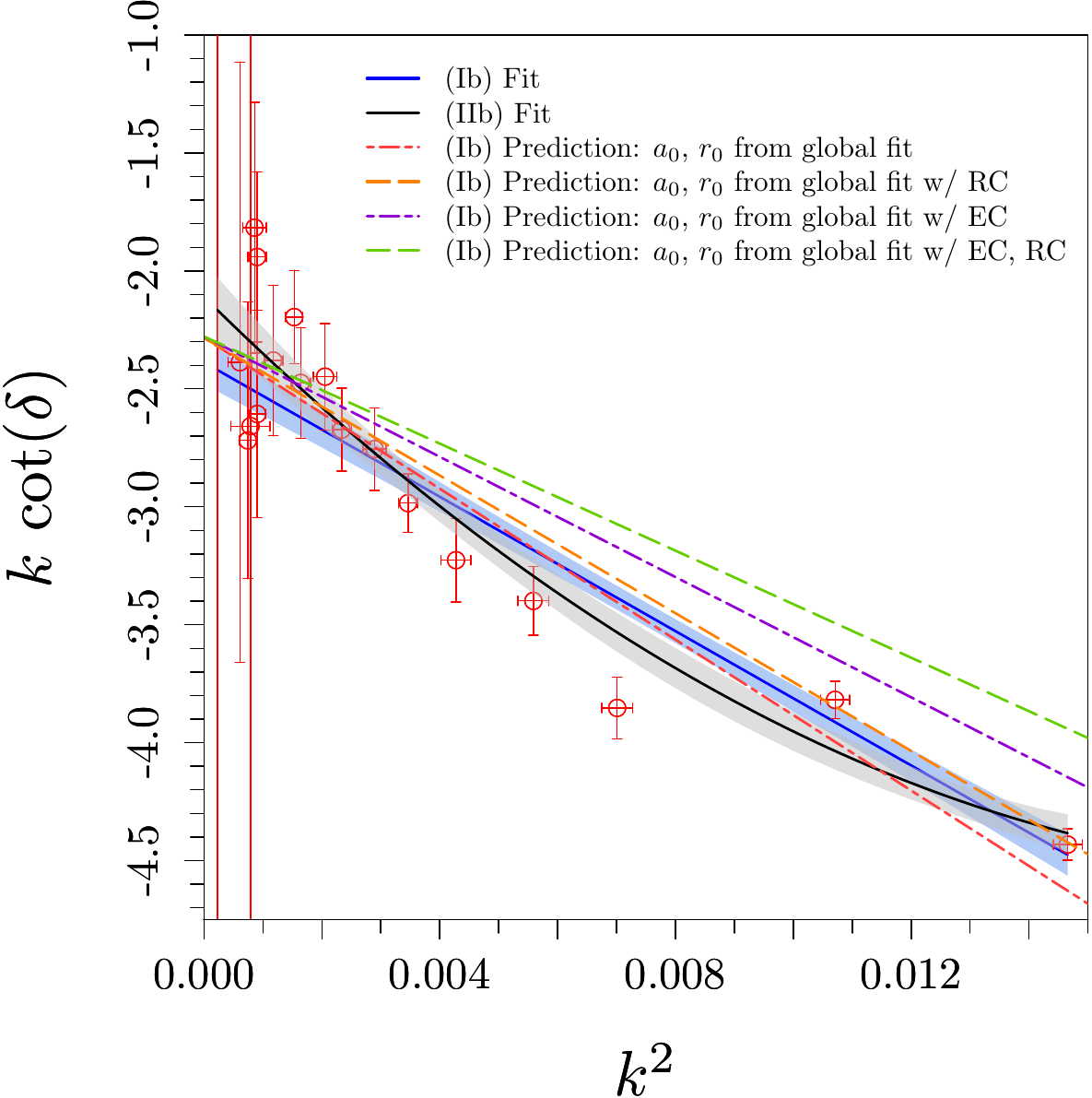}
	\caption{Comparison of the phase-shift fit, which is using the model (Ib) as well as
          (IIb), with the predictions from the global fit without priors. All predictions,
          displayed here, are based on the model (Ib) and involve the scattering parameters,
          obtained from the global fits shown in Tab.~\ref{tab:global_Eshift_resultsLmin6}.
          Both fit curves are plotted with error bands. The corresponding fit results are listed
          in Tab.~\ref{tab:delta_fit_results}.}
	\label{fig:kcot_delta_1}
\end{figure}

As it can be seen, the results from fits, which are using model (Ia) and (Ib), are in good agreement with the previous findings, which are obtained  from the perturbative expansion
of the the $N$-particle ground-state energies, cf. with Tabs.~\ref{tab:Eshift_results}, \ref{tab:Eshift_resultsLmin6} and \ref{tab:global_Eshift_resultsLmin6}. In contrast, the results
from the fits with (IIa) and (IIb) do not seem to match those from the energy shift fits.
A potential explanation could be provided by exponentially-suppressed effects that
have been seen to be relevant before---see e.g. Tab. \ref{tab:global_Eshift_resultsLmin6}---and are not considered in the phase-shift data points. It is important to note that we
must include points with large $k^2$ to fit the parameter $P$ reliably. However, these
points originate from simulations with $L<8$, and exponentially-suppressed effects can
be very significant. To check this, in Fig.~\ref{fig:kcot_delta_1} we plot the results from
the fits (Ib) and (IIb) along with the predictions, based on the results for $a_0$ and $r_0$,
obtained from the global energy shift fits in Tab. \ref{tab:global_Eshift_resultsLmin6}.
If EC are not included in the expansion of the energy levels, the resulting line describes
the red data points very well. From this, one can conclude two things: (i) the convergence
of the $1/L$ is not of a concern here, and (ii) EC produce a visible impact. A side
observation is that RC are also noticeable in the phase-shift curve.
\begin{table}[h!]
	\centering
	\begin{tabular}{cccccc}
		\hline\hline
		fit model & $a_0$ & $r_0$ & $P$ & $\chi^2 / \text{ndof}$ & $p$-value\\
		\hline\hline
		(Ia) & 0.407(17) & $-274(23)$ & $-$ & $14.63 / 17 = 0.86$ & 0.62\\
		(Ib) & 0.419(16) & $-285(21)$ & $-$ & $28.20 / 17 = 1.66$ & 0.04\\
		(IIa) & 0.483(36) & $-534(93)$ & $0.000049(19)$ & $10.13 / 16 = 0.63$ & 0.86\\
		(IIb) & 0.474(33) & $-494(89)$ & 0.000052(35) & $19.96 / 16 = 1.25$ & 0.22\\
		\hline\hline
	\end{tabular}
	\caption{Results from the different $S$-wave phase-shift fits to the complete set of data points.}
	\label{tab:delta_fit_results}
\end{table}

\subsubsection{$N$-particle ground-state fits with perturbative constraints}
\label{subsec:pt_fits}

As explained in Section \ref{sec:PT}, in the $\varphi^4$ theory, the different scattering parameters are determined by a single coupling constant.  A rough estimate for
$r_0$ and $\bar{\cT}$ in perturbation theory, which can be obtained
with $m \approx 0.2$ and $a_0 \approx 0.4$, are:
\begin{equation}
r_{0,\text{pt}} \approx -50 \qquad \text{and} \qquad \bar{\mathcal{T}}_\text{pt} \approx 57000.
\label{eq:perturb_res}
\end{equation}
It is however surprising to see that the perturbative results are $\sim 9\sigma$  away from the best global fit in Tab. \ref{tab:global_Eshift_resultsLmin6}. As an attempt to understand the reason for this discrepancy, one could enforce the constraints of Eqs.~(\ref{eq:lambda_r}) and (\ref{eq:Tau}) in the threshold expansion [Eq.~(\ref{eq:deltaEn})]. This way, the fits would have a reduced number of free parameters.
In that context, different \textit{ansätze} for a modification of Eq.~(\ref{eq:deltaEn}) are studied. We can either simultaneously constrain both $r_0$ and $\bar{\mathcal{T}}$ by their one-loop values, or just one of them.  The best results, stemming from global fits with $L\in [12,24]$ are shown in Tab.~\ref{tab:global_Eshift_resultsLmin6_subst}. There we also show the perturbative expectations for $r_0$ and $\bar{\cT}$, labeled $r_{0,\text{pt}}$ and $\bar{\mathcal{T}}_\text{pt}$.
\begin{table}[h!]
	\centering
	\begin{tabular}{ccccc|cc}
		\hline\hline 
		fit[PC; PR] & $a_0$ & $r_0$ & $\bar{\mathcal{T}}$ & $p$-val. & $r_{0,\text{pt}}$ & $\bar{\mathcal{T}}_\text{pt}$\\
		\hline\hline
		$\text{global} [r_0; \times]$ & 0.412(11) & $-$ & 42559(12695) & 0.46 & $-$48(2) & 57101(3052)\\
		$\text{global} [r_0; \checkmark]$ & 0.414(11) & $-$ & 40462(13274) & 0.34 & $-$48(2) & 57656(3067)\\
		\hline
		$\text{global} [\bar{\mathcal{T}}; \times]$ & 0.412(12) & $-37(10)$ & $-$ & 0.43 & $-48(2)$ & 57101(3329)\\
		$\text{global} [\bar{\mathcal{T}}; \checkmark]$ & 0.415(13) & $-40(12)$ & $-$ & 0.24 & $-48(2)$ & 57935(3633)\\
		\hline
		$\text{global} [r_0, \bar{\mathcal{T}}; \times]$ & 0.423(11) & $-$ & $-$ & 0.14 & $-47(1)$ & 60190(3134)\\
		 \hline
		$\text{global} [\times ; \times]$ & 0.491(24) &$-334(48)$ &$-443857(134298) $& 0.89 & $-39(2)$ & 81074(7925)\\ 
		\hline\hline
	\end{tabular}
	\caption{Results from the different energy shift fits with perturbative constraints (PC).
          Both fixed-$N$ fits ($\Delta E_N, N=2,3,4,5$) and global fits to order $L^{-6}$ are listed,
          without as well as with $a_0$ priors (PR), obtained from the respective
          $\Delta E_2[r_0;\times]$-fit. In case of combined fits without $a_0$ priors, $\Delta E_2$ contributes, in contrast to global fits with priors. All fits are performed in the interval $[12, 24]$ for $L$, and they include EC as well as RC. The perturbative predictions $r_{0,\text{pt}}$ and $\bar{\mathcal{T}}_\text{pt}$ are calculated from Eqs.~(\ref{eq:lambda_r}) and (\ref{eq:Tau}), using the corresponding values of $a_0$.}
	\label{tab:global_Eshift_resultsLmin6_subst}
\end{table}
If only $r_0$ is constrained, the best fit result for $\bar{\cT}$ is about $1.5\sigma$ away from the perturbative prediction. Alternatively, if  $\bar{\cT}$ is constrained, $r_0$ is also in $\sim 1 \sigma$ agreement with perturbation theory. Constraining both also produces a similar result. All these fits are seemingly reasonable.  Yet, if no constraints are applied, the fit quality---see the $p$-value---moderately improves, while the best fit results for $\bar{\cT}$ and $r_0$ strongly differ from the perturbative predictions. This remains an open question, that will be addressed in the next section.

\subsubsection{Final assessment of results}

\begin{figure}[h!]
	\centering
	\includegraphics[width=0.6\textwidth]{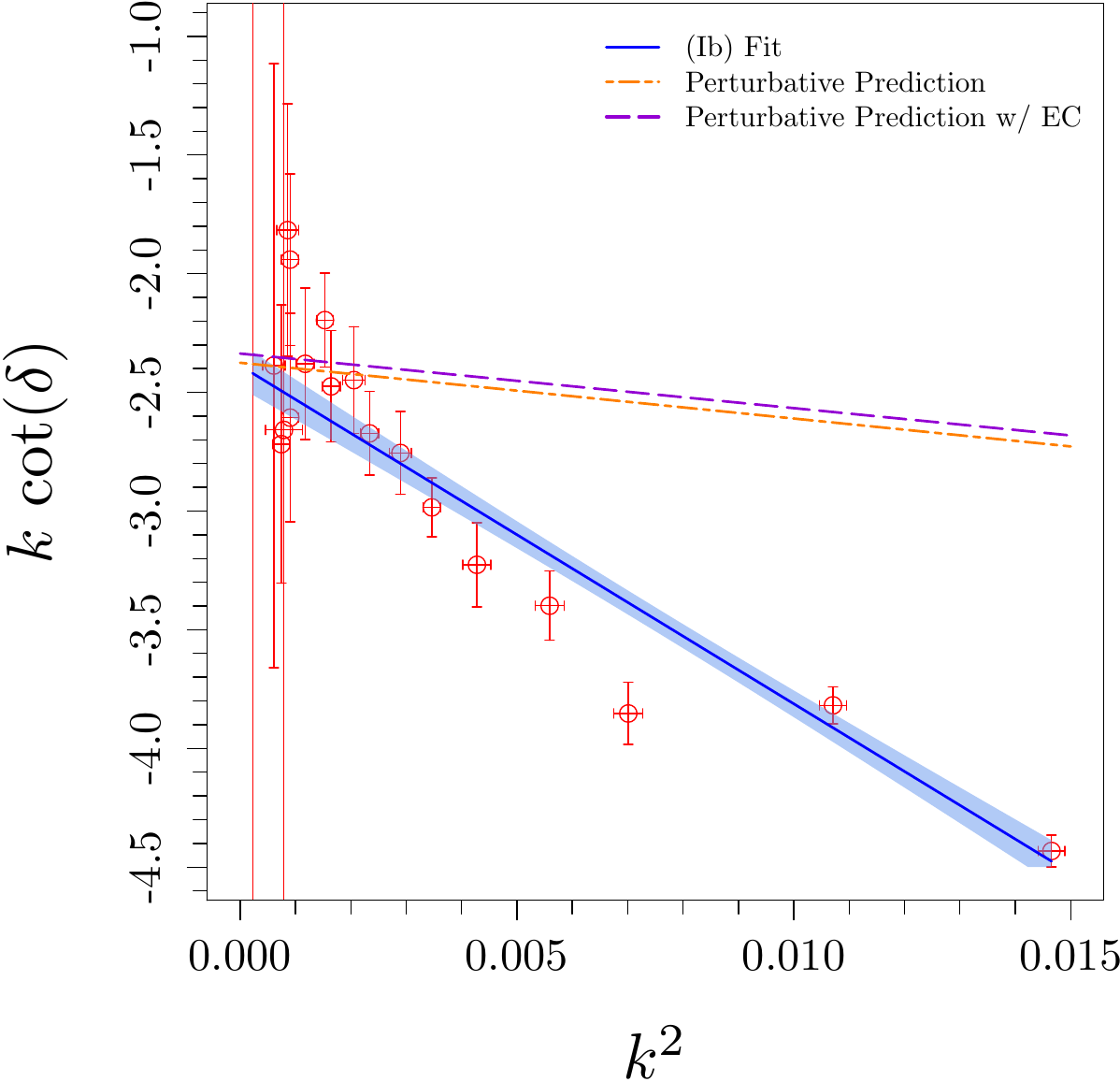}
	\caption{Comparison of the phase-shift fit using model (Ib) with perturbative predictions. The latter are computed by the scattering length extracted from the $\Delta E_2$ energy shift fits to order $L^{-5}$.}
	\label{fig:kcot_delta_2}
\end{figure}

We have extracted the two- and three-particle scattering parameters using the $N$-particle ground state of simulations in complex $\varphi^4$ theory in the symmetric phase. Reviewing the results discussed in the previous sections, one can see that the
different fit strategies seem to agree with each other. More specifically, it can be
stated that $a_0$ is consistent, independently of the applied fit method. Nevertheless,
we have noticed a concerning tension of $\sim 9\sigma$ between our unconstrained
fit results and the predictions of the perturbation theory.  This deviation affects two
parameters, $r_0$ and $\bar{\cT}$, whose contribution is very subleading---they appear
first at $O(L^{-6})$.  To illustrate this discrepancy, a comparison between the phase
shift and the perturbative prediction is shown in Fig.~\ref{fig:kcot_delta_2}. In this plot,
the perturbative prediction uses the value of $a_0$ from the fit to $\Delta E_2$ at
order $L^{-5}$ (with and without EC), and $r_0$ from Eq.~(\ref{eq:perturb_res}),
corresponding to the same $a_0$. At this stage, it is obvious that the strong deviations
cannot emerge from the statistical fluctuations.

In Section \ref{subsec:phase_shift_fits}, we have also checked that the $1/L$ expansion seems to be converging properly.
In addition, we have explicitly accounted for the leading exponentially-suppressed effects.
Besides this, we note that the impact of EC in the perturbative prediction is minimal---see Fig.~\ref{fig:kcot_delta_2}. Even though subleading EC effects could contribute,
it seems unlikely that they are sizable. 

At this point, there remain very few explanations for this discrepancy. One possibility could be discretization effects. In order to study this, one could perform a Symanzik analysis \cite{Symanzik:1979ph}. The idea is to write down the effective Lagrangian of the theory, including higher-dimensional operators that account for the cutoff effects. The contribution
to the Lagrangian in the Euclidean space from the dimension-6 operators is \footnote{We omit $O(4)$-breaking operator, $(\partial_\mu \partial_\mu \varphi) (\partial_\mu \partial_\mu \varphi^*) $, since it does not contribute at tree-level to the scattering amplitude.}
\begin{equation}
\Delta \mathcal{L}_6 = a^2 c_1 \left( \varphi^* \partial_\mu \varphi \,  \varphi^*  \partial_\mu \varphi + \text{ h.c.} \right) + a^2 c_2 \partial_\mu \varphi^*  \partial_\mu \varphi   \varphi^* \varphi + a^2 c_3 \left(\varphi^* \varphi \right)^3.
\end{equation}
As can be seen, both $c_1$ and $c_2$ contribute at tree-level to the two-particle scattering amplitude, and particularly to $a_0$ and $r_0$. Moreover, all three coefficients $c_1,c_2,c_3$
contribute at the three level to $\bar{\cT}$. The crucial point is that the
discretization effects are of the order $(a M)^2 \sim 0.04$. However, a measure of the strength of the interactions, $M a_0  \sim 0.08$, is of the same order. Thus, it is plausible that cutoff effects could be significant, and so the continuum perturbation theory would not be an appropriate description of the lattice $\varphi^4$ theory with our set of parameters. 
\begin{figure}[h!]
	\centering
	\includegraphics[width=1\textwidth]{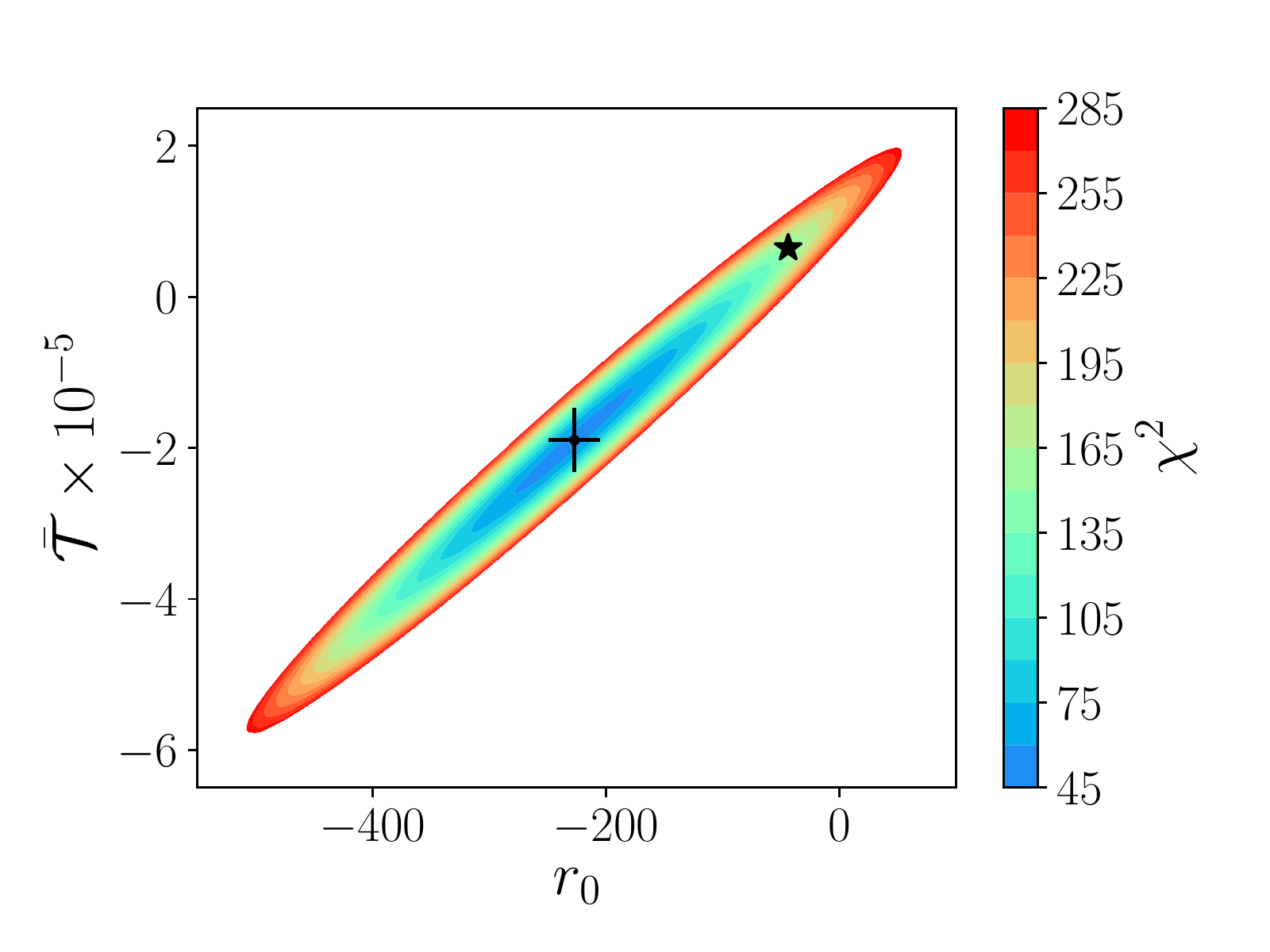}
	\caption{$\chi^2$ heat map that corresponds to the global fit at order $L^{-6}$ without priors, taking both EC and RC into account. The fit interval is $[10, 24]$, and $a_0$ is fixed to 0.439. The black data point with error bars marks the corresponding fit results for $r_0$ and $\bar{\mathcal{T}}$, cf. Tab.~\ref{tab:global_Eshift_resultsLmin6}. The black star marks the perturbative prediction.}
	\label{fig:global_chisqr_plot}
\end{figure}

To gain further insight into the above discrepancy and the fit systematics, we study the shape of the $\chi^2$ function for the previously discussed energy shift fits to order $L^{-6}$. For this, we use heat map plots to localize emerging minima and identify potential quasi-flat directions. Furthermore, the best fit parameters can then be easily compared with perturbative predictions.  For a two-dimensional visualization in the $r_0$, $\bar{\cT}$ plane,
$a_0$ is fixed to the best fit result.  We show the heat map plot for a global fit with EC and RC in Fig.~\ref{fig:global_chisqr_plot}. There, one can see that the $\chi^2$ function grows much slower in one particular direction in the $r_0$, $\bar{\cT}$ plane. This happens,
because in Eq.~(\ref{eq:deltaEn}), the dependence on $r_0$ and $\bar{\cT}$ can only be disentangled by considering different values of $N$.  This feature increases the difficulty of measuring $\bar{\cT}$ with statistical significance. In the Appendices \ref{app:heat_map1} and \ref{app:heat_map2} we discuss further heat map plots.

Since the inconsistency between perturbation theory and the unconstrained fit could not be resolved even after many different tests, we conclude that the perturbative results do not describe the lattice $\varphi^4$ theory properly.  Beyond that, Fig.~\ref{fig:global_chisqr_plot} displays the challenges that arise, when extracting low energy scattering parameters on the lattice. Namely, is clearly demonstrates the difficulty of extracting subleading
effects and separating two- and three-body effects in the spectrum.

\section{Conclusions and outlook}
\label{sec:conclu}

  In the present paper, we discuss the derivation of the formula for the finite-volume energy shift of $N$ identical relativistic bosons in detail, up to and including $O(L^{-6})$.
  Moreover, in case of two- and three-particle systems, the derivation is extended
  to the first excited state, which is contained in the irreducible representations $A_1^+$
  and $E^+$ of the cubic group. The derivation is performed in the framework of a
  non-relativistic effective theory, and relativistic effects are included perturbatively.
  Moreover, if needed, using the same effective Lagrangian, the derivation can be straightforwardly extended to higher excited states, and to excited states of systems containing more particles. This paper contains the necessary prescription in order to carry out such calculations.

  We have also calculated finite-volume
  energy levels in the sectors with one, two, three, four and five
  identical bosons in the symmetric phase of the complex $\varphi^4$ theory on the lattice for many different
  values of the box size $L$. The wealth of the available lattice data enables us
  to carry out fits using different fit strategies, aiming at finding an optimal
  way to extract infinite-volume observables from the measured spectrum.
  
  The main goal of lattice calculations is to establish few parameters (effective couplings),
  which describe the interactions in multiparticle systems. Only three parameters
  contribute at order in $1/L$ we are working. These are: the two-body scattering
  length $a_0$ and the effective range $r_0$, as well as the non-derivative three-body 
  coupling that can be traded for the threshold amplitude $\bar\cT$. The latter is of
  our primary concern. It turns out that $a_0$ can be reliably determined in all fits,
  whereas $r_0$ and $\bar\cT$ are correlated in the fixed-$N$ fits. Only performing
  a global fit, involving sectors with different $N$, enables one to fix these parameters
  at a reasonable precision. The three-particle amplitude, $\bar\cT$, turns out to be different from
  zero at $7\sigma$, taking into account only statistical error.
  Still, the $\chi^2$ heat map in the parameters $r_0,\bar\cT$
  represents a rather elongated ellipse that indicates a strong correlation.
  It is important to stress also that this property stems merely from the way these two parameters
  enter the expression of the ground-state energy shift of $N$-particles at order $L^{-6}$.
  In this sense, it does not depend on the dynamics and will show up in QCD as well.
  
  From this point of view, the study of the excited states becomes particularly interesting.
  As it is clear from the explicit expressions that are contained in the present paper,
  in excited states the relativistic and the effective-range corrections contribute already
  at order $L^{-5}$, whereas the contribution from $\bar\cT$ emerges again first at order
  $L^{-6}$. Moreover, the contribution from $\bar \cT$ is present in certain irreducible
  representations and is absent in others. All this gives hope that the study of the excited
  states might help one to break the spell of large correlations between extracted
  parameters. It would be certainly
  interesting to study this issue in the future.

  The study of the role of relativistic effects, as well as leading-order exponentially-suppressed polarization effects was one of the main aims of the present work. We have shown that at the
  values of $L$ considered,  both effects
  are substantial in the sense that including them in the fit leads to the changes in some
  observables beyond one standard deviation. This circumstance must be kept in mind
  when analyzing data from lattice QCD.

  We have explicitly observed that finite lattice spacing effects might have a substantial
  impact on the values of the extracted parameters. In our example, the continuum theory
  badly fails to predict the values of $r_0$ and $\bar\cT$, which are determined on the
  lattice. This fact should be also remembered when extracting the corresponding observables
  in lattice QCD that contribute at the subleading order(s) in the $1/L$ expansion.

\acknowledgments

We thank Silas Beane and Martin Savage for providing us with the detailed notes
on the derivation of the ground-state energy shift, as well as Christian Lang for the
helpful discussion of the different phases in the $\varphi^4$-theory.

FRL acknowledges the support provided by the European projects
H2020-MSCA-ITN-2015/674896-ELUSIVES, H2020-MSCA-RISE-2015/690575-InvisiblesPlus,
the Spanish project FPA2017-85985-P, and the Generalitat Valenciana grant
PROMETEO/2019/083. The work of FRL also received funding from the European Union
Horizon 2020 research and innovation program  under the Marie Sk{\l}odowska-Curie
grant agreement No. 713673 and ``La Caixa'' Foundation  (ID 100010434,
LCF/BQ/IN17/11620044).

AR acknowledges the support from the DFG (CRC 110 ``Symmetries 
and the Emergence of Structure in QCD''), Volkswagenstiftung under
contract no. 93562 and the Chinese Academy of Sciences (CAS) President's
International Fellowship Initiative (PIFI) (grant no. 2021VMB0007).

The authors gratefully acknowledge the Gauss Centre for Supercomputing
e.V.\@ (www.gauss-centre.eu) for funding this project by providing
computing time on the GCS Supercomputer JUQUEEN~\cite{juqueen} and the
John von Neumann Institute for Computing (NIC) for computing time
provided on the supercomputers JURECA~\cite{jureca} and
JUWELS~\cite{juwels} at Jülich Supercomputing Centre (JSC). The open
source software packages R~\cite{R:2019} and hadron~\cite{hadron:2020}
have been used.

\appendix

\appendix

\section{The finite parts of the two-loop diagrams}
\label{app:delta}

The finite parts of the two-loop diagrams in Eq.~(\ref{eq:T3de}) in the chosen
configuration of the external momenta are given by the following
integrals over the Feynman parameters:
\eq
\delta^{(d)}&=&\frac{1}{2}\,\ln 3-\ln 2+\frac{1}{9}\,\sum_{(ijk)}\sum_{(lmn)}
\int_0^1dx\int_0^1\frac{dy}{2\sqrt{y}}\,\biggl(\ln g-\frac{1-y}{g}\,\frac{dg}{dy}\biggr)\simeq -1.08964,\,
\nonumber\\[2mm]
  g&=&\frac{1}{4}\,(1-y)^2+\frac{1}{2}\,(1-y)+2y+\frac{1}{2}\,(1-y)^2x(1-x)(a_{kn}-1)\, ,
  \en
  and
  \eq
  \delta^{(e)}&=&-\frac{1}{9}\,\sum_{(ijk)}\sum_{(lmn)}
  \int_0^1 dx\int_0^1\frac{ydy}{h^{3/2}}\,\biggl(-\ln G+\frac{3}{2}\,\ln h\biggr)\,  \simeq  3.92587 ,
  \nonumber\\[2mm]
  h&=&(1-y)-\frac{1}{4}\,(1-y)^2+x(1-x)y^2\, ,
  \nonumber\\[2mm]
  G&=&-\frac{3}{8}\,(1-y)(y-3)-y^2x(1-x)\biggl(-1+\frac{y}{4}+\frac{1}{4}\,(1-y)a_{kn}\biggr)\, .
  \en
In these equations, $a_{kn}={\bf p}_k'{\bf p}_n/\lambda^2$. 
According to Eq.~(\ref{eq:scalarproducts}), 
\begin{equation}
a_{kn}= \frac{1}{2} -\frac{3}{2} \delta_{kn}.
\end{equation}

\section{$\chi^2$ heat map for the $\Delta E_2$-fit to order $L^{-6}$}
\label{app:heat_map1}

\begin{figure}[htbp!]
	\centering
	\includegraphics[width=0.9\textwidth]{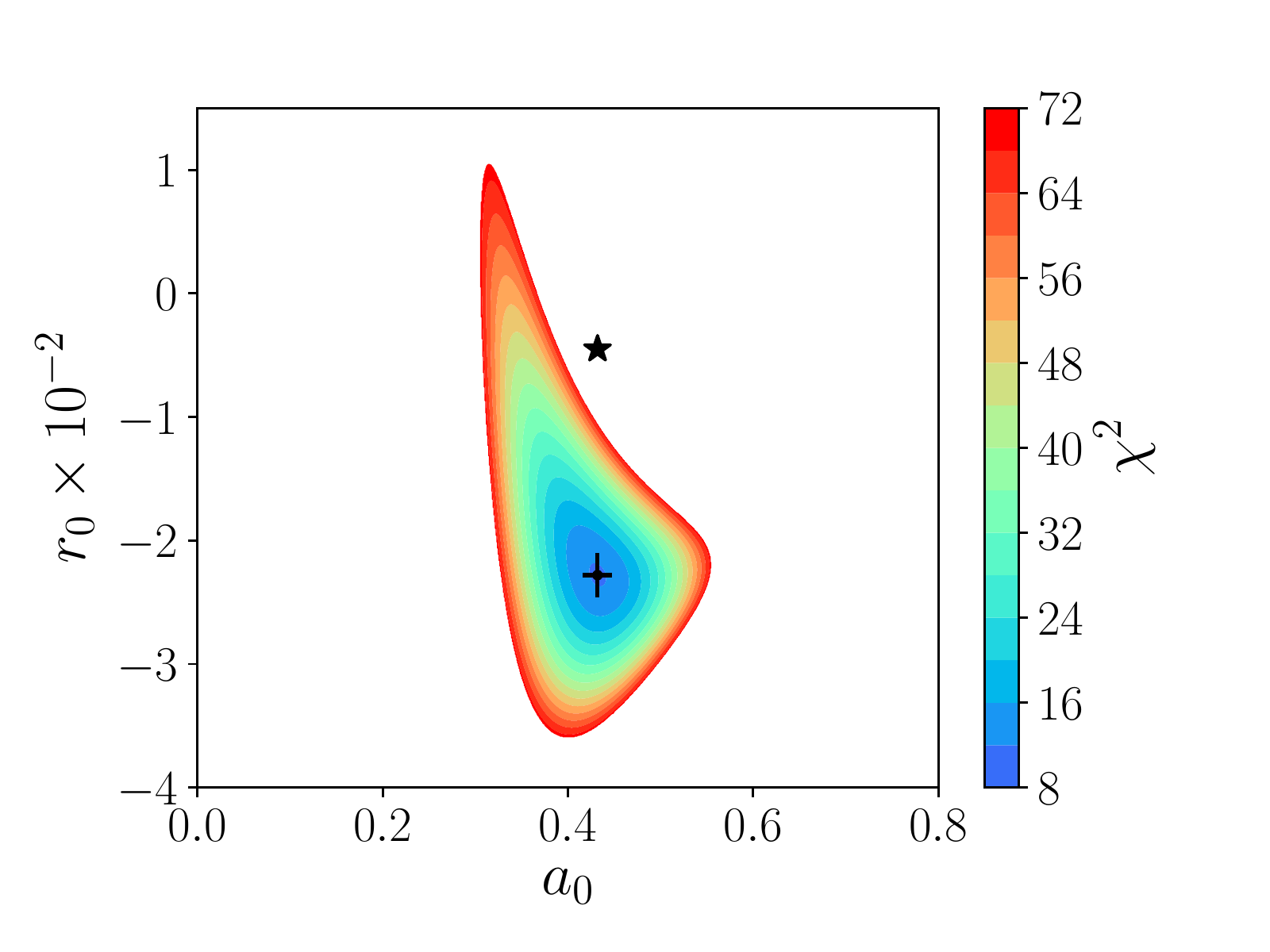}
	\caption{$\chi^2$ heat map that corresponds to the $\Delta E_2$-fit to order $L^{-6}$ taking both EC and RC into account. The fit interval is $[9, 24]$, and $a_0$ is fixed to 0.432. {The black data point with error bars marks the corresponding fit result for $r_0$,}
          cf. Tab.~\ref{tab:Eshift_resultsLmin6}. The black star corresponds to the perturbative prediction.}
	\label{fig:DeltaE2_chisqr_plot}
\end{figure}

Fig.~\ref{fig:DeltaE2_chisqr_plot} depicts the $\chi^2$ heat map, corresponding to the $\Delta E_2$-fit to order $L^{-6}$ in the interval $[9, 24]$, {including}
EC as well as RC. Here, we follow the visualization convention of Fig. \ref{fig:global_chisqr_plot}. As expected, a distinct minimum that can be observed, is in agreement with
the parameters obtained from the corresponding fit (black data point with error bars,
cf. Tab.~\ref{tab:Eshift_resultsLmin6}). The black star marks the corresponding
perturbative prediction.
Similarly to other fits in this work, the perturbative prediction is far from the best fit results.

\section{$\chi^2$ heat map for the $\Delta E_N$-fits to $L^{-6}$ }
 \label{app:heat_map2}

\begin{figure}[htb!]
	\begin{minipage}{0.5\textwidth}
		\centering
		\includegraphics[width=1\textwidth]{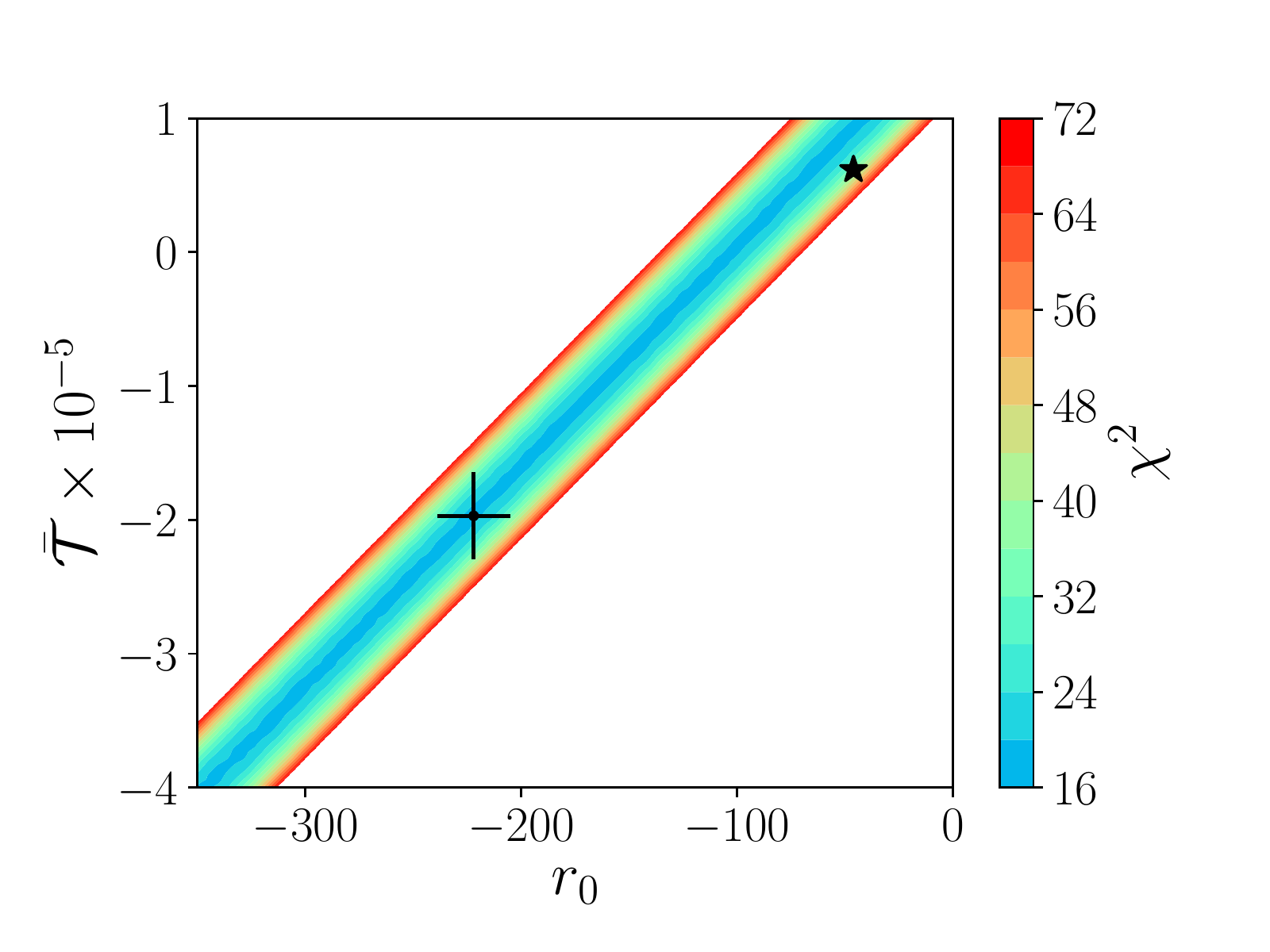}
		\subcaption{Heat map for $\Delta E_3$-fit to order $L^{-6}$\\without priors; $a_0$ is fixed to 0.428. \label{fig:heatmapa}}
	\end{minipage}
	\begin{minipage}{0.5\textwidth}
		\centering
		\includegraphics[width=1\textwidth]{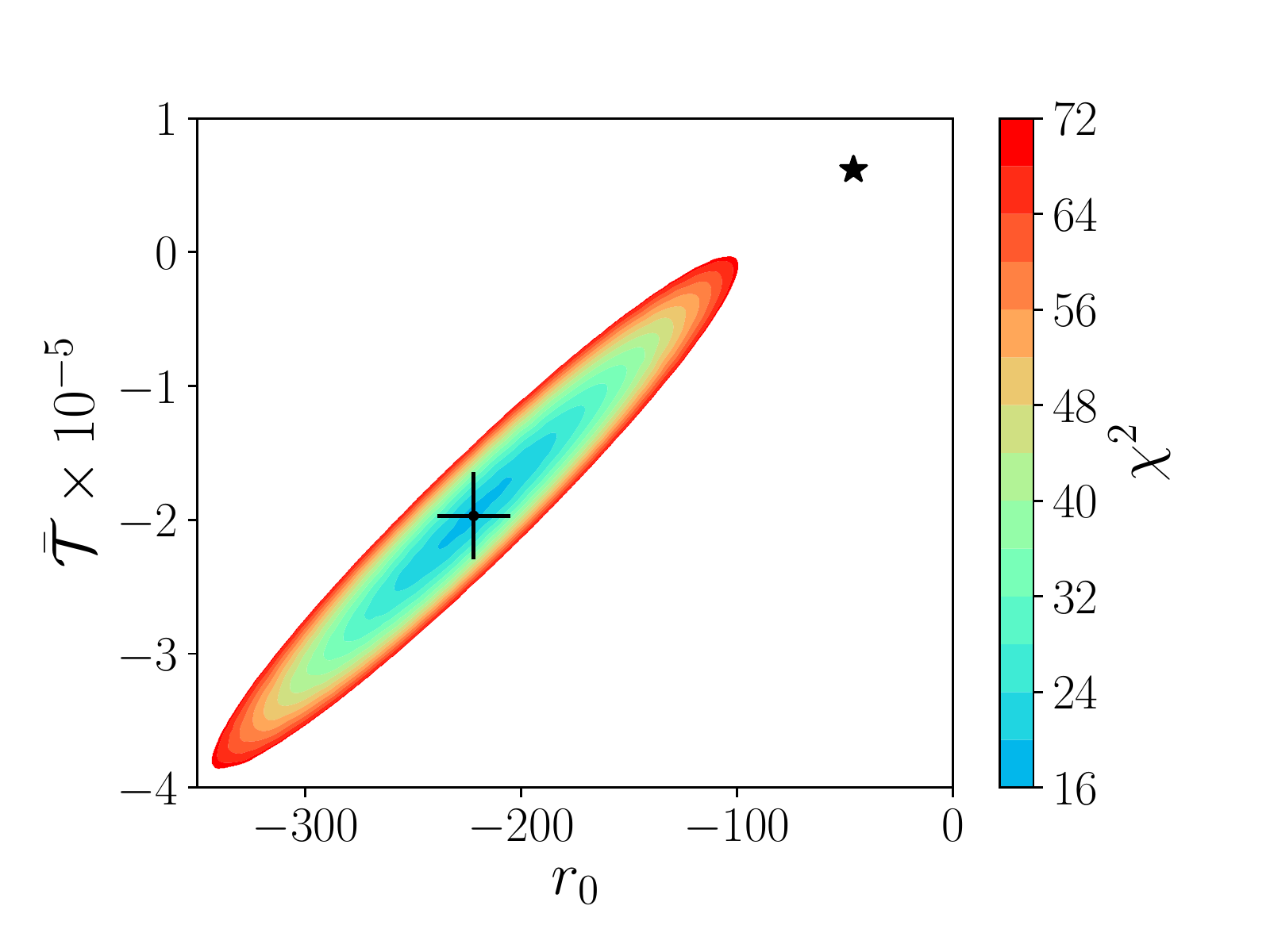}
		\subcaption{Heat map for $\Delta E_3$-fit to order $L^{-6}$\\with priors, cf. Tab.~\ref{tab:Eshift_resultsLmin6}; $a_0$ is fixed to 0.428. \label{fig:heatmapb}}
	\end{minipage}
	\begin{minipage}{0.5\textwidth}
		\centering
		\includegraphics[width=1\textwidth]{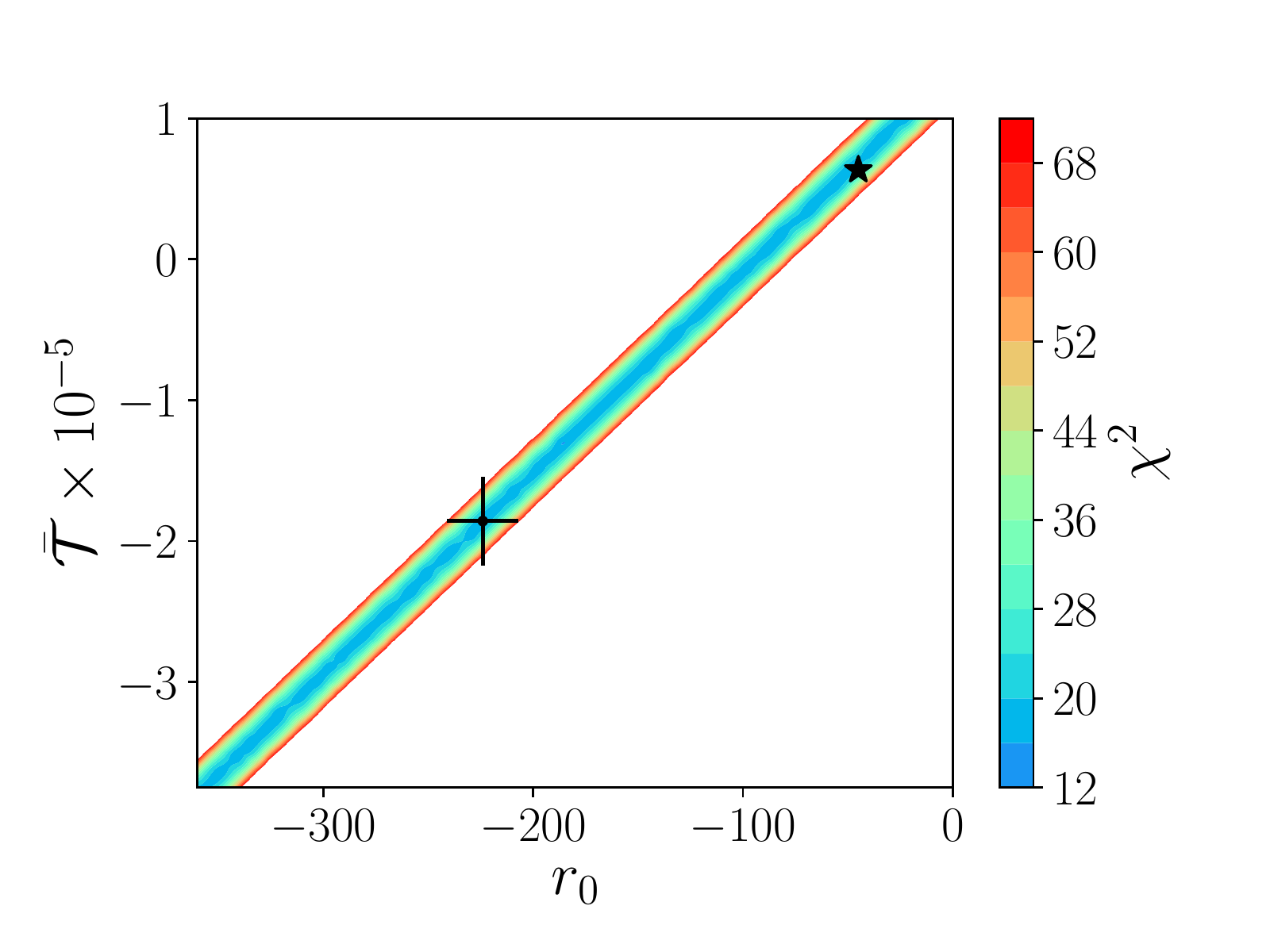}
		\subcaption{Heat map for $\Delta E_4$-fit to order $L^{-6}$\\without priors; $a_0$ is fixed to 0.434. \label{fig:heatmapc}}
	\end{minipage}
	\begin{minipage}{0.5\textwidth}
		\centering
		\includegraphics[width=1\textwidth]{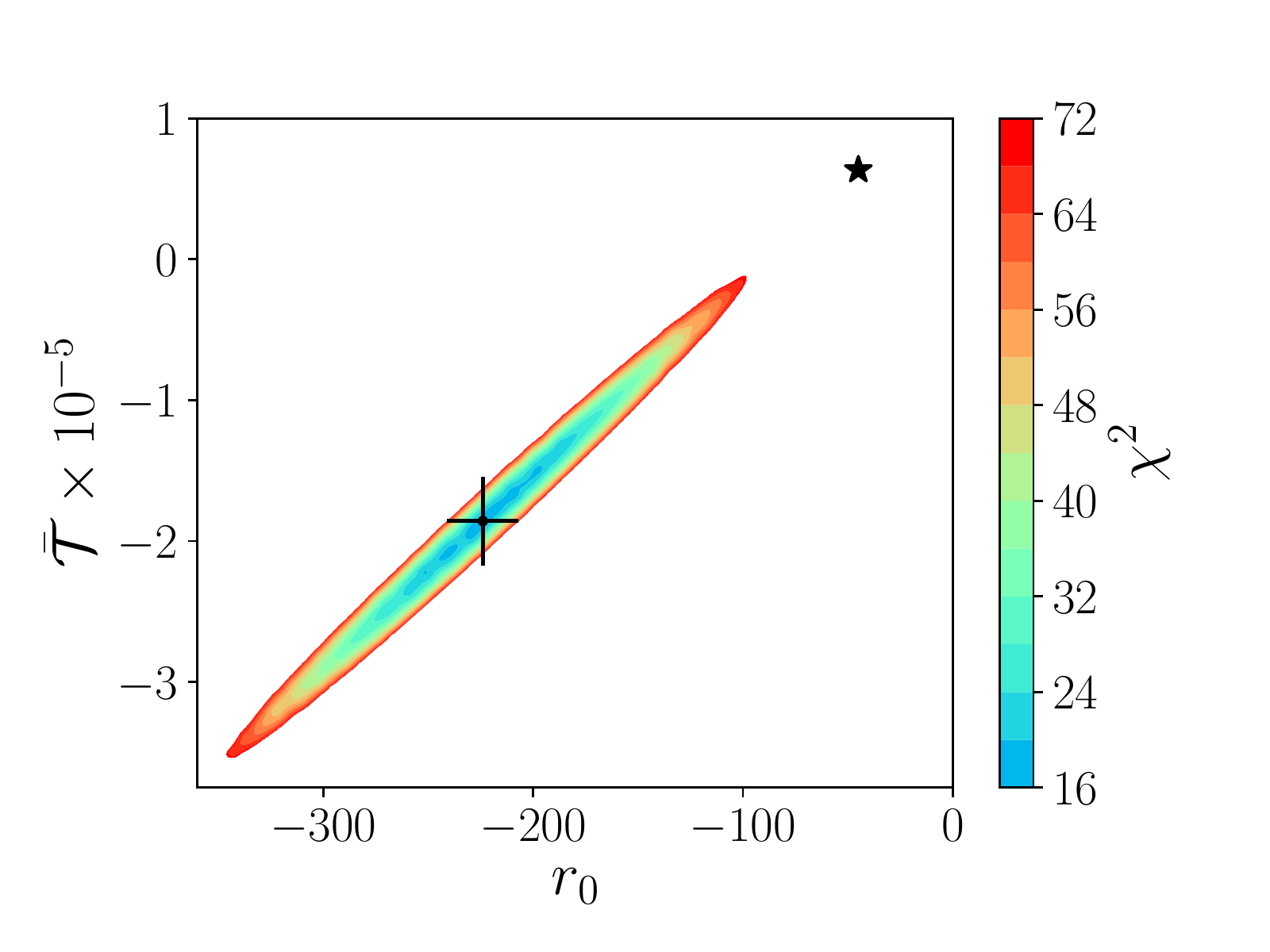}
		\subcaption{Heat map for $\Delta E_4$-fit to order $L^{-6}$\\with priors, cf. Tab.~\ref{tab:Eshift_resultsLmin6}; $a_0$ is fixed to 0.434. \label{fig:heatmapd}}
	\end{minipage}
	\begin{minipage}{0.5\textwidth}
		\centering
		\includegraphics[width=1\textwidth]{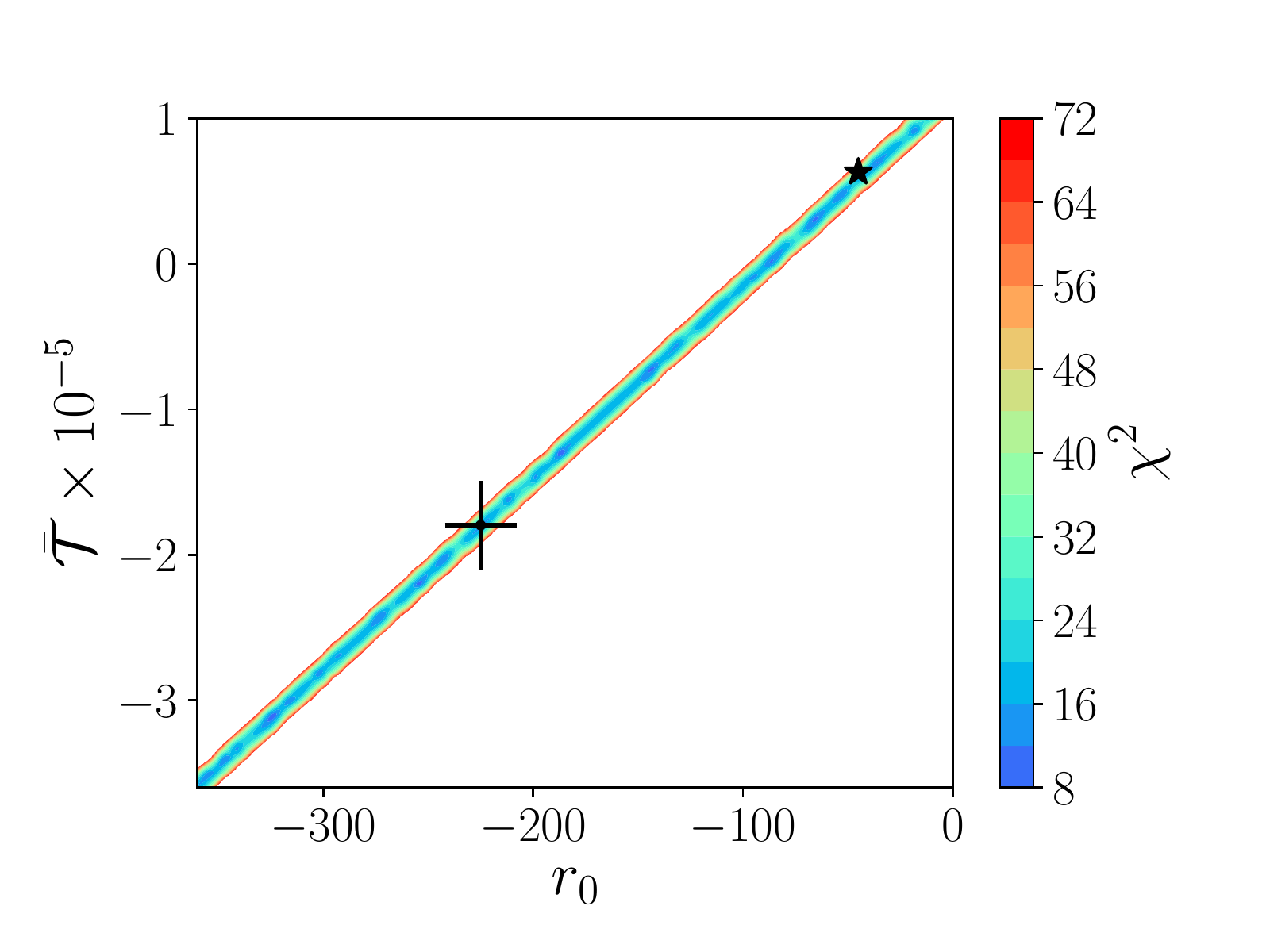}
		\subcaption{Heat map for $\Delta E_5$-fit to order $L^{-6}$\\without priors; $a_0$ is fixed to 0.433. \label{fig:heatmape}}
	\end{minipage}
	\begin{minipage}{0.5\textwidth}
		\centering
		\includegraphics[width=1\textwidth]{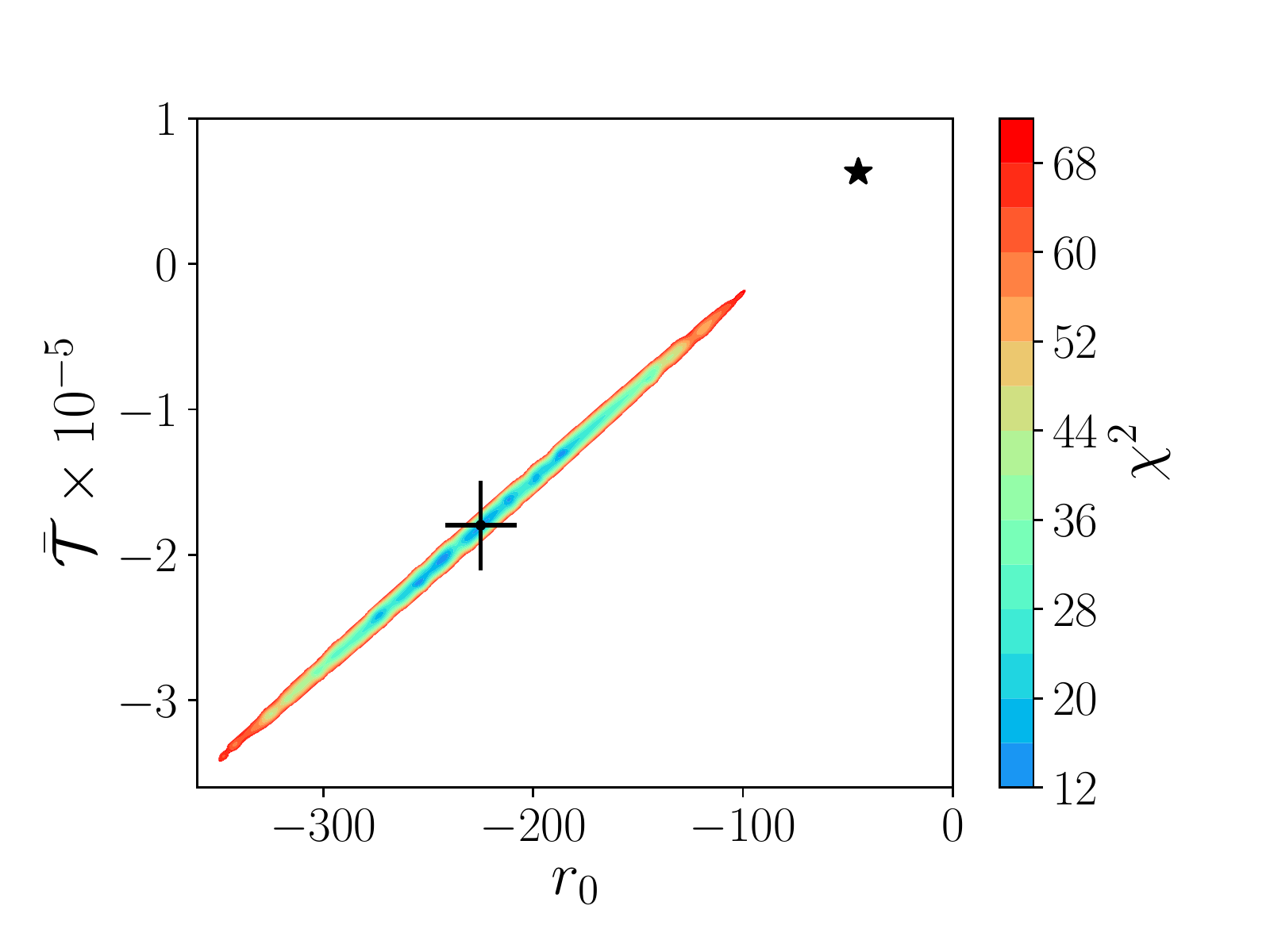}
		\subcaption{Heat map for $\Delta E_5$-fit to order $L^{-6}$\\with priors, cf. Tab.~\ref{tab:Eshift_resultsLmin6}; $a_0$ is fixed to 0.433. \label{fig:heatmapf}}
	\end{minipage}
        \caption{$\chi^2$ heat maps a) to f) that correspond to the fits with EC as well as RC
          in the interval $[9,24]$ for $L$. The black data points with error bars mark the
          respective results for $r_0$ and $\bar{\mathcal{T}}$ from prior constrained fits,
          cf. Tab.~\ref{tab:Eshift_resultsLmin6}. The black stars correspond to the perturbative predictions, calculated from the fixed $a_0$ values by using Eqs.~(\ref{eq:lambda_r}) and (\ref{eq:Tau}).}
\label{fig:DeltaE3toDeltaE5_chisqr_plot}
\end{figure}

 In this appendix, we show some additional heat map plots in order to explore the
 impact of priors in our fits -- see Fig. \ref{fig:DeltaE3toDeltaE5_chisqr_plot}. As
 already explained, $r_0$ and $\bar{\cT}$ cannot be disentangled in a fixed-$N$ fit
 with $N>2$, because they enter at the same order in the threshold expansion. This
 is seen very well in Figs. \ref{fig:heatmapa}, \ref{fig:heatmapc} and \ref{fig:heatmape},
 as a flat direction with constant $\chi^2$. If priors from the $\Delta E_2$-fit are used for
 $r_0$, the flat direction disappears and $r_0$ and $\bar{\cT}$ can be obtained --
 see Figs. \ref{fig:heatmapb}, \ref{fig:heatmapd}, and \ref{fig:heatmapf}.

\clearpage

\clearpage
\bibliographystyle{JHEP}

\bibliography{ref.bib}

\providecommand{\href}[2]{#2}\begingroup\raggedright\begin{thebibliography}{10}

\bibitem{Beane:2007es}
S.~R. Beane, W.~Detmold, T.~C. Luu, K.~Orginos, M.~J. Savage and A.~Torok,
  \emph{{Multi-Pion Systems in Lattice QCD and the Three-Pion Interaction}},
  \href{https://doi.org/10.1103/PhysRevLett.100.082004}{\emph{Phys. Rev. Lett.}
  {\bfseries 100} (2008) 082004}
  [\href{https://arxiv.org/abs/0710.1827}{{\ttfamily 0710.1827}}].

\bibitem{Horz:2019rrn}
B.~H{\"o}rz and A.~Hanlon, \emph{{Two- and three-pion finite-volume spectra at
  maximal isospin from lattice QCD}},
  \href{https://arxiv.org/abs/1905.04277}{{\ttfamily 1905.04277}}.

\bibitem{Culver:2019vvu}
C.~Culver, M.~Mai, R.~Brett, A.~Alexandru and M.~D\"oring, \emph{{Three pion
  spectrum in the $I=3$ channel from lattice QCD}},
  \href{https://doi.org/10.1103/PhysRevD.101.114507}{\emph{Phys. Rev. D}
  {\bfseries 101} (2020) 114507}
  [\href{https://arxiv.org/abs/1911.09047}{{\ttfamily 1911.09047}}].

\bibitem{Fischer:2020jzp}
M.~Fischer, B.~Kostrzewa, L.~Liu, F.~Romero-L\'opez, M.~Ueding and C.~Urbach,
  \emph{{Scattering of two and three physical pions at maximal isospin from
  lattice QCD}},  \href{https://arxiv.org/abs/2008.03035}{{\ttfamily
  2008.03035}}.

\bibitem{Blanton:2019vdk}
T.~D. Blanton, F.~Romero-López and S.~R. Sharpe, \emph{{$I = 3$ three-pion
  scattering amplitude from lattice QCD}},
  \href{https://arxiv.org/abs/1909.02973}{{\ttfamily 1909.02973}}.

\bibitem{Hansen:2020otl}
M.~T. Hansen, R.~A. Brice\~no, R.~G. Edwards, C.~E. Thomas and D.~J. Wilson,
  \emph{{The energy-dependent $\pi^+ \pi^+ \pi^+$ scattering amplitude from
  QCD}},  \href{https://arxiv.org/abs/2009.04931}{{\ttfamily 2009.04931}}.

\bibitem{Alexandru:2020xqf}
A.~Alexandru, R.~Brett, C.~Culver, M.~D\"oring, D.~Guo, F.~X. Lee et~al.,
  \emph{{Finite-volume energy spectrum of the $K^-K^-K^-$ system}},
  \href{https://arxiv.org/abs/2009.12358}{{\ttfamily 2009.12358}}.

\bibitem{Lee:1957zzb}
T.~D. Lee, K.~Huang and C.~N. Yang, \emph{{Eigenvalues and Eigenfunctions of a
  Bose System of Hard Spheres and Its Low-Temperature Properties}},
  \href{https://doi.org/10.1103/PhysRev.106.1135}{\emph{Phys. Rev.} {\bfseries
  106} (1957) 1135}.

\bibitem{Huang:1957im}
K.~Huang and C.~Yang, \emph{{Quantum-mechanical many-body problem with
  hard-sphere interaction}},
  \href{https://doi.org/10.1103/PhysRev.105.767}{\emph{Phys. Rev.} {\bfseries
  105} (1957) 767}.

\bibitem{Wu:1959zz}
T.~T. Wu, \emph{{Ground State of a Bose System of Hard Spheres}},
  \href{https://doi.org/10.1103/PhysRev.115.1390}{\emph{Phys. Rev.} {\bfseries
  115} (1959) 1390}.

\bibitem{Luscher:1986n2}
M.~L{\"u}scher, \emph{{Volume Dependence of the Energy Spectrum in Massive
  Quantum Field Theories. 2. Scattering States}},
  \href{https://doi.org/10.1007/BF01211097}{\emph{Commun.Math.Phys.} {\bfseries
  105} (1986) 153}.

\bibitem{Beane:2007qr}
S.~R. Beane, W.~Detmold and M.~J. Savage, \emph{{n-Boson Energies at Finite
  Volume and Three-Boson Interactions}},
  \href{https://doi.org/10.1103/PhysRevD.76.074507}{\emph{Phys. Rev.}
  {\bfseries D76} (2007) 074507}
  [\href{https://arxiv.org/abs/0707.1670}{{\ttfamily 0707.1670}}].

\bibitem{Detmold:2008gh}
W.~Detmold and M.~J. Savage, \emph{{The Energy of n Identical Bosons in a
  Finite Volume at $O(L^{-7})$}},
  \href{https://doi.org/10.1103/PhysRevD.77.057502}{\emph{Phys. Rev.}
  {\bfseries D77} (2008) 057502}
  [\href{https://arxiv.org/abs/0801.0763}{{\ttfamily 0801.0763}}].

\bibitem{Tan:2007bg}
S.~Tan, \emph{{Three-boson problem at low energy and implications for dilute
  Bose-Einstein condensates}},
  \href{https://doi.org/10.1103/PhysRevA.78.013636}{\emph{Phys. Rev. A}
  {\bfseries 78} (2008) 013636}
  [\href{https://arxiv.org/abs/0709.2530}{{\ttfamily 0709.2530}}].

\bibitem{Polejaeva:2012ut}
K.~Polejaeva and A.~Rusetsky, \emph{{Three particles in a finite volume}},
  \href{https://doi.org/10.1140/epja/i2012-12067-8}{\emph{Eur. Phys. J. A}
  {\bfseries 48} (2012) 67} [\href{https://arxiv.org/abs/1203.1241}{{\ttfamily
  1203.1241}}].

\bibitem{Hansen:2014eka}
M.~T. Hansen and S.~R. Sharpe, \emph{{Relativistic, model-independent,
  three-particle quantization condition}},
  \href{https://doi.org/10.1103/PhysRevD.90.116003}{\emph{Phys. Rev.}
  {\bfseries D90} (2014) 116003}
  [\href{https://arxiv.org/abs/1408.5933}{{\ttfamily 1408.5933}}].

\bibitem{Hansen:2015zga}
M.~T. Hansen and S.~R. Sharpe, \emph{{Expressing the three-particle
  finite-volume spectrum in terms of the three-to-three scattering amplitude}},
  \href{https://doi.org/10.1103/PhysRevD.92.114509}{\emph{Phys. Rev.}
  {\bfseries D92} (2015) 114509}
  [\href{https://arxiv.org/abs/1504.04248}{{\ttfamily 1504.04248}}].

\bibitem{Hammer:2017uqm}
H.-W. Hammer, J.-Y. Pang and A.~Rusetsky, \emph{{Three-particle quantization
  condition in a finite volume: 1. The role of the three-particle force}},
  \href{https://doi.org/10.1007/JHEP09(2017)109}{\emph{JHEP} {\bfseries 09}
  (2017) 109} [\href{https://arxiv.org/abs/1706.07700}{{\ttfamily
  1706.07700}}].

\bibitem{Hammer:2017kms}
H.~W. Hammer, J.~Y. Pang and A.~Rusetsky, \emph{{Three particle quantization
  condition in a finite volume: 2. general formalism and the analysis of
  data}}, \href{https://doi.org/10.1007/JHEP10(2017)115}{\emph{JHEP} {\bfseries
  10} (2017) 115} [\href{https://arxiv.org/abs/1707.02176}{{\ttfamily
  1707.02176}}].

\bibitem{Mai:2017bge}
M.~Mai and M.~{D{\"o}ring}, \emph{{Three-body Unitarity in the Finite Volume}},
  \href{https://doi.org/10.1140/epja/i2017-12440-1}{\emph{Eur. Phys. J.}
  {\bfseries A53} (2017) 240}
  [\href{https://arxiv.org/abs/1709.08222}{{\ttfamily 1709.08222}}].

\bibitem{Hansen:2019nir}
M.~T. Hansen and S.~R. Sharpe, \emph{{Lattice QCD and Three-particle Decays of
  Resonances}},  \href{https://arxiv.org/abs/1901.00483}{{\ttfamily
  1901.00483}}.

\bibitem{Kreuzer:2010ti}
S.~Kreuzer and H.-W. Hammer, \emph{{The Triton in a finite volume}},
  \href{https://doi.org/10.1016/j.physletb.2010.10.003}{\emph{Phys. Lett. B}
  {\bfseries 694} (2011) 424}
  [\href{https://arxiv.org/abs/1008.4499}{{\ttfamily 1008.4499}}].

\bibitem{Kreuzer:2009jp}
S.~Kreuzer and H.-W. Hammer, \emph{{On the modification of the Efimov spectrum
  in a finite cubic box}},
  \href{https://doi.org/10.1140/epja/i2010-10910-6}{\emph{Eur. Phys. J. A}
  {\bfseries 43} (2010) 229} [\href{https://arxiv.org/abs/0910.2191}{{\ttfamily
  0910.2191}}].

\bibitem{Kreuzer:2008bi}
S.~Kreuzer and H.-W. Hammer, \emph{{Efimov physics in a finite volume}},
  \href{https://doi.org/10.1016/j.physletb.2009.02.035}{\emph{Phys. Lett. B}
  {\bfseries 673} (2009) 260}
  [\href{https://arxiv.org/abs/0811.0159}{{\ttfamily 0811.0159}}].

\bibitem{Kreuzer:2012sr}
S.~Kreuzer and H.~W. Grie{\ss}hammer, \emph{{Three particles in a finite
  volume: The breakdown of spherical symmetry}},
  \href{https://doi.org/10.1140/epja/i2012-12093-6}{\emph{Eur. Phys. J. A}
  {\bfseries 48} (2012) 93} [\href{https://arxiv.org/abs/1205.0277}{{\ttfamily
  1205.0277}}].

\bibitem{Hansen:2015zta}
M.~T. Hansen and S.~R. Sharpe, \emph{{Perturbative results for two and three
  particle threshold energies in finite volume}},
  \href{https://doi.org/10.1103/PhysRevD.93.014506}{\emph{Phys. Rev.}
  {\bfseries D93} (2016) 014506}
  [\href{https://arxiv.org/abs/1509.07929}{{\ttfamily 1509.07929}}].

\bibitem{Hansen:2016fzj}
M.~T. Hansen and S.~R. Sharpe, \emph{{Threshold expansion of the three-particle
  quantization condition}}, \href{https://doi.org/10.1103/PhysRevD.96.039901,
  10.1103/PhysRevD.93.096006}{\emph{Phys. Rev.} {\bfseries D93} (2016) 096006}
  [\href{https://arxiv.org/abs/1602.00324}{{\ttfamily 1602.00324}}].

\bibitem{Briceno:2017tce}
R.~A. Brice\~no, M.~T. Hansen and S.~R. Sharpe, \emph{{Relating the
  finite-volume spectrum and the two-and-three-particle $S$ matrix for
  relativistic systems of identical scalar particles}},
  \href{https://doi.org/10.1103/PhysRevD.95.074510}{\emph{Phys. Rev.}
  {\bfseries D95} (2017) 074510}
  [\href{https://arxiv.org/abs/1701.07465}{{\ttfamily 1701.07465}}].

\bibitem{Sharpe:2017jej}
S.~R. Sharpe, \emph{{Testing the threshold expansion for three-particle
  energies at fourth order in $\phi^4$ theory}},
  \href{https://doi.org/10.1103/PhysRevD.96.054515}{\emph{Phys. Rev.}
  {\bfseries D96} (2017) 054515}
  [\href{https://arxiv.org/abs/1707.04279}{{\ttfamily 1707.04279}}].

\bibitem{Briceno:2018mlh}
R.~A. Brice\~no, M.~T. Hansen and S.~R. Sharpe, \emph{{Numerical study of the
  relativistic three-body quantization condition in the isotropic
  approximation}},
  \href{https://doi.org/10.1103/PhysRevD.98.014506}{\emph{Phys. Rev.}
  {\bfseries D98} (2018) 014506}
  [\href{https://arxiv.org/abs/1803.04169}{{\ttfamily 1803.04169}}].

\bibitem{Briceno:2018aml}
R.~A. Brice\~no, M.~T. Hansen and S.~R. Sharpe, \emph{{Three-particle systems
  with resonant subprocesses in a finite volume}},
  \href{https://doi.org/10.1103/PhysRevD.99.014516}{\emph{Phys. Rev.}
  {\bfseries D99} (2019) 014516}
  [\href{https://arxiv.org/abs/1810.01429}{{\ttfamily 1810.01429}}].

\bibitem{Blanton:2019igq}
T.~D. Blanton, F.~Romero-L\'opez and S.~R. Sharpe, \emph{{Implementing the
  three-particle quantization condition including higher partial waves}},
  \href{https://doi.org/10.1007/JHEP03(2019)106}{\emph{JHEP} {\bfseries 03}
  (2019) 106} [\href{https://arxiv.org/abs/1901.07095}{{\ttfamily
  1901.07095}}].

\bibitem{Briceno:2019muc}
R.~A. Briceno, M.~T. Hansen, S.~R. Sharpe and A.~P. Szczepaniak,
  \emph{{Unitarity of the infinite-volume three-particle scattering amplitude
  arising from a finite-volume formalism}},
  \href{https://arxiv.org/abs/1905.11188}{{\ttfamily 1905.11188}}.

\bibitem{Romero-Lopez:2019qrt}
F.~Romero-López, S.~R. Sharpe, T.~D. Blanton, R.~A. Briceño and M.~T. Hansen,
  \emph{{Numerical exploration of three relativistic particles in a finite
  volume including two-particle resonances and bound states}},
  \href{https://doi.org/10.1007/JHEP10(2019)007}{\emph{JHEP} {\bfseries 10}
  (2019) 007} [\href{https://arxiv.org/abs/1908.02411}{{\ttfamily
  1908.02411}}].

\bibitem{Hansen:2020zhy}
M.~T. Hansen, F.~Romero-López and S.~R. Sharpe, \emph{{Generalizing the
  relativistic quantization condition to include all three-pion isospin
  channels}},  \href{https://arxiv.org/abs/2003.10974}{{\ttfamily 2003.10974}}.

\bibitem{Blanton:2020jnm}
T.~D. Blanton and S.~R. Sharpe, \emph{{Equivalence of relativistic
  three-particle quantization conditions}},
  \href{https://arxiv.org/abs/2007.16190}{{\ttfamily 2007.16190}}.

\bibitem{Blanton:2020gha}
T.~D. Blanton and S.~R. Sharpe, \emph{{Alternative derivation of the
  relativistic three-particle quantization condition}},
  \href{https://arxiv.org/abs/2007.16188}{{\ttfamily 2007.16188}}.

\bibitem{Doring:2018xxx}
M.~{D{\"o}ring}, H.~W. Hammer, M.~Mai, J.~Y. Pang, A.~Rusetsky and J.~Wu,
  \emph{{Three-body spectrum in a finite volume: the role of cubic symmetry}},
  \href{https://doi.org/10.1103/PhysRevD.97.114508}{\emph{Phys. Rev.}
  {\bfseries D97} (2018) 114508}
  [\href{https://arxiv.org/abs/1802.03362}{{\ttfamily 1802.03362}}].

\bibitem{Pang:2019dfe}
J.-Y. Pang, J.-J. Wu, H.~W. Hammer, U.-G. Mei{\ss}ner and A.~Rusetsky,
  \emph{{Energy shift of the three-particle system in a finite volume}},
  \href{https://doi.org/10.1103/PhysRevD.99.074513}{\emph{Phys. Rev.}
  {\bfseries D99} (2019) 074513}
  [\href{https://arxiv.org/abs/1902.01111}{{\ttfamily 1902.01111}}].

\bibitem{Pang:2020pkl}
J.-Y. Pang, J.-J. Wu and L.-S. Geng, \emph{{$DDK$ system in finite volume}},
  \href{https://arxiv.org/abs/2008.13014}{{\ttfamily 2008.13014}}.

\bibitem{Jackura:2019bmu}
A.~Jackura, S.~Dawid, C.~Fernández-Ramírez, V.~Mathieu, M.~Mikhasenko,
  A.~Pilloni et~al., \emph{{Equivalence of three-particle scattering
  formalisms}}, \href{https://doi.org/10.1103/PhysRevD.100.034508}{\emph{Phys.
  Rev. D} {\bfseries 100} (2019) 034508}
  [\href{https://arxiv.org/abs/1905.12007}{{\ttfamily 1905.12007}}].

\bibitem{Mikhasenko:2019vhk}
M.~Mikhasenko, Y.~Wunderlich, A.~Jackura, V.~Mathieu, A.~Pilloni, B.~Ketzer
  et~al., \emph{{Three-body scattering: Ladders and Resonances}},
  \href{https://doi.org/10.1007/JHEP08(2019)080}{\emph{JHEP} {\bfseries 08}
  (2019) 080} [\href{https://arxiv.org/abs/1904.11894}{{\ttfamily
  1904.11894}}].

\bibitem{Jackura:2018xnx}
{\scshape JPAC} collaboration, A.~Jackura, C.~Fernández-Ramírez, V.~Mathieu,
  M.~Mikhasenko, J.~Nys, A.~Pilloni et~al., \emph{{Phenomenology of
  Relativistic $\mathbf{3} \to \mathbf{3}$ Reaction Amplitudes within the
  Isobar Approximation}},
  \href{https://doi.org/10.1140/epjc/s10052-019-6566-1}{\emph{Eur. Phys. J.}
  {\bfseries C79} (2019) 56}
  [\href{https://arxiv.org/abs/1809.10523}{{\ttfamily 1809.10523}}].

\bibitem{Mai:2018djl}
M.~Mai and M.~{D{\"o}ring}, \emph{{Finite-Volume Spectrum of $\pi^+\pi^+$ and
  $\pi^+\pi^+\pi^+$ Systems}},
  \href{https://doi.org/10.1103/PhysRevLett.122.062503}{\emph{Phys. Rev. Lett.}
  {\bfseries 122} (2019) 062503}
  [\href{https://arxiv.org/abs/1807.04746}{{\ttfamily 1807.04746}}].

\bibitem{Mai:2019fba}
M.~Mai, M.~D{\"o}ring, C.~Culver and A.~Alexandru, \emph{{Three-body unitarity
  versus finite-volume $\pi^+\pi^+\pi^+$ spectrum from lattice QCD}},
  \href{https://arxiv.org/abs/1909.05749}{{\ttfamily 1909.05749}}.

\bibitem{Jackura:2020bsk}
A.~W. Jackura, R.~A. Brice\~no, S.~M. Dawid, M.~H.~E. Islam and C.~McCarty,
  \emph{{Solving relativistic three-body integral equations in the presence of
  bound states}},  \href{https://arxiv.org/abs/2010.09820}{{\ttfamily
  2010.09820}}.

\bibitem{Dawid:2020uhn}
S.~M. Dawid and A.~P. Szczepaniak, \emph{{Bound states in the B-matrix
  formalism for the three-body scattering}},
  \href{https://arxiv.org/abs/2010.08084}{{\ttfamily 2010.08084}}.

\bibitem{Kawai:2017goq}
{\scshape HAL QCD} collaboration, D.~Kawai, S.~Aoki, T.~Doi, Y.~Ikeda,
  T.~Inoue, T.~Iritani et~al., \emph{{$I=2$ $\pi\pi$ scattering phase shift
  from the HAL QCD method with the LapH smearing}},
  \href{https://doi.org/10.1093/ptep/pty032}{\emph{PTEP} {\bfseries 2018}
  (2018) 043B04} [\href{https://arxiv.org/abs/1711.01883}{{\ttfamily
  1711.01883}}].

\bibitem{Doi:2011gq}
{\scshape HAL QCD} collaboration, T.~Doi, S.~Aoki, T.~Hatsuda, Y.~Ikeda,
  T.~Inoue, N.~Ishii et~al., \emph{{Exploring Three-Nucleon Forces in Lattice
  QCD}}, \href{https://doi.org/10.1143/PTP.127.723}{\emph{Prog. Theor. Phys.}
  {\bfseries 127} (2012) 723}
  [\href{https://arxiv.org/abs/1106.2276}{{\ttfamily 1106.2276}}].

\bibitem{Agadjanov:2016mao}
D.~Agadjanov, M.~D{\"o}ring, M.~Mai, U.-G. Mei{\ss}ner and A.~Rusetsky,
  \emph{{The Optical Potential on the Lattice}},
  \href{https://doi.org/10.1007/JHEP06(2016)043}{\emph{JHEP} {\bfseries 06}
  (2016) 043} [\href{https://arxiv.org/abs/1603.07205}{{\ttfamily
  1603.07205}}].

\bibitem{Bulava:2019kbi}
J.~Bulava and M.~T. Hansen, \emph{{Scattering amplitudes from finite-volume
  spectral functions}},
  \href{https://doi.org/10.1103/PhysRevD.100.034521}{\emph{Phys. Rev.}
  {\bfseries D100} (2019) 034521}
  [\href{https://arxiv.org/abs/1903.11735}{{\ttfamily 1903.11735}}].

\bibitem{Guo:2018ibd}
P.~Guo, M.~D{\"o}ring and A.~P. Szczepaniak, \emph{{Variational approach to
  $N$-body interactions in finite volume}},
  \href{https://doi.org/10.1103/PhysRevD.98.094502}{\emph{Phys. Rev.}
  {\bfseries D98} (2018) 094502}
  [\href{https://arxiv.org/abs/1810.01261}{{\ttfamily 1810.01261}}].

\bibitem{Guo:2020spn}
P.~Guo, \emph{{Modeling few-body resonances in finite volume}},
  \href{https://arxiv.org/abs/2007.12790}{{\ttfamily 2007.12790}}.

\bibitem{Guo:2020wbl}
P.~Guo, \emph{{Threshold expansion formula of $N$ bosons in a finite volume
  from a variational approach}},
  \href{https://doi.org/10.1103/PhysRevD.101.054512}{\emph{Phys. Rev. D}
  {\bfseries 101} (2020) 054512}
  [\href{https://arxiv.org/abs/2002.04111}{{\ttfamily 2002.04111}}].

\bibitem{Konig:2017krd}
S.~K\"onig and D.~Lee, \emph{{Volume Dependence of N-Body Bound States}},
  \href{https://doi.org/10.1016/j.physletb.2018.01.060}{\emph{Phys. Lett. B}
  {\bfseries 779} (2018) 9} [\href{https://arxiv.org/abs/1701.00279}{{\ttfamily
  1701.00279}}].

\bibitem{Guo:2017crd}
P.~Guo and V.~Gasparian, \emph{{Numerical approach for finite volume three-body
  interaction}}, \href{https://doi.org/10.1103/PhysRevD.97.014504}{\emph{Phys.
  Rev.} {\bfseries D97} (2018) 014504}
  [\href{https://arxiv.org/abs/1709.08255}{{\ttfamily 1709.08255}}].

\bibitem{Guo:2018xbv}
P.~Guo and T.~Morris, \emph{{Multiple-particle interaction in (1+1)-dimensional
  lattice model}},
  \href{https://doi.org/10.1103/PhysRevD.99.014501}{\emph{Phys. Rev.}
  {\bfseries D99} (2019) 014501}
  [\href{https://arxiv.org/abs/1808.07397}{{\ttfamily 1808.07397}}].

\bibitem{Romero-Lopez:2018rcb}
F.~Romero-López, A.~Rusetsky and C.~Urbach, \emph{{Two- and three-body
  interactions in $\varphi ^4$ theory from lattice simulations}},
  \href{https://doi.org/10.1140/epjc/s10052-018-6325-8}{\emph{Eur. Phys. J.}
  {\bfseries C78} (2018) 846}
  [\href{https://arxiv.org/abs/1806.02367}{{\ttfamily 1806.02367}}].

\bibitem{Beane:2020ycc}
S.~Beane et~al., \emph{{Charged multi-hadron systems in lattice QCD+QED}},
  \href{https://arxiv.org/abs/2003.12130}{{\ttfamily 2003.12130}}.

\bibitem{Gasser:2007zt}
J.~Gasser, V.~E. Lyubovitskij and A.~Rusetsky, \emph{{Hadronic atoms in QCD +
  QED}}, \href{https://doi.org/10.1016/j.physrep.2007.09.006}{\emph{Phys.
  Rept.} {\bfseries 456} (2008) 167}
  [\href{https://arxiv.org/abs/0711.3522}{{\ttfamily 0711.3522}}].

\bibitem{Landau-Lifshits}
L.~Landau and E.~Lifschits, \emph{{Quantum Mechanics}}, vol.~Volume 3 of
  \emph{Course of Theoretical Physics}. Pergamon Press, Oxford, 2004.

\bibitem{phi4:2020}
N.~Schlage, ``Relativistic {$N$-particle} energy shift in finite volume:
  auxiliary fits.'' https://github.com/NikSchlage/phi4\_auxiliary, 2020.

\bibitem{Gasser:1986vb}
J.~Gasser and H.~Leutwyler, \emph{{Light Quarks at Low Temperatures}},
  \href{https://doi.org/10.1016/0370-2693(87)90492-8}{\emph{Phys. Lett. B}
  {\bfseries 184} (1987) 83}.

\bibitem{Luscher:1991n1}
M.~L{\"u}scher, \emph{{Two particle states on a torus and their relation to the
  scattering matrix}},
  \href{https://doi.org/10.1016/0550-3213(91)90366-6}{\emph{Nucl.Phys.}
  {\bfseries B354} (1991) 531}.

\bibitem{Symanzik:1979ph}
K.~Symanzik, \emph{{\textcolor{blue}{Cutoff dependence in lattice $\varphi^4$
  theory in four-dimensions}}},
  \href{https://doi.org/10.1007/978-1-4684-7571-5\_18}{\emph{NATO Sci. Ser. B}
  {\bfseries 59} (1980) 313}.

\bibitem{juqueen}
{J\"{u}lich Supercomputing Centre}, \emph{{JUQUEEN: IBM Blue Gene/Q
  Supercomputer System at the J\"{u}lich Supercomputing Centre}},
  \href{https://doi.org/10.17815/jlsrf-1-18}{\emph{Journal of large-scale
  research facilities} {\bfseries 1} (2015) }.

\bibitem{jureca}
{J\"{u}lich Supercomputing Centre}, \emph{{JURECA: Modular supercomputer at
  J\"{u}lich Supercomputing Centre}},
  \href{https://doi.org/10.17815/jlsrf-4-121-1}{\emph{Journal of large-scale
  research facilities} {\bfseries 4} (2018) }.

\bibitem{juwels}
{J\"{u}lich Supercomputing Centre}, \emph{{JUWELS: Modular Tier-0/1
  Supercomputer at the J\"{u}lich Supercomputing Centre}},
  \href{https://doi.org/10.17815/jlsrf-5-171}{\emph{Journal of large-scale
  research facilities} {\bfseries 5} (2019) }.

\bibitem{R:2019}
{R Core Team}, \emph{R: A Language and Environment for Statistical Computing}.
\newblock R Foundation for Statistical Computing, Vienna, Austria, 2019.

\bibitem{hadron:2020}
B.~Kostrzewa, J.~Ostmeyer, M.~Ueding and C.~Urbach, ``hadron: package to
  extract hadronic quantities.'' https://github.com/HISKP-LQCD/hadron, 2020.

\end{thebibliography}\endgroup

\end{document}